\documentclass[prx,showkeys,showpacs,footnoteinbib,twocolumn,
unsortedaddress]{revtex4-2}
\usepackage{amsmath,amssymb,epic,eepic}
\usepackage[binary-units]{siunitx}
\usepackage[utf8]{inputenc}
\usepackage{graphicx}
\usepackage{hyperref}
\usepackage[capitalise]{cleveref}
\usepackage{todonotes}
\usepackage{bbm}
\usepackage{tikz}
\usepackage{circuitikz}
\usepackage{braket}
\usepackage[caption=false]{subfig}
\usepackage{fontawesome}
\usetikzlibrary{%
        shapes.geometric,calc,intersections,%
        decorations.markings,decorations.text,decorations,%
        patterns,decorations.pathmorphing,fpu,%
                decorations.pathreplacing,%
        positioning,automata,shapes.multipart,circuits,%
        circuits.ee, circuits.ee.IEC, shapes.gates.ee,%
                arrows.meta,arrows
}

\newcommand{\comhub}[1]%
{{\todo[inline,backgroundcolor=red!10!white]{#1}
}}

\newcommand{\mytodo}[1]%
{{\todo[inline,backgroundcolor=blue!10!white]{#1}
}}
\newcommand{\me}{\mathrm{e}}
\newcommand{\mi}{\mathrm{i}}
\newcommand{\md}{\mathrm{d}}

\DeclareMathOperator{\sinc}{sinc}
\DeclareMathOperator{\arctanh}{arctanh}

\begin{document}

\title{Quantum sensing of time dependent electromagnetic fields 
with single electron
excitations}

\author{H. Souquet-Basiège$^1$}
\author{B. Roussel$^2$}
\author{G. Rebora$^1$}
\author{G. Menard$^4$}
\author{I. Safi$^3$}
\author{G. Fève$^4$}
\author{P. Degiovanni$^1$}

\today

\affiliation{(1) Univ Lyon, Ens de Lyon, Universit\'e Claude Bernard 
Lyon 1, CNRS, 
Laboratoire de Physique, F-69342 Lyon, France}
\affiliation{(2) 
Departement of Applied Physics, Aalto University, 00076 Aalto, Finland}
\affiliation{(3) Laboratoire de Physique des Solides (UMR 5802), 
CNRS-Universit\'e Paris-Sud and Paris-Saclay, B\^atiment 510, 91405 Orsay,
France}

\affiliation{(4) Laboratoire de Physique de l’Ecole Normale Sup\'erieure,
ENS, Universit\'e PSL, CNRS, Sorbonne Universit\'e, Universit\'e de Paris,
F-75005 Paris, France}

\begin{abstract}
In this study, we investigate the potential of electronic interferometers
for probing the quantum state of electromagnetic radiation on a chip
at sub-nanosecond time scales. We propose to use single electron
excitations propagating within an electronic
Mach-Zehnder interferometer in the Aharonov-Bohm dominated
regime. We discuss
how information about the quantum state of the electromagnetic
radiation is encoded into the interference contribution to the
average outgoing electrical current. By investigating squeezed radiation
and single edge magnetoplasmons probed by Leviton pulses in a
realistic setup, we show 
that single electron interferometers have the potential to probe
quantum radiation in the time domain with 
sub-nanosecond to pico-second time resolution. 
Our research could have significant implications for probing the
fundamental properties of light in the microwave to tera-Hertz
domains at extremely short time scales.
\end{abstract}



\maketitle


\section{Introduction}

Initiated by the demonstration of on-demand single electron sources
in quantum Hall edge channels
\cite{Feve-2007-1,Dubois-2013-2},
electron quantum optics (EQO) \cite{Bocquillon-2014-1} has seen the
demonstration of several important milestones such as electronic Hanbury
Brown and Twiss (HBT) \cite{Bocquillon-2012-1}, Hong Ou Mandel
(HOM)
\cite{Bocquillon-2013-1} experiments, quantitative studies of
electronic decoherence using HOM \cite{Marguerite-2016-1} and Mach Zehnder interferometry 
\cite{Roulleau-2008-2,Roulleau-2009-1,Tewari-2016-1,Rodriguez-2020-1,Jo-2022-1}.
More recently, a full characterization of the quantum state and
coherences of single electron and hole
excitations within a quantum Hall edge channel has been
performed
\cite{Bisognin-2019-2}.
By demonstrating our ability to access electronic quantum states in a
quantum conductor at an
unprecedented level, these achievements strongly suggest that EQO
is now mature for exploring its applications.
The extreme sensitivity of individual electronic excitations
to their electromagnetic environment suggests to explore the potential
of EQO for the sensing of electromagnetic fields at
the $\si{\micro\meter}$ spatial scale and down to a few
$\si{\pico\second}$ time scale~\cite{Sequoia-WP}. 

This would be of great
interest in the current effort 
to extend the paradigms of
quantum coherent nano-electronics to higher frequencies, possibly up to the $\si{\tera\hertz}$
range \cite{Aluffi-2023-1} since the electromagnetic wavelength ($\si{\milli\meter}$)
is  still much larger than the
size of the device (from $1$ to few tens of $\si{\micro\meter}$). For
example, 
interferometric detection of a single electron at sub-nanosecond time
scale is indeed instrumental in a recent proposal for single shot
detection of an electronic flying qubit \cite{Glattli-2020-1}.
In a broader perspective, solid state systems envisioned for 
quantum technologies \cite{Blais-2021-1,Burkard-2023-1,Edlbauer-2022-1} involve quantum
electromagnetic fields in the $\si{\giga\hertz}$ to $\si{\tera\hertz}$
range. Quantum optics in the $\si{\tera\hertz}$ domain also offers interesting
perspectives for studying quantum materials \cite{Viti-2021-1}. This
calls for sensors that can probe the properties of these fields on the
chip and at such short time scales.

However, sensing and analyzing of local quantum
electric fields on very short time scales is
notoriously difficult.
Rydberg atoms based quantum electrometers \cite{Facon-2016-1} 
have reached record sensitivities but
for static fields.
Fast electrical modulation of optical systems enables measuring the
electric field
up to a $\si{\tera\hertz}$-bandwidth
\cite{Calero-2019-1}.
Unfortunately, such optical systems are not suitable for quantum mesoscopic devices due to
the difficulty of combining optics and microwave electronics within
the same cryostat and the large size of the sensing area.
On-chip
systems based on nano-mechanical resonators
\cite{Cleland-1998-1,Chen-2020-1},
rf-capacitive gate based sensing \cite{Ahmed-2018-1},
NV-centers in diamond \cite{Dolde-2014-1} as well as quantum dots
\cite{Vamivakas-2011-1,Arnold-2014-1}
have been demonstrated, often as
charge sensors able to detect a
single electron charge at a few tens of $\si{\nano\meter}$.
But their bandwidth is still
limited to
$1$ to $\SI{10}{\mega\hertz}$ at best, obtained with rf-SETs
\cite{Brenning-2006-1},
Quantum Point Contacts (QPCs) \cite{Reilly-2007-1},
rf-SQUIDS
\cite{Hatridge-2011-1} or quantum dots \cite{Biercuk-2006-1}.
Moreover, they are
not designed to detect quantum features of the electromagnetic field
such as non-classical fluctuations.

In this paper, we discuss the electron quantum radar (EQR) as a way 
to probe the quantum state of an electromagnetic radiation
using a single electron interferometer. This idea is 
dual to the one underlying radars and coherent lidars which are
electromagnetic interferometers
probing a material target. Here, the ``Electron
Quantum Radar'' (EQR) is an electronic Mach-Zehnder
interferometer (MZI) \cite{Ji-2003-1} where one branch is
capacitively coupled to a radiation channel in which the external
electromagnetic radiation propagates as depicted on Fig.
\ref{fig/simple-radar}. The EQR obtains
information on the quantum state of the radiation
by comparing ballistic propagation within
the ``reference branch'' of the MZI and propagation in the
``target branch'' coupled to the radiation. This is different from
the proposal of Ref.~\cite{Souquet-2014-1} in which non-classical radiation
is probed via photo-assisted tunneling. It is closer to an
interferometric probe of a quantum electromagnetic field by 
matter \cite{Guerlin-2007-1} although, here, it
is not a
quantum non-demolition measurement since the coupling operator is a
quadrature of the field \cite{Braginsky-1980-1}.

As shown in the present work, single 
electron excitation are the most promising probes of quantum radiation
involving single to a few photons
because the associated quantum electrical current generates a very small
back-action on this radiation compared to electrical currents involving more
electronic excitations. Moreover, 
the demonstration of single electron tomography protocols
\cite{Jullien-2014-1,Bisognin-2019-2,Fletcher-2019-1} 
supplemented by quantum current analysis
algorithms
\cite{Roussel-2020-1} as well as the 
generation of wave packets in the tens to few 
$\si{\pico\second}$ range in a variety of systems 
\cite{Hohls-2012-1,Fletcher-2013-1,Waldie-2015-1,Roussely-2018-1,Aluffi-2023-1} 
suggest that the versatility of single
electron sources may enable us to probe 
properties of quantum radiation both in the time and frequency
domains. Most importantly, a Levitonic MZI has recently been
demonstrated in graphene \cite{Assouline-2023-1}.

Motivated by this perspective, a full
theory of the single electron quantum radar 
is presented here. 
The central
result of this paper is the single electron radar equation which gives the
interference contribution to the average electrical current in terms of
the electronic single electron wave-packet and the quantum state of the
electromagnetic radiation. All the dynamics is contained in the linear response
properties of the ```radiation coupler'' coupling the target branch
to the radiation.

This equation accounts for the interferometer's back action
on the incident radiation, which leads to electronic decoherence
\cite{Ferraro-2014-2,Cabart-2018-1}.
We show that all information about the state of the incident quantum radiation
is contained in a Franck-Condon recoil factor \cite{Condon-1926-1} describing the change of the incident
radiation upon the propagation of a time resolved single electron
excitation across the interferometer. This result is established using
the plasmon scattering matrix related to finite frequency admittances 
\cite{Safi-1995-1,Safi-1999-1} to describe the coupling between the
external radiation and the plasmon modes within the target branch of the MZI. 

In the
case of a classical radiation, this recoil factor includes a pure phase
corresponding to the classical electrical potential seen by the electron during its
propagation. When the quantum state of the incident quantum radiation is
Gaussian, 
this recoil factor captures the time-dependent fluctuations in the
mode selected by the capacitive coupling between the radiation and the
electrons. This enables us to give an operational criterion to test if the
fluctuations are sub-vacuum. This result may be of interest
in the light of recent theoretical \cite{Ferraro-2018-1,Rebora-2021-2} and
experimental \cite{Bartolomei-2023-1} studies of squeezing in quantum Hall edge
channels as well as in
tunnel junctions
\cite{Gasse-2013-1,Mendes-2015-1}.
As a last example, we discuss the electron radar signature of single to few edge
magnetoplasmons (EMPs) propagating in a nearby 
quantum Hall edge channel and show that, when the radiation coupler
provides a broadband filtering
of the incoming radiation, the recoil factor in the time domain is 
proportional to the instantaneous average heat
current \cite{Moskalets-2004-1} carried by a single EMP.

This paper is structured as follows: Sec.~\ref{sec/radar-physics} presents the
basic ideas for sensing classical or quantum electromagnetic fields with an electronic
interferometer. In particular, we show the importance of the back-action of the
interferometer when probing quantum electromagnetic radiation.
The theory
of a single electron Mach-Zehnder interferometer with one of its branch
irradiated by a quantum electromagnetic field is then presented in
Sec.~\ref{sec/radar-equation} and its core result -- the single electron radar
equation -- is derived.
Explicit
predictions for classical and quantum radiation (squeezed and Fock
states) probed by Levitons are presented in
Sec.~\ref{sec/examples}.


\section{Sensing electromagnetic fields with electrons}
\label{sec/radar-physics}

In this section, we discuss the physics of the
electronic MZI as a probe of an external radiation at a qualitative
level. Secs. \ref{sec/radar-physics/orders-of-magnitude} and
\ref{sec/radar-physics/simple-model} are devoted to estimating 
the classical phase shift induced on the probing electron 
in the radiation-coupler (or radiation coupler) by 
a nearby propagating single electron current and then a classical time
dependent voltage. Then, in Sec.~\ref{sec/radar-physics/quantum-model},
we consider an incident quantum radiation and we show that, to
avoid decoherence effects, a
compromise must be found on the strength of the electron/radiation
coupling. The discussion of back-action effects also
explains why single electron interferometry is appropriate for
probing the quantum state of mesoscopic quantum radiation involving 
a low average number of
photons.

\subsection{Electromagnetic phase shifts}
\label{sec/radar-physics/orders-of-magnitude}

\begin{figure}
	\centering
	\includegraphics[height=5cm]{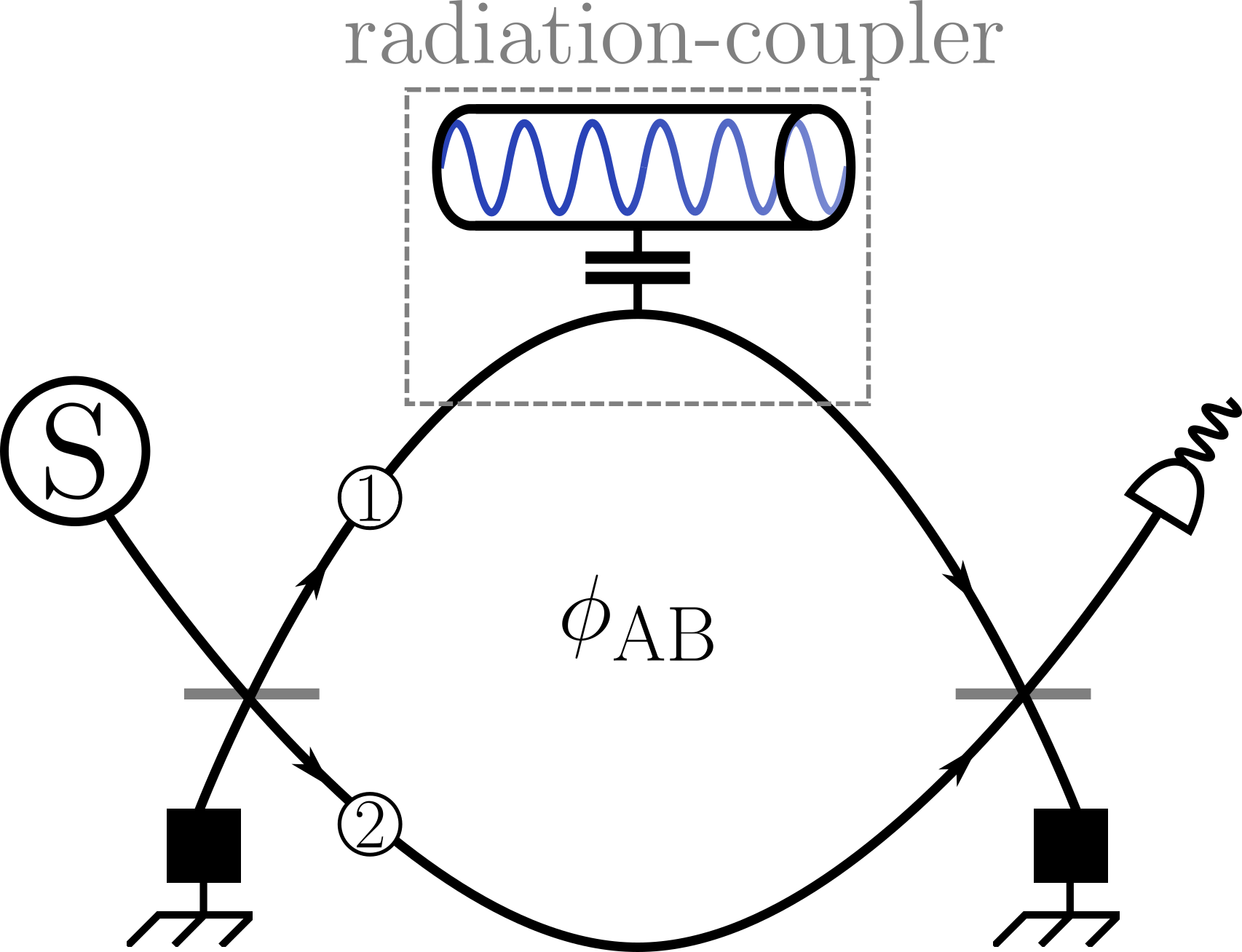}
	\caption{\label{fig/simple-radar} Scheme of principle
	for sensing electromagnetic fields with single electron excitations.
	An electronic Mach-Zehnder interferometer (MZI) is formed by combining two 
	quantum point contacts $A$ and $B$ acting as ideal electronic beam
	splitter. It encloses a magnetic flux $\Phi_{\text{B}}$ inducing an
	Aharonov-Bohm phase $\phi_{\text{AB}}=e\Phi_{\text{B}}/\hbar$. 
	The interferometer is fed by a single electron source $S$
	located right before the first beam splitter $A$. One measures the
	Aharonov-Bohm flux dependent part of the
	average outgoing electrical current $\langle i_{1\text{out}}(t)\rangle $ or
	$\langle \widetilde{\imath}_{1\text{out}}(\omega)\rangle$. The branch $1$ is
	capacitively coupled to external electromagnetic radiation.}
\end{figure}

\begin{figure}
	\includegraphics[width=6cm]{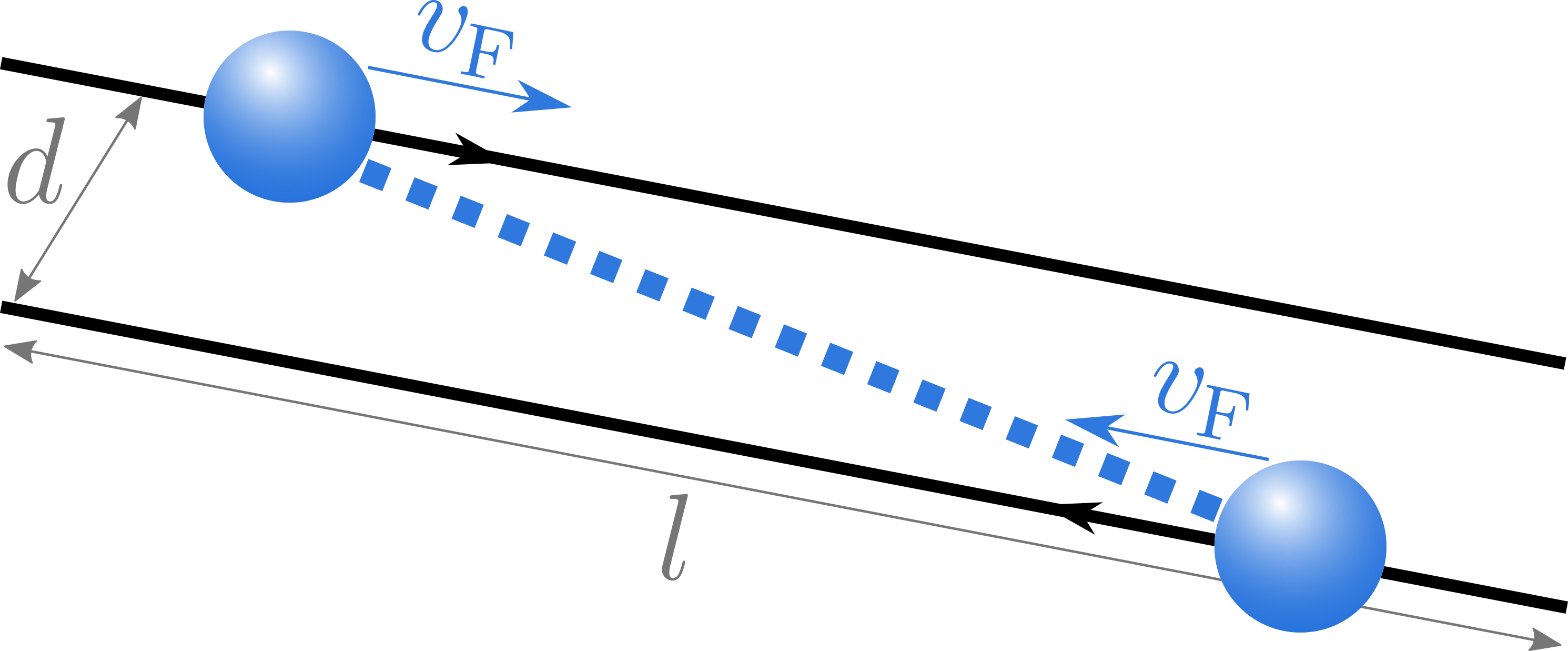}
	\caption{\label{fig/SEC-geometry} Collision of two electrons
	propagating within two counter propagating 1D channels at velocity
	$v_F$, separated by a distance $d$ and feeling Coulomb potential within a region of
	length $l$ in a dielectric material of relative permittivity
	$\varepsilon_{\text{r}}$.}
\end{figure}

Before modeling the MZI, let us estimate the
order of magnitude of the phase shift induced by a nearby electron on
the electron propagating within the MZI. In Ref. \cite{Sequoia-WP}, a ``small electron collider'' 
depicted on Fig. \ref{fig/SEC-geometry} has been considered. 
The
phase shift $\delta\phi_{\text{coll}}$ associated with Coulomb potential for electrons flying
nearby each other at velocity $v_F$ is, up to some
geometric factor, of the order of
the effective fine structure constant within the material
\begin{subequations}
\begin{align}
	\label{eq/delta-phi-coll}
	\delta\phi_{\mathrm{coll}} &= \alpha_{\text{eff}}\,
	\mathrm{arcsinh}\left(l/d\right)\\
	\label{eq/alpha-eff}
	\alpha_{\text{eff}} &= 
	\frac{e^2}{4\pi\varepsilon_0\varepsilon_{\text{r}}\hbar v_F}
	=\frac{\alpha_{\text{qed}}}{\varepsilon_{\text{r}}}\,\frac{c}{v_F}
\end{align}
\end{subequations}
For AlGaAs/AsGa with relative permittivity
$\varepsilon_{\text{r}}=12.9$ and a typical Fermi velocity
$v_F=\SI{e5}{\meter\second^{-1}}$, $\alpha_{\text{eff}}\sim 1.7$ and
therefore, for $l=\SI{1}{\micro\meter}$ and
$d=\SI{100}{\nano\meter}$, $\delta\phi_{\text{coll}}/2\pi\simeq 0.81$
which is not small compared to unity.
Similar phase estimates have also been discussed in two-electron
collision within an HOM interferometer \cite{Ubbelohde-2023-1}. 

This estimate suggests that, using the typical radiation coupler geometry depicted on Fig.
\ref{fig/simple-radar}, a single electron excitation propagating
within the MZI may be able to
detect the time dependent electromagnetic field associated with a single
electron passing by in the radiation channel. 

\subsection{Sensing a classical voltage drive}
\label{sec/radar-physics/simple-model}

To understand how
the shape of the probing electron's wave packet influences its detection
capabilities, we now discuss more precisely the behaviour of the
single electron MZI depicted on 
Fig. \ref{fig/simple-radar}
using single-particle physics.

Two quantum
point contacts (QPC) behave as ideal electronic beam splitters
with scattering matrices
\begin{equation}
S_\alpha=\begin{pmatrix}
\sqrt{T_\alpha} & \mi\sqrt{R_\alpha}\\
\mi\sqrt{R_\alpha} & \sqrt{T_\alpha}
\end{pmatrix}
\end{equation}
for $\alpha=A$ or $\alpha=B$
where $T_\alpha$ and $R_\alpha$ are the transmission and reflection
probabilities ($T_\alpha+R_\alpha=1$). 
A static perpendicular magnetic field $\mathrm{B}$ is applied, generating
an Aharonov-Bohm phase $\phi_{\text{AB}}$ enclosed between the arms $1$ and $2$. 

We assume that
electrons feel a time dependent classical potential $U(t)$ along branch $1$
and propagate ballistically along branch $2$ with time of flight
$\tau_2$. 
The amplitude for a single electron to enter into branch $1$
at time $t'$ and arrive at the second beam splitter at time $t$ is then
given by
\begin{equation}
	\label{eq/classical-voltage/1}
R(t,t')=\delta(t-t'-\tau_1)\,\me^{\frac{\mi
e}{\hbar}\int_{t'}^tU(\tau)\,\md\tau}
\end{equation}
in which $\tau_1$ denotes the ballistic time of flight across branch
$1$. The prefactor is the electric phase accumulated by the electron
during its propagation. We can then compute the probability 
for an electron emitted by source $S$ in a single particle state
$\varphi_e$ to be detected in the
outgoing branch $1$: 
\begin{equation}
	p(1_{\text{out}})=R_AR_B+T_AT_B+\mathcal{P}_q
\end{equation}
in which the quantum interference contribution $\mathcal{P}_q$ is given by
\begin{equation}
	\mathcal{P}_q=\mathcal{K}\Re\left(\me^{\mi\phi_{\text{AB}}}\int_{\mathbb{R}}
	\varphi_e(t-\tau_1)\me^{\frac{\mi
	e}{\hbar}\int_{t-\tau_1}^tU(\tau)\,\md\tau}\varphi_e(t-\tau_2)^*\md
	t\right)
\end{equation}
where
$\mathcal{K}=
\sqrt{R_AR_BT_AT_B}$. 
In the limit of a very short electronic
wave packet emitted at time $t_e$ and for $\tau_1=\tau_2$ we obtain
\begin{equation}
	\mathcal{P}_q \propto\Re\left(\me^{\mi\phi_{\text{AB}}}\me^{\mi\delta\phi_U(t_e)}\right)
\end{equation}
where the phase 
\begin{equation}
\delta\phi_U(t_e)=\frac{e}{\hbar}\int_{t_e}^{t_e+\tau_1}U(\tau)\,\md\tau
\end{equation}
is the electric phase
accumulated during the time interval $[t_e,t_e+\tau_1]$. 
This result
corresponds to our intuition from optics: the interferometer provides a
way to access the phase difference between the two propagation
paths which, in the present case, is directly related to the classical voltage
experienced by the electrons during their propagation along branch $1$.

In order to discuss how an electronic MZI can be used to
probe the quantum state of a radiation, a quantum
description of the radiation/electron coupling is needed. 

\subsection{Interferometric sensing of a quantum system}
\label{sec/radar-physics/quantum-model}

\begin{figure}
\includegraphics[width=6cm]{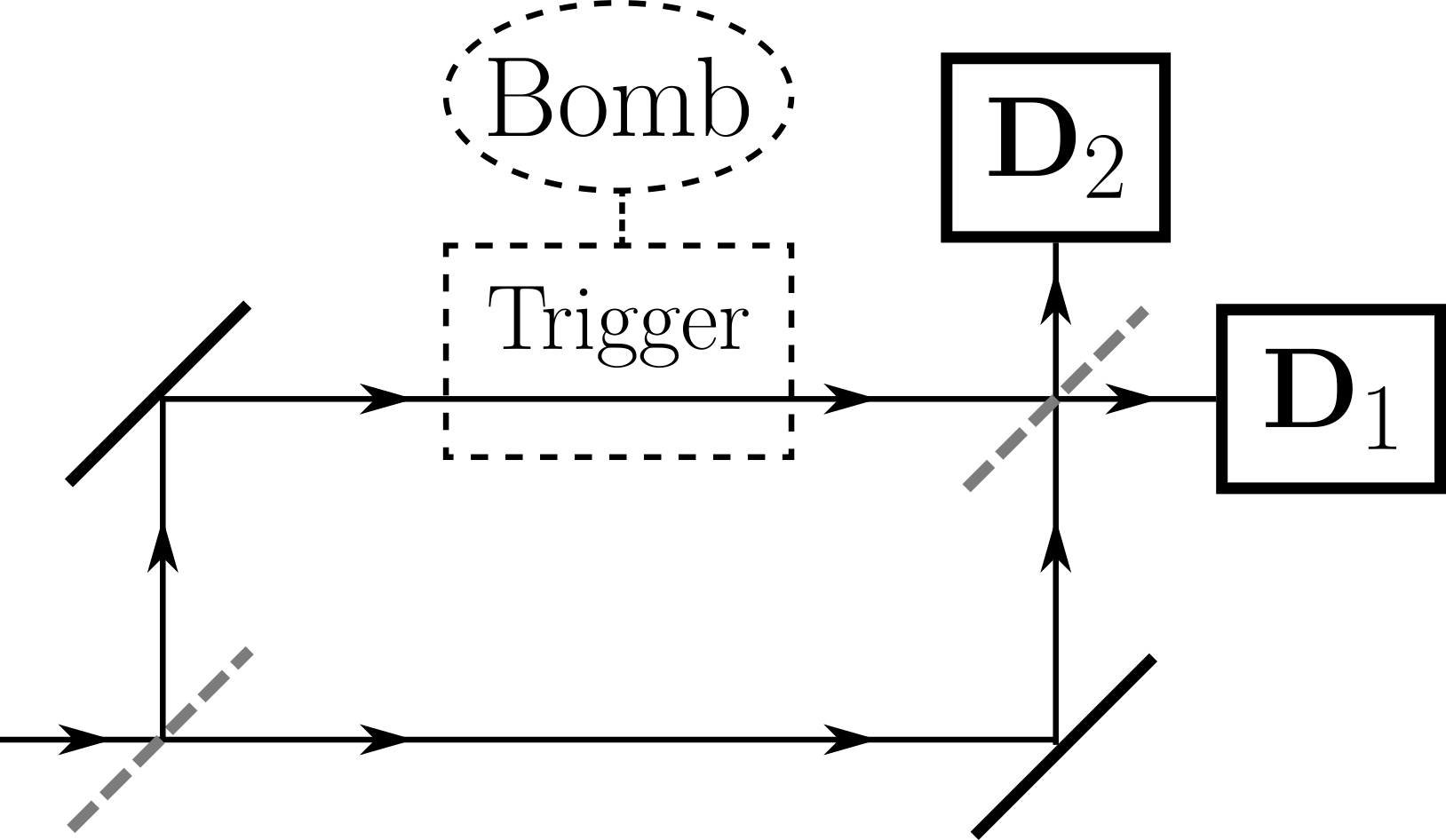}
\caption{\label{fig/elitzur-vaidman} The Elitzur-Vaidman's interferometer: 
in the absence of the bomb,
this single photon Mach-Zehnder interferometer is calibrated so that there 
is no click in $\textbf{D}_2$. In the presence of the bomb, the trigger
absorbs an incident photon with $100\%$ efficiency and detonates the
bomb. Interferences are destroyed and there is a
probability  $1/4$
for the traveling particles to be detected by $\textbf{D}_2$.}
\end{figure}

The simplest way to understand how a quantum interferometer can probe the state
of a quantum system coupled to one of its branches
is to discuss a generalization of the Elitzur-Vaidman 
bomb detector~\cite{Elitzur-1993-1}. 

In this work, a quantum interferometer is used to
detect the presence of a bomb without triggering its explosion. The
bomb's trigger is activated as soon as a particle
travels across branch $1$ of the 
interferometer (see Fig.~\ref{fig/elitzur-vaidman}). This assumes that the bomb's
trigger is a perfectly efficient particle detector. The idea is then to tune
the optical paths of the MZI so that,
in the absence of the bomb,
the particle exits on one of its outgoing branch but not in the other.
In the presence of the bomb, the particle exits with probability
$1/4$ in the branch where it would never exit in the absence of the
bomb.
This provides a sure diagnostic of its presence without
interacting with its trigger, hence the commonly used term ``interaction free
measurement'' to describe this process.

However, in the present context, we are interested into exploiting the
interaction between the particle and the bomb's trigger to gain
information not only about its presence of absence but also about
its quantum state.

The bomb initially being in the
$\ket{\text{Idle}}$ state, interaction with the incident photon leads to
an entangled state involving two alternatives: first, as in Elitzur and
Weidman's work, when the photon is
absorbed, the bomb is detonated. But we also account for a fizzling of the bomb
in which the photon is not absorbed although its quantum state can be
altered. 
The full interaction
between the particle and the bomb is then 
described by the quantum coherent process
\begin{subequations}
    \begin{align}
        \ket{\Psi}\otimes \ket{\text{Idle}} &\longrightarrow
        \mathcal{A}_0\ket{\emptyset}\otimes \ket{\text{Detonated}}
        \nonumber \\
        &+\mathcal{A}_1\ket{\Psi'}\otimes \ket{\text{Fizzled}}
    \end{align}
\end{subequations}
in which $\ket{\emptyset}$ denotes the vacuum state (the particle has been
absorbed) and $\ket{\Psi'}$ the modified state of the photon in case of
fizzling. The 
Elitzur-Vaidman interaction free
measurement corresponds to $|\mathcal{A}_0|^2=1$ but, here both
alternatives are allowed and thus
$|\mathcal{A}_0|^2+|\mathcal{A}_1|^2=1$.
For balanced beam splitters, the conditional output probabilities are then given by the following
expressions :
\begin{subequations}
    \begin{align}
        p(\text{No
        Particle})&=\frac{1}{2}(1-|\mathcal{A}_1|^2)\\
		p(1_\text{out})
        &=\frac{1}{4}(1+|\mathcal{A}_1|^2)
		-\mathcal{P}_{\text{q}}\\
		p(2_{\text{out}})
        &=\frac{1}{4}(1+|\mathcal{A}_1|^2)
		+\mathcal{P}_{\text{q}}
    \end{align}
\end{subequations}
in which
quantum interference effects are contained in 
\begin{equation}
	\mathcal{P}_{\text{q}}=\frac{1}{2}\Re\left(\mathcal{A}_1\me^{\mi\phi_{\text{AB}}}
        \braket{\text{Idle}|\text{Fizzled}}\braket{\Psi|\Psi'}\right)
	\label{eq/p_q}
\end{equation}
which is sensitive to the (Aharonov-Bohm) phase difference $\phi_{\text{AB}}$
associated with free propagation along the two branches of the MZI.
Information
on the quantum state of the bomb is contained in
the product of the two overlaps
$\braket{\Psi|\Psi'}$ and $\braket{\text{Idle}|\text{Fizzled}}$.

From our perspective, the bomb plays the role of the incoming
electromagnetic radiation
and the particle is the quantum electrical current propagating within
the MZI. The
process in which the particle
is absorbed and the bomb is detonated
corresponds to a full electronic decoherence within the MZI. It occurs
whenever Coulomb interactions lead to the
generation of any extra electron/hole pair within the channel $1$ of the
MZI compared to ballistic propagation along channel $2$. 
The resulting many-body state will then have a vanishing overlap with the ballistic
propagation of a single electron excitation within branch $2$ of the
MZI. However, this is not the case of interest for the electron radar
since, in this case, the interference contribution
to the average current vanishes and, as in the Elitzur-Vaidman case,
nothing can be learned on the quantum state of the incoming radiation. 

By contrast, in the absence of generation of extra electron/hole pair particles,
the state of the electron propagating within branch $1$ is altered by
its coupling to the incoming radiation: this corresponds to the
change $\ket{\Psi}\mapsto \ket{\Psi'}$ in the above discussion. For a classical
radiation, this is the phase shift associated with the voltage
experienced by the electrons (see Sec.
\ref{sec/radar-physics/simple-model}).

The alteration of the bomb's state $\ket{\text{Idle}}\mapsto
\ket{\text{Fizzled}}$ in the above discussion corresponds to the effect of the
propagating electron on the incident radiation in the situation
where no extra-electron/hole pairs are created. 
This is the back action of the
interferometer, seen as a measurement device, on the radiation. 
The amplitude
$\braket{\text{Idle}|\text{Fizzled}}$ that measure the ``quantum recoil'' of the
radiation upon propagation of a single electron without
generating extra electron/hole pairs along 
branch $1$ of the MZI.
As we shall
see in Sec. \ref{sec/radar-equation}, this is
precisely the part that will contain information on the quantum
fluctuations of the incoming radiation.

When the
back-action is too important, the overlap $\braket{\text{Idle}|\text{Fizzled}}$ may vanish and the
interference signal is then lost. 
This point explains why quantum electrical
currents carrying a single electronic excitations are relevant for probing mesoscopic
quantum electromagnetic fields which involve a low average number of
photons: besides decoherence which is generically less
important for them, single electron currents lead to a smaller back-action than
currents carrying more excitations. 

In the end, this qualitative discussion suggests that, 
extracting information on the incoming radiation thus requires a
compromise:
electronic decoherence as well as back-action on the quantum
radiation has to be moderate to ensure an experimentally accessible
experimental signal but strong enough to ensure sensitivity to the
incoming quantum radiation.

\section{The single electron radar theory}
\label{sec/radar-equation}

In this section, we derive the central result of this work which is the
single electron radar equation expressing the interference contribution
to the outgoing average current in terms of two distinct quantities:
firstly, the
excess single electron coherence of the injected wave packet which
depends only on the electronic source $S$. Secondly, 
an effective single particle
scattering amplitude describing both the effects of electronic
decoherence and of the incoming radiation. The latter quantity 
reflects the dynamics of
the interferometer coupled to its electromagnetic environment.

Although this result is formally derived in Appendix
\ref{appendix/radar}
using bosonization of chiral quantum Hall edge channels, a more intuitive derivation 
is presented in Secs. \ref{sec/radar-equation/radiation coupler} and
\ref{sec/radar-equation/result}. 

\subsection{Radiation coupler modeling}
\label{sec/radar-equation/radiation coupler}

The radiation coupler involves a capacitive
coupling between a portion of the upper branch of the electronic MZI 
and the external radiation channel which is fed by the incoming electromagnetic radiation we
want to study. Effect of Coulomb interactions within this region
will be described within the framework of edge-magnetoplasmon
scattering. Originally introduced in the context of finite frequency quantum transport in
1D \cite{Safi-1995-1,Safi-1999-1,Safi-1995-2},
it enables us to describe the scattering between the EMP modes
propagating along the MZI edge channel and the bosonic modes within the external radiation
channel \cite{Degio-2009-1}.

\begin{figure}
        \begin{tikzpicture}
[
edge channel/.style={%
			thick
		},
		edge channel dir/.style={%
			thick,
			decoration={markings,mark=at position 0.55 with 
{\arrow{stealth}}},
			postaction={decorate}%
		},
		edge channel dir param/.style={%
			thick,
			decoration={markings,mark=at position #1 with {\arrow{stealth}}},
			postaction={decorate}%
		}%
	]

\def\lc{1.2}

%
\def\sep{0.5}
\begin{scope}[shift={(-3,0)}]
	\draw[thick](0,0) -- (\lc,0);
	\draw[edge channel dir] (-\lc/4,-\lc/4) -- (0,0);
	\node[below] () at (-\lc/4,-\lc/4) {$b_{\text{in}}$};
	\draw[edge channel dir] (\lc,0) -- ({\lc + \lc/4},-\lc/4);
	\node[below] () at ({\lc + \lc/4},-\lc/4) {$b_{\text{out}}$};
	\draw[thick] (0,\sep) -- (\lc,\sep);
	\draw[edge channel dir] (0,\sep) -- (-\lc/4,\sep + \lc/4);
	\node[above] () at (-\lc/4,\sep + \lc/4) {$a_{\text{out}}$};
	\draw[edge channel dir] ({\lc + \lc/4},\sep + \lc/4) -- (\lc,\sep);
	\node[above] () at ({\lc + \lc/4},\sep + \lc/4) {$a_{\text{in}}$};
	\foreach \i in {1,...,5}{
		\begin{scope}[shift={(-\lc/10+\lc/5*\i,0)}]
		\draw[dashed] (0,0) -- (0,\sep);
		\end{scope}
	}
	\node[below] at (\lc/2,-0.75) {(a)};
\end{scope}

\def\sep{0.3}
\def\size{0.5}
\begin{scope}[shift={(0,0)}]
	\draw[thick](0,0) -- (\lc,0);
	\draw[edge channel dir] (-\lc/4,-\lc/4) -- (0,0);
	\node[below] () at (-\lc/4,-\lc/4) {$b_{\text{in}}$};
	\draw[edge channel dir] (\lc,0) -- ({\lc + \lc/4},-\lc/4);
	\node[below] () at ({\lc + \lc/4},-\lc/4) {$b_{\text{out}}$};
	
	\draw[-stealth,thick] (\lc/5,4*\sep) -- (\lc/5,2.5*\sep) 
node[midway,left] {$a_{\text{in}}$};
	\draw[-stealth,thick] (4*\lc/5,2.5*\sep) -- (4*\lc/5,4*\sep) 
node[midway,right] {$a_{\text{out}}$};
	
	\draw[thick] (\lc/3,\sep) -- (2*\lc/3,\sep);

	\foreach \i in {1,...,2}{
		\begin{scope}[shift={(\lc/3-\lc/12+\lc/6*\i,0)}]
		\draw[dashed] (0,0) -- (0,\sep);
		\end{scope}
	}
	
	\draw  ({\lc/2+0.5*cos(120)},{4*\sep + \size + 0.5*sin(-120)}) --
	({\lc/2+0.5*cos(120)},{4*\sep + 0.5*sin(-120)}) 
arc(-120:-60:0.5)
-- ({\lc/2+0.5*cos(60)},{4*\sep + \size + 0.5*sin(-60)});
\draw ({\lc/2+0.5*cos(60)},{4*\sep + 0.5*sin(-60)}) 
arc(60:120:0.5);
	\draw[thick] (\lc/2,\sep) -- (\lc/2,{4*\sep+0.5*sin(-120)});
	
	\node[right] () at (\lc/2,1.7*\sep) {$Z(\omega)$};
	
		\node[below] at (\lc/2,-0.75) {(b)};
\end{scope}

\def\sep{0.5}
\begin{scope}[shift={(3,0)}]
	\draw[edge channel dir] (-\lc/2,0) -- (0,0)
	node[below,midway] {$b_{\text{in}}$};
	\draw[edge channel dir] (\lc,0) -- ({\lc + \lc/2},0)
	node[midway, below] {$b_{\text{out}}$};
	\draw[edge channel dir] (-\lc/2,\sep) -- (0,\sep)
	node[above,midway] {$a_{\text{in}}$};
	\draw[edge channel dir] (\lc,\sep) -- ({\lc + \lc/2},\sep)
	node[midway, above] {$a_{\text{out}}$};
	
	\draw (0,-\sep/3) rectangle (\lc,4*\sep/3);
	\node () at (\lc/2,\sep/2) {$S(\omega)$};

	\node[below] at (\lc/2,-0.75) {(c)};
\end{scope}

\end{tikzpicture}
        \caption{\label{fig/EMP-scattering} (Color online) The edge-magnetoplasmon
scattering
approach describes many situations, such as for example (a) two
counter-propagating edge channels
capacitively coupled over a distance $l$,
(b) a chiral edge channel
capacitively coupled to a linear external circuit described by a
frequency dependent impedance $Z(\omega)$. (c) Solving the equation
of motions leads to a frequency dependent scattering matrix $S(\omega)$
between the channel's edge-magnetoplasmon modes and the bosonic modes of
the other system.}
\end{figure}
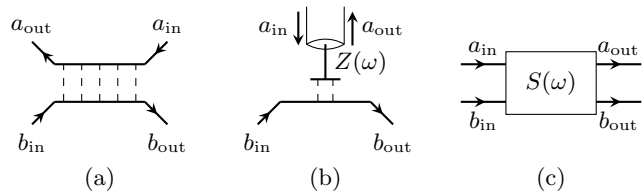

For simplicity, we assume that the whole upper branch of the MZI
is included in the radiation coupler so that the EMP scattering matrix will
account for the detailed geometry of the sample. We also assume 
that only the EMP modes associated with the $b(\omega)$ and
$b^\dagger(\omega)$ destruction and creation operators 
and the external radiation modes associated with
the $a(\omega)$ and $a^\dagger(\omega)$ operators appear in the
scattering matrix $S(\omega)$:
\begin{equation}
\begin{pmatrix}
a_{\text{out}}(\omega) \\
b_{\text{out}}(\omega)
\end{pmatrix}
	=\begin{pmatrix}
		S_{aa}(\omega) & S_{ab}(\omega)\\
		S_{ba}(\omega) & S_{bb}(\omega)
	\end{pmatrix}\,
\begin{pmatrix}
a_{\text{in}}(\omega) \\
b_{\text{in}}(\omega)
\end{pmatrix}\,.
\end{equation}
Depending on the design of the radiation coupler, the $(a,a^\dagger)$ modes may be
photonic (case of Fig. \ref{fig/EMP-scattering}-(b)) or
edge-magnetoplasmonic (see Fig.  \ref{fig/EMP-scattering}-(a)).
These matrix elements are related to finite frequency
admittances 
\cite{Safi-1995-1,Safi-1999-1,Degio-2010-1}. In principle, they could
be inferred from experimental 
measurements~\cite{Bocquillon-2013-2}.
Quantitative predictions can also be made from theoretical models of
the radiation coupler. Experimentally relevant examples include the case of two
counter-propagating edge channels (see Fig.
\ref{fig/EMP-scattering}-(a)) as well as the case of a
capacitive coupling to a transmission line (see
Fig. \ref{fig/EMP-scattering}-(b)). Computations for
the case of two 
counter-propagating edge channels in total mutual influence (see Fig.
\ref{fig/EMP-scattering}-(a)) are detailed in
Appendix \ref{appendix/two-channels-EMP-scattering}. This specific
radiation coupler model will be used all along the present manuscript to
illustrate explicit examples.

\subsection{The single electron radar equation}
\label{sec/radar-equation/result}

\subsubsection{Single particle scattering approach}
\label{sec/radar-equation/result/single-particle}

We now derive the radar equation assuming that electronic
propagation inside the radiation-coupler is described in terms of time
dependent single particle scattering. We are neglecting 
electron-hole pairs creation by Coulomb interactions within the MZI. We also assume that
the two QPCs are ideal electronic
beam splitters with energy independent scattering matrix.
Denoting by $R(t,t')$ the
amplitude for an electron to enter branch $1$ at time $t'$ and exit it
at time $t\geq t'$, we can compute the Aharonov Bohm flux dependent
contribution to the outgoing time dependent average electrical
current $\langle i_{1_{\text{out}}}(t)\rangle$. In a MZI
interferometer, its Aharonov-Bohm phase dependence reduces to
\begin{equation}
	\langle i_{1_{\text{out}}}(t)\rangle =-e\left(
	I_0(t)+\me^{\mi\phi_{\text{AB}}}I_+(t)+
	\me^{-\mi\phi_{\text{AB}}}I_+(t)^*\right)\,.
\end{equation}
where $ \phi_{\text{AB}}=2\pi\Phi_B/\Phi_0$ ($\Phi_0=h/e$ being the
flux quantum).
Denoting by $\mathcal{A}(S\xrightarrow j 1_{\text{out}})$ the amplitude
for an electron to be emitted by the source $\mathrm{S}$, propagating along branch $j$ between
the two QPCs and be detected in $1_\text{out}$, we have
\begin{equation}
	I_+(t)=v_F \mathcal{A}(S\xrightarrow 1 1_{\text{out}})\, 
	\mathcal{A}(S\xrightarrow 2 1_{\text{out}})^*\\
\end{equation}
with
\begin{subequations}
\begin{align}
\mathcal{A}(S\xrightarrow 1 1_{\text{out}})&=
	\sqrt{T_AT_B}\,\int_{\mathbb{R}}R(t,t')\varphi_e(t')\,\md
	t'\\
	\mathcal{A}(S\xrightarrow 2 1_{\text{out}})&=-\sqrt{R_AR_B}\,\varphi_e(t-\tau_2)\,.
\end{align}
\end{subequations}
Note that, in the present context, 
$2\,\Re(I_+(t)\,\me^{\mi\phi_{\text{AB}}})$ plays
the role of $\mathcal{P}_q$ in
Sec.~\ref{sec/radar-physics}.
The quantity $X_+(t)$ defined by
$I_+(t)=-\sqrt{R_AT_AR_BT_B}\,X_+(t)$ does not depends on the
properties of the electronic beam splitters and completely determines
the interference contribution to the average electrical current.
It is equal to
\begin{equation}
	\label{eq/radar-equation/first-occurence}
	X_+(t)=v_F\int_{\mathbb{R}}R(t,t')\varphi_e(t')\,\varphi_e(t-\tau_2)^*\md
	t'\,.
\end{equation}
This result
is called the electronic radar equation by analogy with the
existing signal processing literature \cite{Book-Woodward}.
Its integral over time
\begin{equation}
	X_+^{(\text{dc})}=\int_{\mathbb{R}}X_+(t)\,\md t\,.
\end{equation}
represents the interference contribution to the total charge
detected on the output $1$ of the MZI. For a balanced MZI
($T_A=T_B=1/2$), the average dc current measured when the
experiment is repeated at measurement frequency $f_{\text{m}}$ is given
by:
\begin{equation}
\langle
i_{1_{\text{out}}}^{(\text{dc})}\rangle
	=-\frac{ef_{\text{m}}}{2}\left(1+\Re\left(\me^{\mi\phi_{AB}}X_+^{(\text{dc})}\right)\right)\,.
\end{equation}

\subsubsection{Coulomb interaction effects}
\label{sec/radar-equation/result/intuition}

The discussion of Sec.~\ref{sec/radar-physics/quantum-model} suggests
that the result in 
presence of Coulomb interactions and of an incoming quantum radiation is
of
the same form than Eq.~\eqref{eq/radar-equation/first-occurence} with an effective single electron scattering
amplitude $R_{\text{eff}}(t,t')$ that takes into account decoherence effects as well as the
electron's back-action on the incoming quantum radiation. The formal
derivation of this fact is given in Appendix \ref{appendix/radar} but,
here,  we will
proceed along obtaining the form of the effective scattering amplitude using
semi-qualitative arguments to emphasize its physical meaning.

In the absence of incoming electromagnetic radiation, the amplitude
$R_{\text{eff}}(t,t')$ is the amplitude for a single electron to propagate
elastically across the branch $1$, taking into account the capacitive
coupling to the radiation coupler's radiation channel. Any process leading
to the generation of an extra electron/hole pair within the edge channel
or of an excitation within the electromagnetic environment would lead to
decoherence in the MZI.
This elastic scattering amplitude has already
appeared in studies of electronic decoherence in the MZI 
\cite{Chalker-2007-1,Levkivskyi-2008-1,Neuenhahn-2008-1,Neuenhahn-2009-1}.
Electronic decoherence in MZI has been experimentally simulated using a voltage
probe in Ref.~\cite{Roulleau-2009-1}, thereby showing the reduction of the
interferometer's contrast by the amplitude (square root of the
probability) for the electron to be
transmitted across the probe.

In the absence of external radiation, $R_{\text{eff}}(t,t')$ should
then be
the elastic scattering amplitude 
for an incoming single electron excitation injected at time $t'$ in the
presence of the Fermi sea to exit at time $t$ without experiencing any
inelastic scattering:
$R_{\text{eff}}(t,t')=\Theta(t-t')\,\mathcal{Z}_{1}(t-t')$ where
\begin{equation}
	\label{eq/radar-equation/Z-time-domain}
	\mathcal{Z}_1(\tau)=\int_0^{+\infty}\widetilde{\mathcal{Z}}_{1}(\omega)
	\,\me^{-\mi\omega\tau}\frac{\md\omega}{2\pi}\,.
\end{equation}
is the Fourier transform of the elastic scattering amplitude 
$\widetilde{\mathcal{Z}}_1(\omega)$ for a single electron of energy
$\hbar\omega>0$ across
branch $1$ of the MZI. This quantity, computed in 
Refs.~ \cite{Degio-2009-1,Degio-2011-1,Cabart-2018-1}, has recently been
reconsidered in the light of 
recent experimental studies of electronic relaxation
\cite{Tewari-2016-1,Rodriguez-2020-1,Rebora-2021-1}. 
Going back to the discussion of Sec.
\ref{sec/radar-physics/quantum-model}, $\mathcal{Z}_{1}(\tau)$
corresponds to the product
$\mathcal{A}_1\,
\braket{\Psi|\Psi'}\,\braket{\text{Idle}|\text{Fizzled}}$
when the radiation channel is fed with the vacuum state.

We now have to discuss 
the back-action of the electron in the presence of 
incoming radiation. The back action on the vacuum is already taken into
account by the elastic scattering amplitude $\mathcal{Z}_1(\tau)$ or
$\widetilde{\mathcal{Z}}_1(\omega)$ but here, we consider the effect of the
single electron current on the incoming photonic excitations.

An outgoing electron at time $t$ corresponds, in terms of EMP, to a
localized current pulse coming out of the branch $1$. 
Such a current
pulse comes from an incoming pulse in the edge channel $1$ as well as
some coherent pulse in the radiation channel whose amplitude can be
inferred from the scattering matrix $S(\omega)$. Denoting by
$\Lambda_t(\omega)=-\me^{\mi\omega t}/\sqrt{\omega}$ the corresponding amplitude of the mode
$b_{\text{out}}(\omega)$, the amplitudes of the incoming
pulses are given by:
\begin{subequations}
	\label{eq/backaction-kicks}
	\begin{align}
		\langle b_{\text{in}}(\omega)\rangle &=
		S_{bb}^*(\omega)\,\Lambda_t(\omega)\label{eq/backaction-kicks/1}\\
		\langle a_{\text{in}}(\omega)\rangle &=
		S_{ba}^*(\omega)\,\Lambda_t(\omega)\,.
		\label{eq/backaction-kicks/2}
	\end{align}
\end{subequations}
Eq. \eqref{eq/backaction-kicks/2} gives the amplitude of the
``back-action kick'' on the incoming radiation associated
with the detection of a electron localized at time $t$ at the
end of the upper MZI branch. Consequently, the unitary operator
representing the back-action is the infinite dimensional displacement
operator $D_{a_{\text{in}}}\left[S_{ba}^*\Lambda_t\right]$ (see
Eq.~\eqref{eq/bosonization/displacement-operator} for its definition) 
associated with the back-action kick
$S_{ba}^*\Lambda_t$. 

The full
back-action factor is then the average value of 
$D_{a_{\text{in}}}\left[S_{ba}^*\Lambda_t\right]$ 
in the reduced density operator
$\rho_{\text{em}}$ describing the state of the incoming quantum radiation fed into
the radiation coupler. But the contribution in the vacuum state $\ket{0}$ for
the $a_{\text{in}}(\omega)$ modes is already included in
$\mathcal{Z}_1(t)$. 
Consequently,
the excess back-action on the incoming radiation
is obtained by dividing $\langle
D_{a_{\text{in}}}\left[S_{ba}^*\Lambda_t\right]\rangle_{\rho_{\text{em}}}$ by $\langle
D_{a_{\text{in}}}\left[S_{ba}^*\Lambda_t\right]\rangle_{\ket{0}}$.
This leaves us with the average value of the bosonic normal ordered
back-action displacement operator for the 
prefactor representing the recoil of the incident radiation
induced by a single electron excitation propagating across the upper
branch of the MZI. 
We thus define the Franck-Condon factor
\begin{equation}
	\label{eq/radar-equation/F-definition}
	\mathcal{F}_{\rho_{\text{em}}}(t)=
	\Big\langle
	:D_{a_{\text{in}}}\left[S_{ba}^*\Lambda_t\right]:\Big\rangle_{\rho_{\text{em}}}\,.
\end{equation}
by analogy with the Franck-Condon factor that appears in
the spectroscopy of complex molecules \cite{Condon-1926-1} or in the
Mössbauer effect \cite{Singwi-1960-1,Tzara-1961-1} where it is called
the Lamb-Mössbauer factor. As noticed in Ref.~\cite{Dasenbrook-2016-3}
and explicitly derived here in Appendix \ref{appendix/FCS}, the Franck-Condon
factor contains
information on the Full Counting Statistics of
charge propagating within the radiation channel. More specifically, when
the radiation channel is a chiral quantum Hall edge channel at $\nu=1$, 
\begin{equation}
    \label{eq/radar/non-linear-1e/FCS-expression}
    \mathcal{F}_{\rho}(t)=\Big\langle :\,\me^{2\pi
    \mi N(t)}\,:\Big\rangle_{\rho_{\text{em}}}\,.
\end{equation}
in which the
$N(t)$ operator represents a filtering of the charge operator flowing
across the interaction region within the radiation channel. 

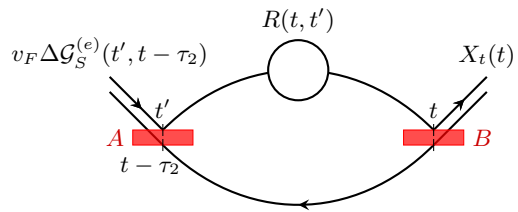
\begin{figure}
\centering
	\begin{tikzpicture}[beamsplit/.style={red,fill=red,fill opacity=0.7},
		ray/.style={thick},
		arrowat/.style={
		decoration={
			markings,
			mark=at position #1 with {\arrow{stealth}}},
		postaction={decorate}}]
	\def\bsthick{0.1}
	\def\bswidth{0.4}
	\def\bssplit{1.8}

	\def\raylength{1}
	\pgfmathsetmacro{\ticklength}{0.2*\bswidth}

	\node[left,above=+3pt,font=\small] at ($ (-\bssplit, 0) + (90+45:\raylength) $)
		{$v_F \Delta \mathcal{G}^{(e)}_{S}(t', t - \tau_2)$};
	\node[right,above=+3pt,font=\small] at ($ (\bssplit, 0) + (45:\raylength) $)
		{$X_t(t)$};

	\draw[ray,arrowat={0.1},arrowat={0.93}]
		(-\bssplit, \bsthick) +(90+45:\raylength) --
		(-\bssplit, \bsthick) .. controls +(45:1.5) and +(90+45:1.5) ..
		(\bssplit, \bsthick) node[midway,fill=white,circle,draw,minimum height=0.8cm] (R) {}
		-- +(45:\raylength)
		;
	\node[above=-1pt] at (R.north) {$R(t,t')$};

	\draw[ray,xscale=-1,arrowat={0.5}] (-\bssplit, -\bsthick) +(+90+45:\raylength) --
		(-\bssplit, -\bsthick) .. controls +(-45:1.5) and +(-90-45:1.5) ..
		(\bssplit, -\bsthick) -- +(45:\raylength)
		;

	\draw[beamsplit]
		(-\bswidth-\bssplit,-\bsthick) rectangle (+\bswidth-\bssplit,+\bsthick); 
	\node[left, color=red!80!black] at (-\bswidth-\bssplit, 0) {$A$};
	\draw[beamsplit]
		(-\bswidth+\bssplit,-\bsthick) rectangle (+\bswidth+\bssplit,+\bsthick); 
	\node[right, color=red!80!black] at (+\bswidth+\bssplit, 0) {$B$};

	\draw (-\bssplit, \bsthick -\ticklength) -- ++(0, 2*\ticklength)
		node[above=-1pt] {$t'$}; 
	\draw (-\bssplit, -\bsthick -\ticklength) -- ++(0, 2*\ticklength)
		node[at start,below left=-1pt and -10pt] {$t - \tau_2$}; 
	\draw (\bssplit, \bsthick -\ticklength) -- ++(0, 2*\ticklength)
		node[above=-1pt] {$t$}; 
	\draw (\bssplit, -\bsthick -\ticklength) -- ++(0, 2*\ticklength);
\end{tikzpicture}
\caption{\label{fig/electron-radar/interpretation/time}
Physical interpretation of the linear radar equation
\eqref{eq/radar-equation/time-domain} in
the time domain showing the contribution of incoming single
electron coherence to the signal
the temporal signal
$X_+(t)$.
The figure presents the product of two
quantum amplitude: the one with an arrow oriented away from the source
corresponds for a direct amplitude whereas the one, with the arrow
arriving to the source corresponds to the complex conjugated amplitudes
contributing to $X_+(t)$.
}
\end{figure}

The quantum interference contribution to the average outgoing electrical
current then has the form
\begin{equation}
		\label{eq/radar-equation/time-domain}
		X_+(t)=v_F\int_{-\infty}^t
	R_{\text{eff}}(t,t')\,
	\varphi_e(t')\,
	\varphi_e(t-\tau_2)^*\,\md
	t'
\end{equation}
in which the effective single particle scattering amplitude is given by
\begin{equation}
	\label{eq/radar-equation/time-domain/2}
		R_{\text{eff}}(t,t')=
	\mathcal{Z}_1(t-t')\mathcal{F}_{\rho_\text{em}}(t)\,.
\end{equation}
In the case where independent repeated single electron experiments are
performed, one can replace the product
$\varphi_e(t')\varphi_e(t-\tau_2)^*$ by the excess single electron
coherence $\Delta\mathcal{G}^{(e)}_S(t'|t-\tau_2)$ emitted by the
imperfect single electron sources. This takes into account the statistical
fluctuations in the imperfect emission of single electron excitations by
the source $S$.
Eq.~\eqref{eq/radar-equation/time-domain}
can then be
interpreted as described on
Fig.~\ref{fig/electron-radar/interpretation/time}. 
Appendix \ref{appendix/frequency-domain} discusses the frequency domain
form of the single electron radar equation and its interpretation.

Expression \eqref{eq/radar-equation/time-domain/2} takes into account
electronic decoherence and the effect of the incoming quantum radiation.
Together with the expressions for 
$\mathcal{Z}_1(\tau)$ and for the Franck-Condon factor,
Eqs.~\eqref{eq/radar-equation/time-domain} and
\eqref{eq/radar-equation/time-domain/2} form the central result of the
single electron radar theory. 
The factorized form of the r.h.s. of Eq.~\eqref{eq/radar-equation/time-domain/2} has an important
consequence: the Franck-Condon factor appears as the ratio the time-resolved electron radar
signal in the presence of the external radiation to the one in the
absence of it:
\begin{equation}
    \label{eq/radar-equation/time-domain/relative-contrast}
    \frac{\left[X_+(t)\right]_{\rho_{\text{em}}}}{\left[X_+(t)\right]_{\ket{0}}}
    =\mathcal{F}_{\rho_{\text{em}}}(t)\,.
\end{equation}
However, high precision time domain measurements of the average current with very large
bandwidth are difficult to perform. We thus have to discuss
how Eqs.~\eqref{eq/radar-equation/time-domain} and
\eqref{eq/radar-equation/time-domain/2} can be used to interpret the
experimental data obtained by recording interference fringes on the dc
average current which is the usual quantity measured
with high precision.

\subsubsection{The dc-current interference contrast}
\label{sec/electron-radar/Levitons}

Assuming that the source could inject a very short electronic
wave packet of duration $\tau_e$ at time $t_e$, the electron radar signal 
in the presence of external radiation would (naively) be given by (see Appendix
\ref{appendix/limiting-regimes}) $[X_+^{(\text{dc})}]_{\rho_{\text{em}}}\simeq
\tau_e\,\mathcal{Z}(\tau_2)\mathcal{F}_{\rho_{\text{em}}}(t_e+\tau_2)$.
Comparing the result in the presence and in the absence of
electronic radiation ($\mathcal{F}_{\ket{0}}(t)=1$) injected into the radiation channel then leads to
\begin{equation}
	\label{eq/radar-equation/time-domain/time-resolved}
	\left[X_+^{(\text{dc})}\right]_{\rho_{\text{em}}}\simeq
	\left[X_+^{(\text{dc})}\right]_{\ket{0}}
	\mathcal{F}_{\rho_{\text{em}}}(t_e+\tau_2)\,.
\end{equation}
Consequently, sweeping the
emission time $t_e$ of the infinitely short probe wave packet
would sample $\mathcal{F}_{\rho_{\text{em}}}(t_e+\tau_2)$
thereby providing us with a time resolved probe of the electromagnetic
radiation.
Note that the $\tau_2$ time delay follows from 
ballistic propagation along the lower branch of the MZI.
Unfortunately, Eq.~\eqref{eq/radar-equation/time-domain/time-resolved} can only
be seen as an heuristics
since electronic wave packets with only positive energy
components cannot be arbitrarily localized. 
Nevertheless, short duration
Levitons
\cite{Dubois-2013-1} which are purely electronic provide us with a close analogue to a
perfectly localized electronic excitation. But
the issue of electronic decoherence 
must be addressed carefully. 

\subsection{Leviton excitations}
\label{sec/radar-equation/Leviton}

\subsubsection{General result}

As shown in Appendix \ref{appendix/levitons}, 
the simple expression for the Leviton wave packet in the frequency
domain enables us to derive a
convenient form for $X_+^{(\text{dc})}$ suitable
for numerical evaluations:
\begin{equation}
    \label{eq/radar-equation/Leviton/X+dc}
    X_+^{(\text{dc})}=\int_{\mathbb{R}}
    \widetilde{\mathcal{F}}_{\rho_{\text{em}}}(\Omega)\,
    \me^{-\mi\Omega(t_e+\tau_2)}\,f_{\tau_e,\tau_2}(\Omega)
    \frac{\md\Omega}{2\pi}
\end{equation}
in which $\widetilde{\mathcal{F}}_{\rho_{\text{em}}}(\Omega)$ denotes the Fourier transform of the
Franck-Condon factor.
The filter
\begin{equation}
    \label{eq/radar-equation/Leviton/filter}
    f_{\tau_e,\tau_2}(\Omega)=4\pi \tau_e\int_{|\Omega|/2}^{+\infty}
    \widetilde{\mathcal{Z}}_1\left(\omega-\frac{\Omega}{2}\right)\,
    \me^{-2\omega\tau_e}\me^{-\mi(\omega-\frac{\Omega}{2})\tau_2}
    \frac{\md\omega}{2\pi}
\end{equation}
contains the effects of electronic decoherence along branch $1$ and ballistic
propagation along branch $2$ of the interferometer for a Leviton
excitation of duration $\tau_e$.

\subsubsection{The vacuum baseline}

Let us start by the dc contrast in the
absence of external radiation 
($\rho_{\text{em}}=\ket{0}\bra{0}$) which we call the vacuum baseline:
\begin{equation}
	\label{eq/radar-equation/Leviton/vacuum-baseline}
	\left[X_+^{(\text{dc})}\right]_{\ket{0}}=4\pi\tau_e\int_0^{+\infty}
	\widetilde{\mathcal{Z}}_1(\omega)
	\,\me^{-2\omega\tau_e}\me^{-\mi\omega\tau_2}\frac{\md\omega}{2\pi}\,.
\end{equation}
This quantity which represents the interference contrast, is reduced because
of two different physical effects. 

First of all, even in the absence of electronic decoherence, the
asymmetry of times of flights along the target and the reference branch
reduces the interference signal. For
$\widetilde{\mathcal{Z}}_1(\omega)=\me^{\mi\omega\tau_1}$ which corresponds to ballistic
propagation along the target branch with time of flight $\tau_1$, we
find
\begin{equation}
	\label{eq/radar-equation/Leviton/vacuum-baseline/no-decoherence}
	\left[X_+^{(\text{dc})}\right]_{\ket{0}}=\frac{2\tau_e}{2\tau_e+\mi\tau_{12}}
\end{equation}
where $\tau_{12}=\tau_1-\tau_2$ characterizes the interferometer's
imbalance. As expected, the interference signal decays as soon as
$|\tau_{12}|$ becomes larger than $2\tau_e$. 

In the perspective of the electron radar, when using localized
wavepackets in time such as the Levitons, one wishes to achieve such a
time of flight synchronization to have the strongest information signal.
Depending on the experimental situation, the geometry of the device may be constrained
so that a lower bound $\tau_e\gtrsim \mathrm{min}|\tau_{21}|/2>0$ is imposed, thereby
intuitively limiting the time resolution of the interferometer. 
In other
devices, we may be able to tune $\tau_2$ in order to maximize the contrast
in the absence of external situation. This is what we assume in
the rest of this paper but the general expressions given by
Eqs.~\eqref{eq/radar-equation/Leviton/X+dc} and
\eqref{eq/radar-equation/Leviton/filter} enable us to discuss any situation.

The second cause of interference contrast reduction is electronic decoherence. In
the present formalism, electronic decoherence is associated with the
decay of the electronic quasi-particle which is measured by the
inelastic scattering probability
$\sigma_{\text{in}}(\omega)=1-|\widetilde{\mathcal{Z}}_1(\omega)|^2$. 

\begin{figure}
	\centering
	\includegraphics{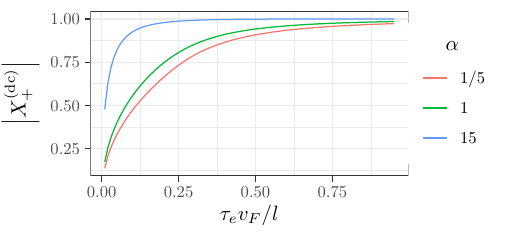}
	\caption{\label{fig/Leviton-decoherence} (Color online) Contrast
	$|X_+^{(\text{dc})}|$ of dc-current
	interference fringes for Leviton of duration $\tau_e$ as function of
	$\tau_e v_F/l$ in the absence of external radiation. 
	The radiation coupler involves two
	counter-propagating edge channels in total mutual electrostatic
	influence over a
	distance $l$ (see Appendix~\ref{appendix/two-channels-EMP-scattering}). 
	Curves have been plotted for $\alpha=1/5$ (red), $1$
	(green) and
	$15$ (blue).}
\end{figure}

Fig.~\ref{fig/Leviton-decoherence}
depicts $\max_{\tau_2}\left|X_+^{(\text{dc})}\right|$
as a function of the
Leviton's width $\tau_e$. as a function of
$v_F\tau_e/l$. These graphs, obtained for a radiation coupler built from two
counter-propagating edge channels in total electrostatic influence over a
distance $l$ via a
geometric capacitance $C_g$
(see Fig.~\ref{fig/EMP-scattering}-(a)), 
are plotted for different values of the 
dimensionless coupling constant 
\begin{equation}
	\label{eq/dimensionless-coupling}
	\alpha=\frac{e^2l}{hv_FC_g}
\end{equation}
encoding the importance of Coulomb interaction effects within this
radiation coupler (see Appendix~\ref{appendix/two-channels-EMP-scattering}).
As expected, at fixed $\alpha$, electronic decoherence is stronger for shorter Leviton
pulses due to their high energy components. 
For a given $\tau_e$, it is also stronger at small $\alpha$ 
compared to the Coulomb dominated regime (large $\alpha$), an
effect already predicted \cite{Idrisov-2018-1} and observed \cite{Sivre-2019-1} 
when an electron of energy $\hbar\omega_e$ propagates across a metallic
island with Coulomb energy $E_C\gg\hbar\omega_e$. Finally, a
50~\% contrast can be achieved when considering Levitons of durations 
$\tau_e\gtrsim l/10\,v_F$ which, in the case of $l=\SI{10}{\micro\meter}$
and $v_F=\SI{e5}{\meter\per\second}$ corresponds to pulses of duration
down to $\SI{10}{\pico\second}$.

\subsubsection{Filtering of the Franck-Condon factor}

We now consider the effect of the Leviton's duration as well as of
electronic decoherence on the Franck-Condon prefactor. Without
electronic decoherence, one expects the duration of the Leviton to be the limiting
time resolution for accessing $\mathcal{F}_{\rho_{\text{em}}}(t)$.
In the absence of electronic decoherence, using
$\mathcal{Z}_1(\omega)=\me^{\mi\omega\tau_1}$ leads to
\begin{equation}
	\label{eq/radar-equation/Leviton/filtering/no-decoherence}
	\left[	X_+^{(\text{dc})}\right]_{\rho_{\text{em}}}=\left[X_+^{(\text{dc})}\right]_{\ket{0}}\,\left(K_{\tau_e,\tau_{21}}\star
	\mathcal{F}_{\rho_{\text{em}}}\right)(t_e+\tau_2)
\end{equation}
where $[X_+^{(\text{dc})}]_{\ket{0}}$ is given by 
Eq.~\eqref{eq/radar-equation/Leviton/vacuum-baseline/no-decoherence} and
the kernel
\begin{equation}
	\label{eq/radar-equation/Leviton/filtering-kernel/no-decoherence}
	K_{\tau_e,\tau_{21}}(\tau)=\frac{1}{\pi}\,
	\frac{\tau_e+\mi\frac{\tau_{21}}{2}}{\left(\tau_e+\mi\frac{\tau_{21}}{2}\right)^2+
	\left(\tau-\frac{\tau_{21}}{2}\right)^2}
\end{equation}
encodes the way the Leviton's duration and the MZI's imbalance blur the
temporal resolution on the Franck-Condon factor
$\mathcal{F}_{\rho_{\text{em}}}(t)$. Eqs. \eqref{eq/radar-equation/Leviton/filtering/no-decoherence} and
\eqref{eq/radar-equation/Leviton/filtering-kernel/no-decoherence} show that
the duration $\tau_e$ of the Leviton
limits the time resolution for accessing 
$\mathcal{F}_{\rho_{\text{em}}}(t)$. 

Electronic decoherence alters the expression of the kernel in
Eq.~\eqref{eq/radar-equation/Leviton/filtering/no-decoherence}. Let us
now define 
$\tau_1$ as the Wigner-Smith time delay associated
with the elastic scattering amplitude
$\widetilde{\mathcal{Z}}_1(\omega)$:
$\tau_1=-\mi[\md(\ln(\widetilde{\mathcal{Z}}_1(\omega)))/\md\omega]$ at
$\omega=0$. We have to resort to the general filtering expressions
given by Eqs.~\eqref{eq/radar-equation/Leviton/X+dc} and \eqref{eq/radar-equation/Leviton/filter}.
As explained in
Appendix \ref{appendix/levitons}, the filtering function $f_{\tau_e,\tau_2}(\Omega)$
always satisfies 
\begin{equation}
    f_{\tau_e,\tau_2}(\Omega\geq 0) 
    =\me^{-\Omega\tau_e}f_{\tau_e,\tau_2}(0) \, .
    \label{eq/Leviton/filter/positive-Omega}
\end{equation}
which only requires the numerical computation of
$f_{\tau_e,\tau_2}(0)$.
But for $\Omega<0$, one has
\begin{align}
	f_{\tau_e,\tau_2}(\Omega & <0)
	=\me^{-|\Omega|\tau_e}\me^{\mi\Omega\tau_{21}}\nonumber \\
	&\times 4\pi
	\tau_e\int_0^{+\infty}\me^{-2\omega\tau_e}\me^{-\mi\omega\tau_{21}}
	\widetilde{\mathcal{Z}}_{1\star}(|\Omega|+\omega)\,\frac{\md\omega}{2\pi}\,.
    \label{eq/Leviton/filter/negative-Omega}
\end{align}
in which, for convenience, we have isolated the $\tau_1$ dependence by
introducing
$\widetilde{\mathcal{Z}}_{1\star}(\omega)=\me^{-\mi\omega\tau_1}
\widetilde{\mathcal{Z}}_{1}(\omega)$. Its
$\Omega$ dependence is a priori more 
involved than in Eq.~\eqref{eq/Leviton/filter/positive-Omega} and must
be evaluated numerically.

Nevertheless,
whenever
the typical time scale associated with the radiation is much slower than
$\tau_e$ and $\tau_D$, defined as the inverse of the energy scale over
which the inelastic scattering probability $\sigma_{\text{in}}(\omega)$
starts to increase significantly, one may neglect the $\Omega$-dependence in
$\widetilde{\mathcal{Z}}_{1\star}(|\Omega|+\omega)$, thereby leading to
\begin{equation}
    f_{\tau_e,\tau_2}(\Omega<0)
	\simeq\me^{-|\Omega|\tau_e}\me^{-\mi|\Omega|\tau_{21}}f_{\tau_e,\tau_2}(0)\,.
    \label{eq/Leviton/filter/negative-Omega/adiabatic}
\end{equation}
Eqs.~\eqref{eq/Leviton/filter/positive-Omega} 
and \eqref{eq/Leviton/filter/negative-Omega/adiabatic} then lead to
Eq.~\eqref{eq/radar-equation/Leviton/filtering/no-decoherence} with the
same kernel $K_{\tau_e,\tau_{21}}$ as in
Eq.~\eqref{eq/radar-equation/Leviton/filtering-kernel/no-decoherence}.
Electronic decoherence is fully encapsulated in the vacuum
baseline $[X_+^{(\text{dc})}]_{\ket{0}}$. 

Eq.~\eqref{eq/Leviton/filter/negative-Omega/adiabatic} is an
``adiabatic radiation approximation'' assuming that electronic decoherence can be
neglected for the frequencies involved
in the Franck-Condon factor. 
Note that this approximation does not assume that Leviton wavepackets, which involve
much higher frequencies, experience a weak
electronic decoherence. Electronic decoherence is indeed accounted for by the vacuum
baseline contrast whose modulus can be much lower 
than the modulus of the r.h.s of
Eq.~\eqref{eq/radar-equation/Leviton/vacuum-baseline/no-decoherence}
which is only valid for ballistic propagation along the target
branch. 

Interestingly, for very short Levitons and a balanced interferometer
($\tau_{21}=0$), the Kernel converges to the $\delta(\tau)$ distribution
and we 
recover the heuristic expression obtained in
Eq.~\eqref{eq/radar-equation/time-domain/time-resolved}. The
point is that we have an exact expression for the vacuum
baseline given by Eq.~\eqref{eq/radar-equation/Leviton/vacuum-baseline}.
Moreover, we know that this is the
result of an approximation which breaks down when electronic decoherence
effects manifest themselves at frequencies similar to the inverse time
scale of variation of $\mathcal{F}_{\rho_{\text{em}}}(t)$. More
importantly, the results presented here enable us to go beyond this
approximation and to consider an imbalanced ($\tau_1\neq \tau_2$)
interferometer.

\section{Predictions for classical and quantum radiation}
\label{sec/examples}

We now consider various types of radiation directly relevant for
forthcoming experiments. This includes classical radiation
for which the physics of a time dependent phase seen by independent
electrons of Sec.~\ref{sec/radar-physics/simple-model}
will be recovered in the voltage locked regime of the
radiation coupler. 

We then discuss the ability of the electron
radar to probe non-classical radiation by considered squeezed states 
as well as single EMPs. Such quantum states of radiation cannot be
described in terms of a classically fluctuating voltage and therefore,
the full quantum approach of the previous section will be necessary to
obtain quantitative predictions.

\subsection{Classical radiation}
\label{sec/examples/classical}

\begin{figure}
	\centering
	\includegraphics[width=6cm]{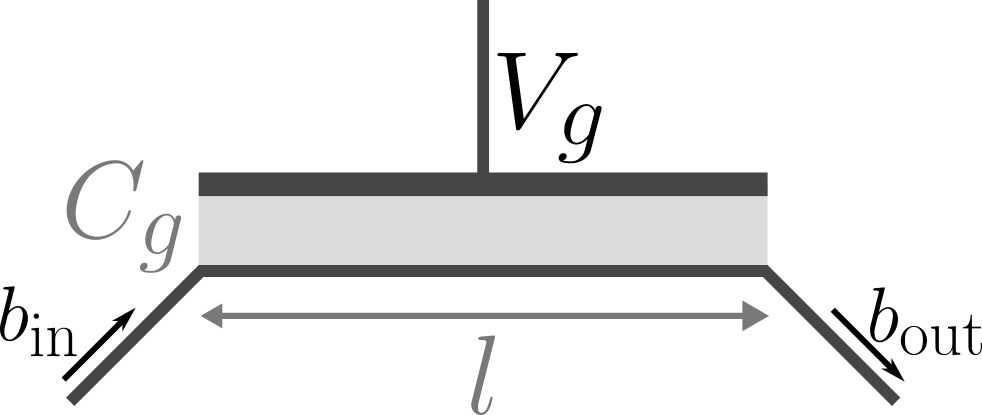}
    \caption{\label{fig/top-gate} A top gate is capacitively coupled to
    the $|x|\leq l/2$ region of a chiral edge channel. It is driven by a
    time dependent gate voltage $V_g(t)$. $C_g$ denotes the geometric
	capacitance between the two conductors.}
\end{figure}

We consider a classical drive $V_g(t)$ applied to a top gate capacitively coupled to the
edge channel along the target branch $|x|\leq l/2$ of the MZI via a geometric capacitance
$C_g$ as depicted on Fig. \ref{fig/top-gate}. This  corresponds to
Fig. \ref{fig/EMP-scattering}-b without dynamical degrees of
freedom coupled to the top gate ($Z(\omega)=0$). This is the limit of
an infinite number of electronic channels in the top gate's lead in Ref.
\cite{Misiorny-2018-1}. 

Because there are no external dynamical degrees of freedom in this
top gate model, its dynamics can be described by an
input/output relations for the edge channel's EMP modes of the form:
\begin{equation}
    \label{eq/EMP-scattering/only-transmission}
    i_{\text{out}}(\omega)=t(\omega)\,i_{\text{in}}(\omega)+Y(\omega)\,V_g(\omega)
\end{equation}
where $t(\omega)$ denotes the EMP transmission amplitude across the
region $|x|\leq l/2$ and the admittance $Y(\omega)$ describes the response
of the outgoing current
to the top gate potential $V_g(\omega)$. Assuming total
mutual influence between the top gate and target branch of the MZI,
$Y(\omega)$ is the finite frequency admittance of the
electrical dipole formed by the top gate and the edge channel. It is
related to $t(\omega)$ by \cite{Safi-1999-1,Degio-2010-1,Cabart-2018-1}:
\begin{equation}
	\label{eq/EMP-scattering/Y-from-t}
	Y(\omega)=\frac{e^2}{h}(1-t(\omega))\,.
\end{equation}
Predictions for these coefficients can be obtained from
the discrete element model corresponding to
Fig.~\ref{fig/top-gate} presented in Appendix
\ref{appendix/top-gate-EMP-scattering}. 

Since the gate voltage translates the EMP operator
$b_{\text{in}}(\omega)$ by $-Y(\omega)V_g(\omega)/e\sqrt{\omega}$, it
introduces a phase factor 
\begin{equation}
	\label{eq/top-gate/F-factor}
\mathcal{F}_{V_g}(t)=
	\me^{\frac{\mi e}{\hbar}\int_{\mathbb{R}}
	\Gamma(t-\tau)V_g(\tau)\,
	\md\tau}
\end{equation}
in front of the fermionic field coming out of
the radiation coupler. The kernel
$\Gamma(\tau)$ is the inverse Fourier transform of
$-R_KY(\omega)/\mi\omega$.
This shows that the effective single particle scattering amplitude retains the
form given by 
Eq.~\eqref{eq/radar-equation/time-domain/2} with $\mathcal{Z}_1(\tau)$ given by
Eq.~\eqref{eq/radar-equation/Z-time-domain} with
$t(\omega)=1-R_KY(\omega)$ and $\mathcal{F}_{V_g}(t)$ playing the role
of $\mathcal{F}_{\rho_{\text{em}}}(t)$ defined by \eqref{eq/radar-equation/F-definition} in
the full quantum theory. In particular, because the only effect of the
external drive $V_g(t)$ is to introduce a phase, it does not contribute
to electronic decoherence
in the time domain.

The importance of Coulomb interactions within the $|x|\leq
l/2$ region depends on the dimensionless coupling constant
$\alpha=C_q/C_g$ defined by Eq.~\eqref{eq/dimensionless-coupling}.
For $\alpha\ll 1$, the voltage drop at the
capacitance vanishes and therefore the electrons directly see
the gate voltage $V_g(t)$. This is the voltage locked regime. 
On the other hand, for
the $\alpha\gtrsim 1$, Coulomb interactions are so strong that 
they tend to block charge accumulation below the top gate. This is the
Coulomb blocked regime. 

\subsubsection{The voltage locked regime}
\label{sec/examples/classical/weak-coupling}

At very small $\alpha$, the electrochemical
capacitance $C_\mu=C_gC_q/(C_g+C_q)\simeq C_q$ is dominated by the quantum
capacitance $C_q\ll C_g$. The voltage drop at the capacitance is
negligible and electrons within the $|x|\leq l/2$ region experience the gate potential
$V_g(t)$. Moreover, in this limit
$t(\omega)\simeq\me^{\mi \omega l/v_F}$ 
Consequently, in this limit, the elastic scattering amplitude is
$\mathcal{Z}_{1}(\omega)\sim \me^{\mi\omega l/v_F}$ and electronic
decoherence can thus be neglected
\cite{Cabart-2018-1}. 
In other terms, there is no back-action of the incoming electron
on the edge channel mediated by the top-gate which could lead to
electronic decoherence. 
The effective scattering amplitude can be approximated by
\begin{equation}
	\label{eq/top-gate/voltage-locked/R-eff}
	R_{\text{eff}}(t,t')\simeq \delta\left(\tau-\frac{l}{v_F}\right)\,
	\mathcal{F}_{V_g}(t)\,.
\end{equation}
At very small $\alpha$,
$\Gamma(\tau)=\mathbf{1}_{[0,l/v_F]}(\tau)$ and therefore, we recover
Eq.~\eqref{eq/classical-voltage/1} with $\tau_1=l/v_F$ and 
$U(t)=V_g(t)$ as
expected. 

\subsubsection{Intermediate regime}
\label{sec/examples/classical/increaded-coupling}

As $\alpha$ is increased, electronic decoherence starts to appear but,
as long as we keep the energy of the electronic excitation low enough,
it can be neglected \cite{Cabart-2018-1}. 
Nevertheless, increasing $\alpha$ 
modifies the voltage seen by the electrons below the top gate which
becomes 
a filtering of $V_g(\omega)$ (see Eq.~
\eqref{eq/capacitive-gate/filtering}). Consequently, 
as long as electronic decoherence can be neglected, the
MZI detects the accumulated electric phase associated with
this filtered effective time dependent potential. 

Nevertheless, for Leviton pulses, the discussion of Sec.~\ref{sec/radar-equation/Leviton} 
shows that as long as the external voltage only involves low frequencies
compared to the ones for which electronic decoherence starts to be
significative, the adiabatic approximation of 
Eq.~\eqref{eq/Leviton/filter/negative-Omega/adiabatic} can still be
applied. 
The ratio of the contrast
in the presence of $V_g(t)$ and without can thus be obtained as 
\begin{align}
	\frac{[X_+^{(\text{dc})}]_{V_g(t)}}{[X_+^{(\text{dc})}]_{0}}&=
	\sum_{n\in\mathbb{Z}}
	\mathcal{F}_n[V_g(t)]\,\me^{-2\pi nf\tau_e}\nonumber\\
	&\times \me^{-2\pi i
	nf(t_e+\tau_2)}
	\me^{-2\pi in\Theta(-n)f\tau_{21}}
\end{align}
in which $\mathcal{F}_n[V_g(t)]$ are the photo-assisted amplitudes
associated with $V_g(t)$ mediated through the top-gate, {\it id est} 
the Fourier coefficients of the
Franck-Condon factor given by Eq.~\eqref{eq/top-gate/F-factor}. Most
importantly, the vacuum baseline $[X_+^{(\text{dc})}]_{0}$ takes into
account electronic decoherence effects. 

Therefore, for an slow enough drive, even when electronic decoherence
effects are strong, quantitative predictions
for the relative dc contrast
$[X_+^{(\text{dc})}]_{V_g(t)}/[X_+^{(\text{dc})}]_{0}$ for Leviton
pulses
can still be obtained from the simple formalism discussed
in Sec.~\ref{sec/radar-physics/simple-model}, provided one properly takes into
account the filtering of $V_g(t)$ by the top gate.

\subsubsection{The Coulomb blocked regime}
\label{sec/examples/classical/strong-coupling}

The limit of
large $\alpha$ corresponds to the regime dominated by Coulomb
interactions: they are
so strong that no charge can accumulate on either
on the $|x|\leq l/2$ interaction region of the
edge channel nor on the top gate. As explained in Appendix
\ref{appendix/EMP-scattering/top-gate}, the outgoing EMP mode
$b_{\text{out}}(\omega)$ has a very small
response to the external time dependent voltage. Consequently, the
electron radar is weakly responding to the external voltage applied to the
top gate. Consequently, this is not the proper regime of operation for
the electron radar. 

Note that electronic decoherence is not a problem in this
regime. This echoes the results obtained recently in Ref.
\cite{Duprez-2019-1} in which a similar phenomenon occurs in a MZI where
the top gate in the strong coupling regime is replaced by a metallic
island with small enough capacitance.

\subsection{Squeezed radiation}
\label{sec/examples/squeezing}

Let us now discuss 
the ability of the electron radar to detect squeezing by
considering a Gaussian squeezed state. 

\subsubsection{General squeezing criterion}
\label{sec/examples/squeezing/general-criterion}

The key observation is that the radiation coupler is only sensitive to
the radiation in a specific filtered mode. 
The Franck-Condon factor can then be computed
using a Gaussian approximation for this mode. This leads to:
\begin{equation}
	\label{eq/targets/squeezing/Gaussian-result}
		\mathcal{F}_{\rho_{\text{em}}}(t)=\me^{i\phi(t)}
		\, \me^{-\left(\langle (\Delta
	Y_t)^2\rangle_{\rho_{\text{em}}}-\langle(\Delta
	Y_t)^2\rangle_{\ket{0}}\right)}\,.
\end{equation}
where 
the phase $\phi(t)$ is due to the average value 
can be viewed as arising from the classical voltage
felt by the electrons within the upper branch of the MZI. The other
contribution comes from the Gaussian fluctuations of the quadrature
\begin{equation}
	Y_t=\frac{\mi}{\sqrt{2}}
	\int_0^{\omega_c}\left(
	\frac{S_{ba}(\omega)}{\sqrt{\omega}}\,\me^{-\mi\omega t}
	\,a_\omega-\mathrm{h.c.}\right)\,\md\omega
\end{equation}
of the mode filtered by the radiation coupler. Here, $\omega_c$ is any UV
cutoff below which all the incoming
radiation is emitted. 

Because this second contribution
changes $|\mathcal{F}_{\rho_{\text{em}}}(t)|$, it
provides a sufficient criterion for squeezing: as soon as
$|\mathcal{F}_{\rho_{\text{em}}}(t)|>1$ there is squeezing in the
filtered mode 
since this is a signature of the fact that, for these values of $t$, 
fluctuations of the quadrature $Y_t$
are smaller than in the vacuum
state:
$\langle (\Delta
Y_t)^2\rangle_{\rho_{\text{em}}}<\langle (\Delta
Y_t)^2\rangle_{\ket{0}}$.

\subsubsection{Squeezing around a given frequency}
\label{sec/examples/squeezing/squeezed-vaccuum}

Because of its relevance for experiments, let us discuss the ability of 
the electron radar to detect squeezed
radiation in an electromagnetic mode around $\omega\simeq \omega_0$
within a bandwidth $\gamma_0$ such that $Q_0=\omega_0/\gamma_0$ is
significantly larger than one so that $S_{ba}(\omega)$ can be taken as
constant for $|\omega-\omega_0|\leq \gamma_0/2$. In Appendix \ref{appendix/squeezing-detection}
we obtain show that for a time-periodic squeezed quantum noise whose power is
concentrated in the narrow band $|\omega-\omega_0|\leq \gamma_0/2$, the
Franck-Condon factor takes the form
\begin{equation}
	\label{eq/examples/squeezing/Franck-Condon/narrowband}
	\mathcal{F}_{\mathrm{Sq}_z}(t)\simeq
	\me^{\frac{|S_{ba}(\omega_0)|^2}{Q_0}\sinh(2|z|)\left[
	\cosh(2|z|)\cos(2\omega_0t-\phi_0)-\sinh(2|z|)\right]}
\end{equation}
where $z$ is a squeezing parameter and 
$\phi_0=2\text{Arg}(S_{ba}(\omega_0))+\text{Arg}(z)$. 
Its minimum value is 
\begin{equation}
	\label{eq/examples/squeezing/nattowband/min}
    \min_t\left|\mathcal{F}_{\text{Sq}_z}(t)\right|=\me^{
		-\frac{|S_{ba}(\omega_0)|^2}{2Q_0}
		(\me^{4|z|}-1)}<1\,.
\end{equation}
Let us recall that the compression factor for quadrature fluctuations
for a squeezed mode in $\ket{\mathrm{Sq}_z}$ is given by
$e^{-4|z|}$. As of today, a compression of $18\,\%$ 
($\SI{0.86}{\decibel}$) with respect to
vacuum fluctuations has been achieved in quantum Hall edge channels
\cite{Bartolomei-2023-1}.
Assuming a quality factor $Q_0=5$ and no losses ($|S_{ba}(\omega_0)|^2=1$) for the
radiation coupler leads to a maximal
increase of $\left|\mathcal{F}_{\text{Sq}_z}(t)\right|$ 
by $1.8\,\%$ whereas a $3\ \mathrm{dB}$ squeezing would lead to 
an increase by $5.1\,\%$. 

\subsubsection{Numerical results and discussion}

Let us now discuss whether or not this is observable using Levitons. We
have done it by evaluating numerically the filtering function
$f_{\tau_e,\tau_2}(\Omega)$ as well as the exact Fourier series of
the Franck-Condon factor given by 
Eq.~\eqref{eq/examples/squeezing/Franck-Condon/narrowband}.

To have an intuition of the results, we have obtained analytical
expressions 
within the
adiabatic radiation approximation (Eq.~\eqref{eq/Leviton/filter/negative-Omega/adiabatic}) 
by retaining only the
first harmonics in the Fourier series of the Franck-Condon
factor\footnote{This is justified by observing that
$S_{ba}(\omega_0)|^2/\mathcal{Q}_0$ is significantly smaller than unity
and experimentally achieved squeezing parameters are below $z\simeq
0.0719$. Even a $\SI{3}{\deci\bel}$ noise reduction corresponds to $z\simeq
0.1733$.}.
Using this approximation, an analytical expression for the
maximum over $t_e$ of the relative contrast with respect to the vacuum
baseline as a function of $|z|$ 
can be derived at the lowest order in $\Lambda=|S_{ba}(\omega_0)|^2/Q_0$:
\begin{equation}
     \label{eq/examples/squeezing/approximated-optimized-contrast/z-function}
    \max_{t_e}\left|\frac{\left[X_+^{(\text{dc})}\right]_{\mathrm{Sq}_z}}{\left[X_+^{(\text{dc})}\right]_{\ket{0}}}
		 \right| \simeq 1+\Lambda F_\eta(z)+\mathcal{O}(\Lambda^2)
\end{equation}
where
\begin{equation}
	F_\eta(z)=
		 \eta \cosh(2|z|)\sinh(2|z|)-\sinh^2(2|z|)
\end{equation}
with
 $\eta=\me^{-2\omega_0\tau_e}\left|\cos(2\omega_0\tau_2)\right|<1$. 
The behavior of $F_\eta(z)$ governs the observability of
squeezing as an increase of the interference contrast. It starts from $0$ for $z=0$ and increases to a
maximum positive value 
\begin{equation}
    \label{eq/examples/squeezing/approximated-optimized-contrast}
    \max_{|z|}\max_{t_e}\left|\frac{\left[X_+^{(\text{dc})}
	\right]_{\mathrm{Sq}_z}}{\left[X_+^{(\text{dc})}\right]_{\ket{0}}}
    \right|\simeq
    1+\frac{\Lambda}{2}\left(1-\sqrt{1-\eta^2}\right)+\mathcal{O}\left(\Lambda^2\right)\,.
\end{equation}
reached for $|z|_{\text{opt}}=\arctanh(\eta)/4$. It is the
maximum experimental signal expected for squeezing since
$F_\eta(z)$ decreases for $|z|\geq |z|_{\text{opt}}$. It becomes negative for
$|z|\geq 2|z_{\text{opt}}|$ meaning that, above $2|z_{\text{opt}}|$,
the interference contrast for Levitons appears
smaller than the vacuum baseline. 
This follows from the increase of the average
photon number in $\ket{\text{Sq}_z}$ with increasing $z$: as shown in
Appendix \ref{appendix/squeezing-detection} (see
Eq.~\eqref{eq/examples/squeezing/time-averaged-contrast}),
randomly injected Levitons of width $\tau_e$ will experience, on
average, an increasing noise within the radiation coupler. For 
$|z|\geq 2|z|_{\text{opt}}$,
this effect compensates the gain in contrast associated with 
sub-vacuum fluctuations experienced by the Leviton thereby
leading to a decrease of the maximum relative contrast over $t_e$ below
unity.

\begin{figure}
     \centering
	\includegraphics[width=85mm]{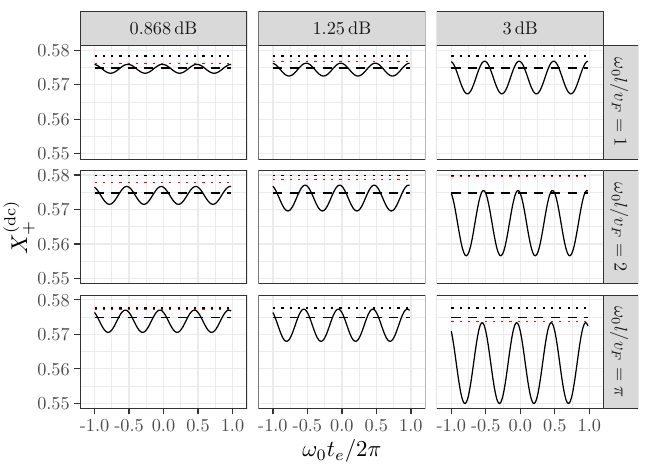}
    \caption{\label{fig/squeezing-numerics/weak-coupling} The various
	plots depict 
	the interference contrast
	$\left[X_+^{(\text{dc})}\right]_{\text{Sq}(z)}$ for
    Leviton excitations of width $\tau_e=\SI{15}{\pico\second}$
    assuming a $\SI{10}{\micro\meter}$ long radiation coupler with
    $v_F=\SI{e5}{\meter\per\second}$ with coupling strength
    $\alpha=1/5$. The plots show the absolute contrast as a function of
    the dimensionless injection time $v_Ft_e/l$ for different values of
    the squeezing (expressed in $\si{\decibel}$) and
    $\omega_0 l/v_F=1$, $2$ and $\pi$.
The dashed black
horizontal lines correspond to the vacuum baseline
$\left[X_+^{(\text{dc})}\right]_{\ket{0}}$. The dotted black lines
correspond to the maximally optimized maximum contrast given by
Eq.~\eqref{eq/examples/squeezing/approximated-optimized-contrast}
whereas
the dotted red lines correspond to the evaluation of
Eq.~\eqref{eq/examples/squeezing/approximated-optimized-contrast/z-function}
for the actual value of the squeezing parameter considered in the
example.}
\end{figure}

Figure \ref{fig/squeezing-numerics/weak-coupling} 
displays the results of the
numerical evaluation of the contrast in the voltage locked 
regime with $\SI{15}{\pico\second}$ Levitons. Filled black lines
correspond to the full numerical evaluation of the contrast.
Dashed black
horizontal lines correspond to the vacuum baseline
$|[X_+^{(\text{dc})}]_{\ket{0}}|$. 
Results are plotted for $z\simeq 0.0496$ corresponding
to $18~\%$ ($\SI{0.86}{\decibel}$) noise reduction demonstrated in
Ref.~\cite{Bartolomei-2023-1}, $z\simeq 0.0719$ corresponding to
$\SI{1.25}{\decibel}$ ($25~\%$) noise reduction and $z\simeq 0.1733$ corresponding to
$\SI{3}{\decibel}$ ($50~\%$) noise reduction.

\begin{figure}
   \includegraphics[width=85mm]{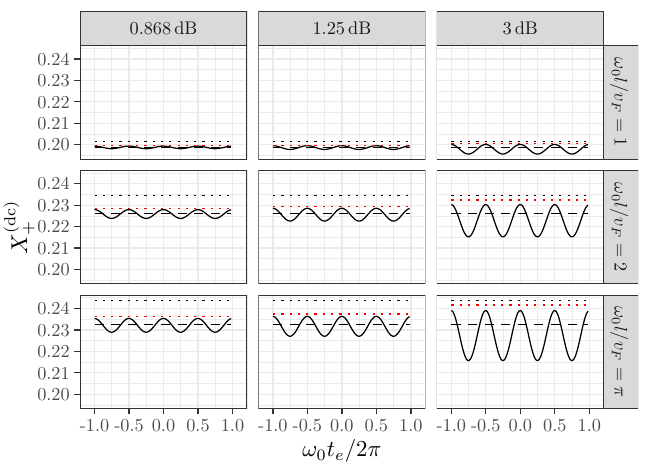}
    \caption{\label{fig/squeezing-numerics/weak-coupling-2} The various
	plots depict
	the interference contrast
	$\left[X_+^{(\text{dc})}\right]_{\text{Sq}(z)}$ for
    Leviton excitations of width $\tau_e=\SI{2.5}{\pico\second}$
	assuming the same parameters and legend than in
	Fig.~\ref{fig/squeezing-numerics/weak-coupling}.}
\end{figure}

At some operating points, the
contrast in the presence of the squeezed vacuum does exceed the vacuum
baseline. It is a positive signature of squeezing but the overshoot is small: 
$0.3\ \%$ on a baseline of $57.5\ \%$. The best numbers are obtained in the
voltage locked regime for
$\SI{1.25}{\decibel}$ squeezing and $\omega_0l/v_F=\pi$. This frequency
choice corresponds to the optimal coupling between the two edge
channels. As expected from the analytics, increasing $|z|$ too much leads to a decrease
of the maximum relative relative contrast and even bring it below
the vacuum baseline. Note also the
good agreement between
Eq.~\eqref{eq/examples/squeezing/approximated-optimized-contrast/z-function}
and the numerical evaluation 
of the maximum absolute contrast. Note also that the best operating
point ($\omega_0l/v_F=\pi$ and $\SI{1.25}{\decibel}$ squeezing) is
close to the optimal maximum contrast given by
Eq.~\eqref{eq/examples/squeezing/approximated-optimized-contrast}.
Shorter Levitons are expected to lead to a lower vacuum
baseline (see Fig.~\ref{fig/Leviton-decoherence}) and also to shift the detrimental
effect of the average noise 
towards higher values of $|z|$.
This can be seen on
Fig.~\ref{fig/squeezing-numerics/weak-coupling-2} where
$\SI{2.5}{\pico\second}$ Levitons are considered, all other parameters
being the same than for Fig.~\ref{fig/squeezing-numerics/weak-coupling}.
We see that the maximum absolute contrast still increases with
increasing $|z|$ up to $\SI{3}{\decibel}$ squeezing instead of
decreasing at $\SI{1.25}{\decibel}$ for $\SI{15}{\pico\second}$
Levitons. Even if the absolute contrast
overshoot over the vacuum baseline is not greater than 
for $\SI{15}{\pico\second}$ Levitons, it comes over a $0.225$
vacuum baseline instead of $0.57$ and thus represents a two times
increase in terms of the relative contrast increase. This clearly shows
the pros and cons of shorter Leviton pulses: the filtering associated
with their duration is better but they also lead to a lower vacuum
baseline.

\subsection{Fock states}
\label{sec/examples/Fock}
 
We now consider the problem of detecting Fock states in a specific EMP
mode.
This state being non-Gaussian, the Gaussian
result given by Eq.~\eqref{eq/targets/squeezing/Gaussian-result} 
breaks down. 

\subsubsection{The Franck-Condon factor}
\label{sec/examples/Fock/Franck-Condon}

We consider $\ket{N;\chi}$, the $N$ photon Fock state in the normalized mode
$\chi$. An explicit computation detailed in  
Appendix \ref{appendix/bosons/modes} shows that the Franck-Condon
factor
$\mathcal{F}_{\ket{N;\chi}}(t)$ is then given by
\begin{equation}
	\label{eq/targets/Fock/result}
	\mathcal{F}_{\ket{N;\chi}}(t)=L_N\left(2\pi\left|\braket{\chi
		|S_{ba}^*\Lambda_t}\right|^2\right)
\end{equation}
in which $L_N$ denotes the $N$-th Laguerre polynomial and
\begin{equation}
	\label{eq/targets/Fock/result/1}
	\braket{S_{ba}^*\Lambda_t|\chi}=-\int_0^{+\infty}\frac{S_{ba}(\omega)}{\sqrt{\omega}}
	\me^{-\mi\omega
	t}\chi(\omega)\,\frac{\md\omega}{2\pi}\,.
\end{equation}
We now focus on
the specific
example of an EMP mode chose wavefunction in the frequency domain
$\chi(\omega)$ is centered at $\omega_0$ with Lorentzian lineshape
of width $\gamma_0$. We assume that $\omega_0/\gamma_0$ is significantly larger
than unity, as expected from spontaneous emission by a two level emitter.

\subsubsection{Narrow bandwidth approximation}

Denoting by 
$x(t)=2\pi \left|
	\braket{\chi|S_{ba}^*\Lambda_t}
	\right|^2$
the argument of the Laguerre polynomial in Eq.~\eqref{eq/targets/Fock/result}, and
assuming that $\chi(\omega)$ is concentrated around $\omega_0$ with a
small bandwidth $\gamma_0\ll \omega_0$, $x(t)$ can be rewritten as (see
Appendix \ref{appendix/single-EMP}):
\begin{equation}
	x(t)\simeq 2\pi\,\frac{\gamma_0}{\omega_0}\,
	\left|S_{ba}(\omega_0)\right|^2
	\frac{\langle
	\mathcal{J}_Q(t)\rangle_{\ket{1;\chi}}}{\gamma_0\hbar\omega_0}\,.
	    \label{eq/target/Fock/analytical-expression/b}
\end{equation}
in which 
\begin{equation}
	\mathcal{J}_Q(t)=\frac{R_K}{2}\,:i(t)^2:\,.
\end{equation}
denotes the instantaneous heat current
operator 
injected in the radiation channel expressed in terms of the electrical
current $i(t)$. 
Consequently, for $\gamma_0/\omega_0\lesssim 1$, the expansion 
$L_N(x)\simeq 1-Nx+\mathcal{O}(x^2)$ for the Laguerre polynomial leads
to: 
\begin{equation}
	\label{eq/Fock/contrast-reduction}
	\mathcal{F}_{\ket{N;\chi}}(t)
	\simeq 1-2\pi
	N\,\frac{\omega_0}{\gamma_0}\,\left|S_{ba}(\omega_0)\right|^2
\frac{\langle
    \mathcal{J}_Q(t)\rangle_{\ket{1;\chi}}}{\gamma_0\hbar\omega_0}\,.
\end{equation}
Therefore, within this approximation\footnote{The only hypothesis for deriving 
Eqs.~\eqref{eq/target/Fock/analytical-expression/b} and
\eqref{eq/Fock/contrast-reduction} 
is that 
$|S_{ba}(\omega)|^2$ does not vary significantly on the bandwidth of
$\chi$.}, the Franck-Condon factor provides a
direct measurement of the average heat current carried by the single
EMP. Provided $2\pi N\omega_0|S_{ba}(\omega_0)|^2/\gamma_0$ does not take
large values for the values of $N$ with non negligible $p_N$ and
noticing
that
$\langle
\mathcal{J}_Q(t)\rangle_{\ket{N;\chi}}=N\langle
\mathcal{J}_Q(t)\rangle_{\ket{1;\chi}}$ enables us to average over $N$
and show that
\begin{equation}
    \label{eq/Fock/contrast-reduction/2}
	\mathcal{F}_{\rho_{\text{mix}}}(t)
    \simeq 1-2\pi
    \,\frac{\omega_0}{\gamma_0}\,\left|S_{ba}(\omega_0)\right|^2
\frac{\langle
	\mathcal{J}_Q(t)\rangle_{\rho_{\text{mix}}}}{\gamma_0\hbar\omega_0}\,.
\end{equation}
for
$\rho_{\text{mix}}=\sum_{N=0}^{+\infty}p_N\ket{N;\chi}\bra{N;\chi}$.
For such
radiation, the electron radar acts as a time resolved ``quantum
bolometer'' by converting the incident energy flux within the radiation
channel into the Franck-Condon factor.
As before, assessing the potential 
for single EMP detection requires discussing experimentally realistic
signals 
in experimentally
realistic situations. 

\subsubsection{Results and discussion}
\label{sec/examples/Fock/results}

We consider 
lorentzian EMP modes
of duration
$\gamma_0^{-1}=\SI{1}{\nano\second}=10\,l/v_F$ centered on
$\omega_0$ such that $\omega_0l/v_F=2$, $5.5$ and $10$
corresponding to respective frequencies $\SI{3.2}{\giga\hertz}$,
$\SI{8.75}{\giga\hertz}$ and $\SI{15.9}{\giga\hertz}$. 
Levitons of duration $\tau_e=\SI{10}{\pico\second}$ have
a time resolution of the order of $1\ \%$ of the EMP wave packet
duration. 

\begin{figure}
    \centering
    \includegraphics{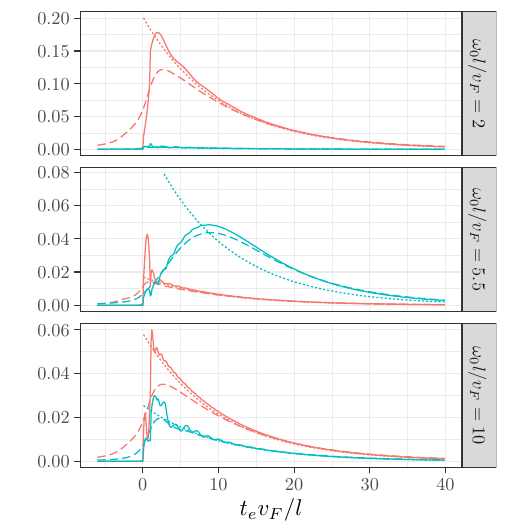}
    \caption{\label{fig/contrast-single-EMP} (Color online) Plot of the
	relative contrast decrease
	$1-\left|\left[X_+^{(\text{dc})}\right]_{\ket{1,\chi}}/\left[X_+^{(\text{dc})}\right]_{\ket{0}}\right|$
	(dashed line)
    associated with the detection of a single EMP with energy
	$\hbar\omega_0$ and Lorentzian lineshape of width
	$\gamma_0=\SI{e9}{\second^{-1}}$ when
	using Leviton pulses of width
	$\tau_e=\SI{10}{\pico\second}$ injected at times $t_e$. The
	radiation coupler is described 
	in Appendix~\ref{appendix/two-channels-EMP-scattering}.
	Red curves
	correspond to the voltage locked regime ($\alpha=1/10$) whereas the
	blue curves correspond to the Coulomb blocked regime ($\alpha=15$).
	Three different values of
    $\omega_0l/v_F$ have been considered. 
	The dotted line corresponds $1-\mathcal{F}_{\ket{1;\chi}}(t)$ using
	the analytic expression given by
	Eq.~\eqref{eq/Fock/contrast-reduction} whereas the full
	line represents $x(t)=2\pi\left|\langle
	\chi|S_{ba}\Lambda_t\rangle\right|^2$.}
\end{figure}

Figure \ref{fig/contrast-single-EMP} depicts the various numerical estimates for the
relative contrast decrease on the dc average current in the presence of the single EMP with respect to the
vacuum baseline. 
With the
parameters considered here, the latter contrast is expected to be above
50~\% (see Sec.~\ref{sec/electron-radar/Levitons}).
The results, presented on
Fig.~\ref{fig/contrast-single-EMP}, confirm that an observable relative
contrast decrease of a few $\%$ can be
expected.
The dominant
effect in the amplitude of the effect comes from the
transparency of the radiation coupler in the bandwidth of the incident
radiation. 

The precise comparison with numerical evaluations shows 
that the analytical result given by 
Eq.~\eqref{eq/target/Fock/analytical-expression/b} is 
semi-quantitatively recovered
except on short time scales where higher frequency
contributions are expected to matter. It also departs from the numerical
evaluation of $x(t)$
because of the 
filtering of high frequencies associated with the finite duration of the
Leviton. 

\section{Conclusions and perspectives}
\label{sec:conclusion}


In this work, we have discussed single electron interferometric sensing
of classical and quantum electromagnetic fields. The idea is to probe a
time dependent, possibly quantum, electromagnetic field by coupling it
to one of the branches of an electronic MZI fed by single electron
excitations which provides a very sensitive although minimally invasive
probe. Information about the field is then obtained by measuring the
interference contribution to the
average outgoing dc current from the interferometer. 

We have developed a general framework describing such a single electron
interferometer in the
Aharonov-Bohm regime for any type of ``radiation coupler''
realizing the capacitive coupling between the radiation to be probed and
the electrons propagating within the MZI. In the limit where the
radiation coupler is strongly dominated by its quantum capacitance,
the electrons directly
experience the voltage imposed by the external radiation. 
A coherent state of the incoming radiation
just imprints a time-dependent phase on the single electron 
excitations injected into the MZI, as expected when Coulomb interaction
effects between electrons can be neglected. However, our framework goes
beyond this simple picture of time dependent single particle scattering 
and fully incorporates
electronic decoherence effects associated with
Coulomb interactions. Although our formalism enables discussing any
single electron excitations, we have for simplicity focused on Leviton pulses which,
when sufficiently short, can be used to sample the external radiation 
when sweeping their injection time.

Besides the case of a classical drive, we have explored
the potential for sensing quantum features of the probed
radiation by considering two examples of quantum radiation.
First, for
Gaussian states for the incident radiation, an operational
criterion for squeezing detection via short duration wave packets is given. 
Realistic estimates suggest that a $\SI{1.15}{\deci\bel}$ squeezing within a
nearby quantum Hall edge channel could, in principle, lead to an observable 
increase of the contrast of Aharonov-Bohm interference
fringes compared to the vacuum baseline when using $\SI{15}{\pico\second}$ Levitons.
Secondly, we have shown that a single EMP propagating in the radiation
edge channel could be detected via
the transient interference
contrast decrease induced by its quantum noise. Moreover, we have seen
that, as a function of their injection time, the interference contrast
for short enough Levitons semi-quantitatively images the average instantaneous
heat current carried by the single edge-magnetoplasmon.

These two examples suggest that
the single electron interferometer fed by 
suitably short Levitons may be used as a time resolved quantum
noise sensor
on sub-nanosecond, 
possibly down to few picoseconds,
time scales. 

Several perspectives are opened by the present work. First of all,
obtaining more quantitative predictions for forthcoming
experiments may require completing the
present study to account for finite temperature and charging effects as
in Refs.~\cite{Halperin-2011-1,Frigeri-2019-1,Frigeri-2020-1}. Secondly,
recent progresses in the shaping \cite{Misiorny-2018-1} and characterization 
\cite{Bisognin-2019-2,Roussel-2020-1} of electronic
excitations emitted by single electron sources \cite{Bauerle-2018-1} suggest investigating
the potential of well known techniques from classical radar
engineering to improve
our ability to probe very short dynamical time
scales of the incident electromagnetic radiation
\cite{Souquet-Basiege-2024-2}. An important related question is to
design a set of single electron excitations enabling a full
reconstruction of the effective single-particle amplitude induced by the
coupling to the external radiation, thereby mimicking the use of
infinite chirps \cite{Klauder-1960-1} in radars and sonars to
reconstruct the target's scattering amplitude by 
inverse Radon transform, a technique
commonly used in computed tomography scans \cite{Radon-1986-1}, 
quantum optics \cite{Smithey-1993-1,Lvovsky-2009-1} and for
reconstructing the quantum state of solitary electrons
\cite{Fletcher-2019-1}.  
However, in the presence of a Fermi sea,
infinite chirps cannot be generated because of the Fermi sea, 
thereby leaving the problem of tomographic
reconstruction of the effective single electron scattering amplitude
open for further investigation.

Another important issue consists in studying the signal to noise radio in such
an electronic interferometer. Quantitative predictions for the current
noise would open the way to discussing the basic question of quantum
metrology in the present context: how to choose the source and the
measured quantity in order to discriminate between various quantum
states of the incoming radiation in an optimal way. This is analogous to the
problem of super-resolution encountered in optics an astronomy
\cite{Tsang-2016-1,Tsang-2017-1,Rehacek-2017-1,Paur-2018-1}. The standard methods
used in this field can certainly be transposed but this requires an
evaluation of the outgoing current noise. 
Motivated by the recent work
\cite{Benatti-2014-1}, one could also investigate whether or not injecting few electron
states into the MZI 
would lead to some
quantum advantage as entangled states do in quantum metrology
\cite{Giovannetti-2011-1}.

Finally, the present work discusses the information extracted from
repeated experiments in which a precise synchronization 
between the source and the external radiation can be
established. The potential emergence of one-shot single electron
detection \cite{Freise-2020-1,Glattli-2020-1,Thiney-2022-1} calls for an adaptation of 
the present work to these forthcoming detection methods.
These very interesting questions, which go well beyond the
scope of the present work, are left for future investigations.


\acknowledgements{
This work was supported by the ANR grant ``QuSig4QuSense''
(ANR-21-CE47-0012) and by the Joint Research Project “SEQUOIA” (17FUN04)
within the European  Metrology  Programme  for Innovation and
Research (EMPIR) co-financed by the Participating States and from the European Union’s
Horizon 2020 research and innovation programme. P.~D. acknowledges
funding by the Latvian Quantum Initiative during his stay at Universtity
of Latvia in Riga. We warmly thank
C.~Bauerle, A. Feller, D.~Ferraro, 
V.~Kashcheyevs,
for useful discussions as well as C. Altimiras and C. Flindt for
pointing out relevant references.}


\appendix

\section{Notations and normalizations}
\label{appendix/normalizations}

This appendix recalls the basic conventions used for fermionic modes and
electronic single particle states within the present work.

\subsection{Electronic modes and wave packets}
\label{appendix/normalizations/modes}

The mode decomposition for fermionic fields is defined by
\begin{equation}
	\label{eq:normalizations:fermion-modes}
	\psi(t)=\int_{\mathbb{R}} c(\omega)\,\me^{-\mi\omega
	t}\frac{\md\omega}{\sqrt{2\pi v_F}}
\end{equation}
so that these modes obey the canonical anticommutation relations
\begin{equation}
	\label{eq:normalizations:canonical-anticommutations}
	\{c(\omega),c^\dagger(\omega')\}=\delta(\omega-\omega')\,.
\end{equation}
Equivalently, we have
\begin{equation}
	\label{eq:normalizations:fermion-modes:inverted}
        c(\omega) =
        \sqrt{\frac{v_F}{2\pi}}
		\int_{\mathbb{R}} \psi(t)\, \me^{\mi \omega t} \md t\,.
\end{equation}
Given an electronic wave packet described by a normalized wave function
$\varphi_{\text{e}}(x)$
such that
\begin{equation}
	\label{eq:normalizations:spatial}
	\int_{\mathbb{R}}|\varphi_{\text{e}}(x)|^2 \, \md x = 1
\end{equation}
Throughout this paper,
we will use the notation $\varphi_{\text{e}}(t)$ for $\varphi_{\text{e}}(-v_Ft)$ so that
\begin{equation}
	v_F \int_{\mathbb{R}} |\varphi_{\text{e}}(t)|^2 \, \md t = 1\,.
\end{equation}
We define
\begin{equation}
	\label{eq:normalizations:frequency-definition}
	\widetilde{\varphi}_{\text{e}}(\omega)
        =
		v_F \int_{\mathbb{R}} \varphi_{\text{e}}(t)\, \me^{\mi \omega t} \, \md t
\end{equation}
so that 
\begin{equation}
	\label{eq:normalizations:frequency}
	\frac{1}{v_F}\int_{\mathbb{R}}
	|\widetilde{\varphi}_{\text{e}}(\omega)|^2 \, \frac{\md
	\omega}{2\pi} = 1\,.
\end{equation}
The creation operator for the electronic wave packet $\varphi_{\text{e}}$
is then defined as
\begin{subequations}
	\begin{align}
\psi^\dagger[\varphi_{\mathrm{e}}]
&=
v_F \int_{\mathbb{R}} \varphi_{\mathrm{e}}(t) \psi^\dagger(t) \, \md t\\
&=
\int_{\mathbb{R}}
		\widetilde{\varphi}_{\mathrm{e}}(\omega) c^\dagger(\omega)
		\, \frac{\md \omega}{\sqrt{2 \pi v_F}}
\end{align}
\end{subequations}

\section{Bosonization and bosonic modes}
\label{appendix/bosons}

Here, we briefly recall useful formulae in the bosonization of
quantum Hall edge channels as well as for dealing with bosonic
excitations.

\subsection{Bosonization of a quantum Hall edge channel}
\label{appendix/bosons/EMP}

For a right moving quantum Hall edge channel at filling fraction
$\nu=1$, the charge and current densities are related to the free chiral
right moving field $\phi_R(x,t)$ by
\begin{subequations}
\label{eq/bosonization/charge-current-densities}
\begin{align}
\rho_R(x,t)=-\frac{e}{\sqrt{\pi}}\,(\partial_x\phi_R)(x,t)\\
i_R(x,t)=\frac{e}{\sqrt{\pi}}\,(\partial_t\phi_R)(x,t)
\end{align}
\end{subequations}
and the right moving field is decomposed in terms of bosonic
destruction and creation operators $b(\omega)$ and $b^\dagger(\omega)$
for $\omega>0$
satisfying
\begin{equation}
\label{eq/bosonization/canonical-commutators}
\left[ b(\omega),b^\dagger(\omega')\right]=\delta(\omega-\omega')
\end{equation}
by 
\begin{equation}
\label{eq/bosonization/mode-expansion/phi}
	\phi_R(x,t)=\frac{-\mi}{\sqrt{4\pi}}\int_{\mathbb{R}^+}
\left(b(\omega)\,\me^{\mi\omega(x/v_F-t)}-\mathrm{h.c.}\right)\,
\frac{\md\omega}{\sqrt{\omega}}
\end{equation}
This leads to the following mode decompositions for the charge and
current densities
\begin{subequations}
\begin{align}
\label{eq/bosonization/mode-expansion/charge-density}
	   \rho_R(x,t) &=-e\int_{\mathbb{R}^+}
    \sqrt{\omega}\left(b(\omega)\me^{\mi\omega
    x/v}+b^\dagger(\omega)\me^{-\mi\omega
    x/v}\right)\,\frac{\md\omega}{2\pi v}\\
	\label{eq/bosonization/mode-expansion/current}
        i_R(x,t) &=-\frac{e}{2\pi}
	\int_{\mathbb{R}^+}\sqrt{\omega}\left(b(\omega)\me^{\mi\omega
    x/v}-b^\dagger(\omega)\me^{-\mi\omega
    x/v}\right)\,\md\omega\,.
\end{align}
\end{subequations}
thereby connecting the finite frequency current to the EMP creation and
destruction operators: $i_R(\omega>0)=-e\sqrt{\omega}\,b(\omega)$.

The right moving electronic field is then expressed in terms of the
chiral bosonic field by
\begin{equation}
\label{eq/bosonization/fermion}
\psi_R(x,t)=\frac{\widehat{U}}{\sqrt{2\pi a}}\,\me^{\mi\sqrt{4\pi}\,\phi_R(x,t)}
\end{equation}
in which $a$, which has the dimension of a length, is an UV cut-off and
$\widehat{U}$ is the operator lowering the total fermion number in the
edge channel. Using the mode expansion
\eqref{eq/bosonization/mode-expansion/phi}, one arrives at the
expression for $\psi_R(x=0,t)$ in terms of an infinite dimensional
displacement operator:
\begin{equation}
\label{eq/bosonization/fermion/2}
\psi_R(0,t)=\frac{\widehat{U}}{\sqrt{2\pi a}}\,D_b\left[\Lambda_t\right]
\end{equation}
in which $\Lambda_t(\omega)=-\me^{\mi\omega t}/\sqrt{\omega}$.
The infinite dimensional displacement operator with complex valued
functional parameter $\Lambda:\omega\mapsto \Lambda(\omega)$ is defined
as:
\begin{equation}
\label{eq/bosonization/displacement-operator}
D_b\left[\Lambda\right]=\exp\left[\int_0^{+\infty}
	\left(\Lambda(\omega)\,b^\dagger(\omega)-\mathrm{h.c.}\right)\,\md\omega\right]\,.
\end{equation}

\subsection{Conventions for bosonic excitations}
\label{appendix/bosons/modes}

A normalized excitation is described by $\chi(\omega)$ for $\omega>0$
such that
\begin{equation}
\int_0^{+\infty}\left|\chi(\omega\right|^2\,\frac{\md\omega}{2\pi}=1
\end{equation}
so that the single particle state
\begin{equation}
\ket{\chi}=\int_0^{+\infty}\chi(\omega)\,b^\dagger(\omega)\ket{\emptyset}\,\frac{\md\omega}{\sqrt{2\pi}}
\end{equation}
is normalized. If we define the corresponding creation operator
\begin{equation}
b^\dagger[\chi]=\int_0^{+\infty}
\chi(\omega)\,b^\dagger(\omega)\,\frac{\md\omega}{\sqrt{2\pi}}
\end{equation}
and its adjoint $b[\chi]$, these operators obey the commutation
relations
\begin{equation}
\left[b[\chi_1],b^\dagger[\chi_2]\right]=\braket{\chi_1|\chi_2}\,\mathbf{1}
\end{equation}
where $\braket{\chi_1|\chi_2}$ denotes the scalar product 
\begin{equation}
\braket{\chi_1|\chi_2}=\int_0^{+\infty}
\chi_1(\omega)^*\chi_2(\omega)\,
\frac{\md\omega}{2\pi}\,.
\end{equation}
on the space
$\mathcal{L}_2(\mathbb{R}^+)$ of square-summable functions on
$\mathbb{R}^+$. Note that with these conventions
$\chi(\omega)=\sqrt{2\pi}\braket{\omega|\chi}$ where $\ket{\omega}=
b^\dagger(\omega)\ket{\emptyset}$ is the single photon state resolved in
energy.

Finally, when given an orthonormal basis of normalized single particle
states $\ket{\chi_n}$ indexed by $n$, the mode operators $b(\omega)$ can
be expressed in terms of the $b_n=b[\chi_n]$ as
\begin{equation}
b(\omega)=\frac{1}{\sqrt{2\pi}}\sum_n\chi_n(\omega)\,b_n\,.
\end{equation}
Starting from a normalized single particle state $\ket{\chi}$, we can
express $b^\dagger(\omega)$ in terms of $b^\dagger[\chi]$ and of
$b^\dagger[\chi_\perp]$ where $\ket{\chi^\perp_\omega}$ denotes the normalized
projection of $\ket{\omega}$ on the space of single particle excitations
orthogonal to $\ket{\chi}$:
\begin{equation}
	\label{eq/mode-decomposition}
b^\dagger(\omega)= \braket{\chi|\omega}\,b^\dagger[\omega]
+\sqrt{1-|\braket{\chi|\omega}|^2}\,b^\dagger[\chi_\omega^\perp]\,.
\end{equation}


\section{The classically driven edge channel}
\label{appendix/top-gate-EMP-scattering}

In this appendix, we discuss two models for a classically driven edge
channel. In the first model discussed in Sec.
\ref{appendix/top-gate-EMP-scattering/single-particle}, we consider that a classical drive $U(t)$
is directly applied to the electrons propagating within the edge channel.
However, in experiments, the external drive is not directly applied to
the edge channels but to a top gate. This leads us to the second model
discussed in Sec. \ref{appendix/EMP-scattering/top-gate} 
which involves a capacitive coupling between a finite length region of
the edge channel and a top gate to which the time-dependent classical
voltage is applied. We then discuss how this second model reduces
to the first one in the regime of weak Coulomb interactions.

\subsection{Direct coupling to the external voltage}
\label{appendix/top-gate-EMP-scattering/single-particle}

Let us first consider electrons propagating within a chiral edge channel
with Fermi velocity $v_F$ and experiencing in the $0\leq x\leq l$ a
time dependent $U(x,t)$. 

\subsubsection{EMP scattering}

The starting point is the equation of motion for the chiral bosonic
field $\phi_R(x,t)$ built from the EMP modes of the
chiral edge channel:
\begin{equation}
	\label{eq/classical-drive/1}
	(\partial_t+v_F\partial_x)\phi_R(x,t)=\frac{e\sqrt{\pi}}{h}\,U(x,t)
\end{equation}
This equation can be solved using the method of characteristics:
\begin{align}
	\phi_R(x+v_F&\tau,t+\tau)=\phi_R(x,t)\nonumber \\
	&+\frac{e\sqrt{\pi}}{h}\int_0^\tau
	U(x+v_F\tau',t+\tau')\,\md\tau'
	\label{eq/classical-drive/2}
\end{align}
which gives the outgoing field $\phi_{R,\text{out}}(t)=\phi_R(l,t)$ in
terms of the incoming field $\phi_{R,\text{in}}(t)=\phi_R(0,t)$ and
of the time and space dependent potential $U(x,t)$:
\begin{align}
	\phi_{R,\text{out}}(t)&=\phi_{R,\text{in}}(t-l/v_F)\nonumber \\
	&+
	\frac{e\sqrt{\pi}}{h}\int_0^{l/v_F}U(v_F\tau',t+\tau'-l/v_F)\,\md\tau'\,.
	\label{eq/classical-drive/3}
\end{align}
The field and therefore the EMP modes propagate ballistically at
velocity $v_F$ and the time and space dependent voltage adds a source
term. 

For comparison with subsequent models, let us consider the case where
$U(x,t)$ is uniform within the $|x|\leq l/2$ region and equal to the
time-dependent potential $U(t)$ and compute how the
$b(\omega)$ EMP annihilation operator is scattered. 
Specializing Eq. \eqref{eq/classical-drive/3}, re-expressing it in the Fourier domain
and using the $b(\omega)$ operators leads to:
\begin{equation}
	\label{eq/classical-drive/4}
	b_{\text{out}}(\omega)=\me^{i\omega\tau_l}b_{\text{in}}(\omega)
	+\frac{\mi e l\sqrt{\omega}}{\hbar v_F}\,f\left(\frac{\omega
		l}{v_F}\right)\,U(\omega)\,.
\end{equation}
where $\tau_l=l/v_F$ is the electronic time of flight across the region of length
$l$ to which the potential is applied and $f(X)=(\me^{\mi X}-1)/\mi X$.
Rewriting this scattering formula in terms of incoming and outgoing
currents leads to
\begin{equation}
	\label{eq/classical-drive/current-input-output}
	i_{\text{out}}(\omega)=\me^{\mi\omega\tau_l}i_{\text{in}}(\omega)
	-\mi \omega C_q(\omega)\,U(\omega)
\end{equation}
in which we recognize the $\omega$-dependent dynamical quantum
capacitance
\begin{equation}
	\label{eq/classical-drive/finite-frequency-Cq}
	C_q(\omega)=\frac{e^2l}{hv_F}\,f\left(\frac{\omega l}{v_F}\right)
\end{equation}
which, in the present case, reduces to the expected $C_q=e^2l/hv_F$ at low frequencies.

\subsubsection{Electronic phase}

Eq. \eqref{eq/classical-drive/3} can then be used to compute how the
electronic field is propagated using 
Eq.~\eqref{eq/bosonization/fermion/2}:
\begin{subequations}
	\begin{align}
		\psi_{\text{out}}(t)&=\me^{\mi\vartheta_U(t)}\psi_{\text{in}}(t-l/v_F)\\
		\vartheta_U(t)&=\frac{e}{\hbar}\int_0^{l/v_F}U(v_F\tau',t+\tau'-l/v_F)\,\md\tau'
	\end{align}
\end{subequations}
This corresponds to a time dependent single particle scattering matrix
for the electrons:
\begin{equation}
	S(t,t')=\delta(t-t'-l/v_F)\,\me^{\mi \vartheta_U(t)}
\end{equation}
which precisely described the propagation at constant velocity $v_F$ of a
particle of charge $-e$, traveling across the region where the potential $U(x,t)$ is
present between 
space-time coordinates $(0,t-l/v_F)$ and $(l,t)$. For such a
particle, the accumulated phase $\vartheta_U(t)$ is the potential seen along a trajectory
that connects $(0,t-l/v_F)$ to $(l,t)$. 

\subsection{The classically driven top gate}
\label{appendix/EMP-scattering/top-gate}

\subsubsection{EMP scattering}

We now consider a top gate classically driven by a time dependent
voltage $V_g(t)$ and capacitively coupled
to a chiral edge channel. This region is modeled as a capacitor with
geometric capacitance $C_g$ involving the top gate and part of a chiral
edge channel of length $l$ (see Fig.~\ref{fig/top-gate}) in
in a discrete element description. The
two conductors are in total influence.
We want to get the expression of the outgoing EMPs modes $b_\text{out}$
in terms of the incoming ones $b_\text{in}$ and of $V_g$.

The basic equations for this system are the equation of motion for the
chiral bosonic field $\phi_R(x,t)$, field built from the EMP modes of the
chiral edge channel:
\begin{equation}
	\label{eq/capacitive-gate/1}
	(\partial_t+v_F\partial_x)\phi_R(x,t)=\frac{e\sqrt{\pi}}{h}\,U(x,t)
\end{equation}
$U(x,t)$ is the potential seen by the electrons. If we assume that $U(x,t)$ is uniform in the
$|x|\leq l/2$ region, it can be calculated by adding the potential drop
across the capacitor to $V_g(t)$:
\begin{equation}
	\label{eq/capacitive-gate/2}
	U(t)-V_g(t)=\frac{Q(t)}{C_g}
\end{equation}
where $Q(t)$ denotes the excess charge in the $|x|\leq l/2$ of the
chiral edge channel. Using the bosonization expression
\begin{equation}
	\label{eq/capacitive-gate/3}
	Q(t)=\frac{e}{\sqrt{\pi}}\left(\phi_R(-l/2,t)-\phi_R(l/2,t)\right)
\end{equation}
in addition of Eqs.~\eqref{eq/capacitive-gate/1} and
\eqref{eq/capacitive-gate/2}, we can obtain a closed differential
equation for $\phi_R  (x,t)$ within
the $|x|\leq l/2$ region. This method is identical to the one used to
discuss the Coulomb interaction effects within a chiral edge channel in Ref.
\cite{Cabart-2018-1}.

The final result can be expressed as connecting the outgoing EMP modes
$b_{\text{out}}(\omega)$ 
to the incoming EMP
modes $b_{\text{in}}(\omega)$ and
$V_G(\omega)$. This relation is best expressed in terms of the incoming and
outgoing currents using
$i(\omega)=-e \sqrt{\omega}\,
	b(\omega)$:
\begin{equation}
	\label{eq/capacitive-gate/current/input-output}
	i_\text{out}(\omega)=t(\omega)
	i_\text{in}(\omega)+\frac{e^2}{h}(1-t(\omega))\,V_G(\omega)\,.
\end{equation}
in which, using $X=\omega l/v_F$:
\begin{equation}
		\label{eq/capacitive-gate/transmission}
		t(\omega) =\me^{\mi X}\frac{1+\alpha\, f^*(X)}{1+\alpha\,f(X)}
\end{equation}
where as before $f(X)=(\me^{\mi X}-1)/\mi X$ and
	\begin{equation}
		\label{eq/capacitive-gate/coupling-constant}
		\alpha=\frac{e^2l}{hv_FC_g}
\end{equation}
a dimensionless coupling constant whose physical meaning is discussed in
Sec.~\ref{sec/radar-equation/Leviton}.

\subsubsection{Impedance 
\& discussion}

\begin{figure}[h]
	\centering
	\includegraphics[width=70mm]{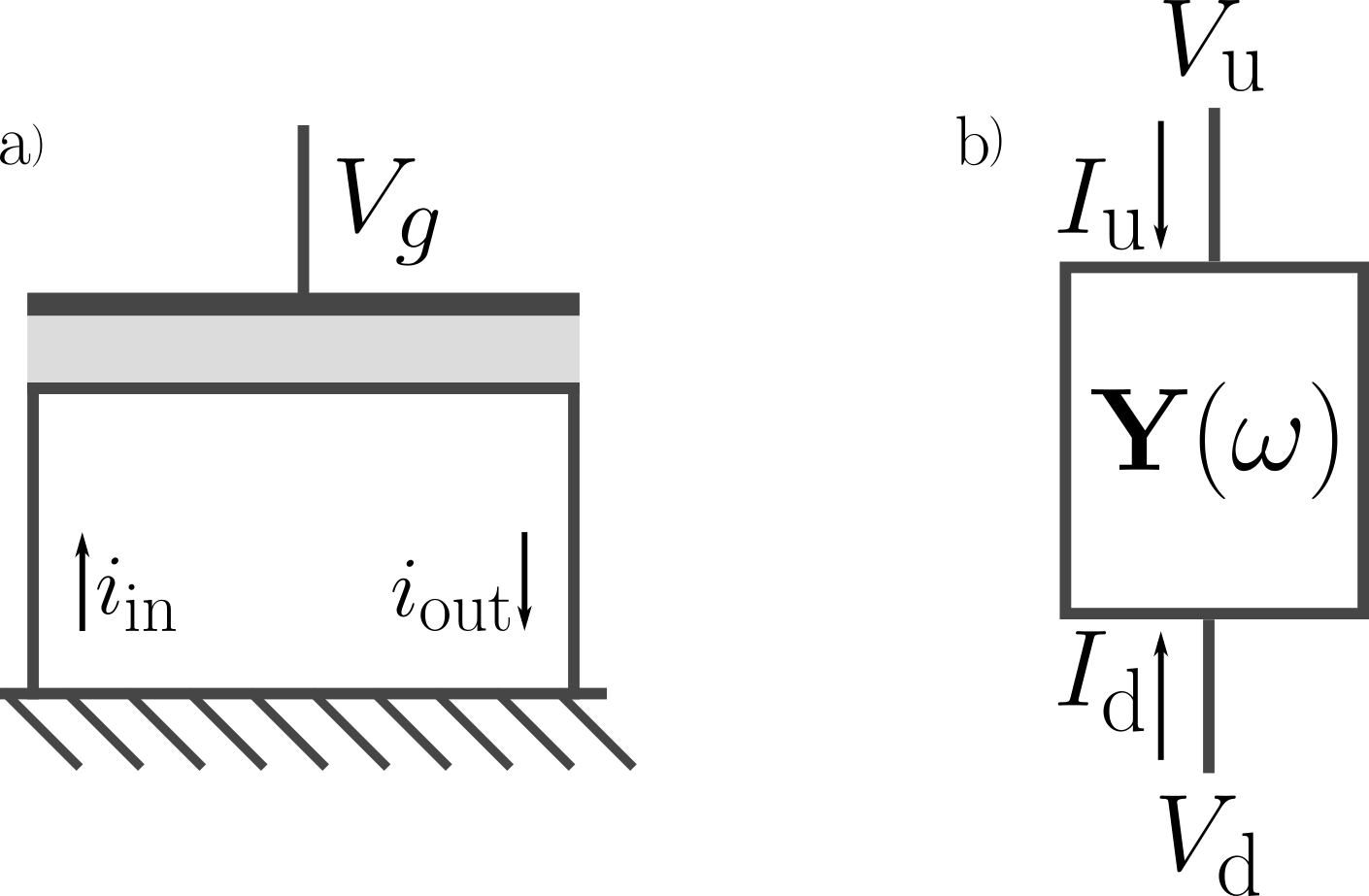}
	\caption{\label{fig/dipole} (a) Folding of the edge channel
	capacitively coupled to the classical top gate to obtain an
	electrical
	dipole. (b) Drawing of an electrical dipole with
	$\mathbf{Y}(\omega)$ its admittance matrix, $V_\text{u}$,
	$V_\text{d}$, $I_\text{u}$ and $I_\text{d}$ respectively denote the
	voltage and the average electrical current in the two connecting
	leads. In panel (a), 
	$V_\text{u}=V_g$, $V_\text{d}=0$ and
	$I_\text{d}=i_\text{in}-i_\text{out}$.}
\end{figure}

Folding the edge channel as shown on Fig.~\ref{fig/dipole}-a enables us to
consider the quantum $RC$ circuit formed by the top gate and the folded
edge channel. Because of the total mutual influence, its admittance
matrix defined by
\begin{equation}
\mathbf{Y}_{\alpha,\beta}(\omega)=\frac{\partial \langle
i_{\alpha}(\omega)\rangle}{\partial V_\beta(\omega)}\,
\end{equation}
where $(\alpha,\beta)\in\{u,d\}^2$ is both
gauge invariant and charge conserving \cite{Buttiker-1993-1} and therefore
determined by a single finite frequency admittance
	\begin{equation}
		\mathbf{Y}(\omega)=Y(\omega)
	\begin{pmatrix}
		1 & -1\\
		-1 & 1
	\end{pmatrix}
\end{equation}
with
\begin{equation}
Y(\omega)=\frac{e^2}{h}\left(1-t(\omega)\right)=\frac{e^2}{h}
\frac{1-\me^{\mi X}}{1+\alpha f(X)}\,.
\label{eq/capacitive-gate/impedance}
\end{equation}
Since $t(\omega)=\me^{\mi\theta(\omega)}$ because of energy conservation and
of the absence of dynamical degrees of freedom in the classically driven
top-gate, $R_KY(\omega)=1-t(\omega)$ spans a circle of radius $1$ centered on $1$
in the complex plane\footnote{The radius one reflects the relaxation
resistance $R_q=R_K/2$ of the circuit predicted in 
Ref.~\cite{Buttiker-1993-2}.}. Its low frequency expansion 
\begin{equation}
	R_KY(\omega)\simeq -\mi \frac{\omega l/v_F}{1+\alpha}
	+\frac{1}{2}\frac{(\omega l/v_F)^2}{(1+\alpha)^2}
	+\cdots
\end{equation}
up to second order
corresponds to an $RC$ circuit with resistance $R_K/2$ and
electrochemical capacitance $C_\mu$ such that
$R_KC_\mu=l/(1+\alpha)v_F$. 

\begin{figure}
    \includegraphics[width=75mm]{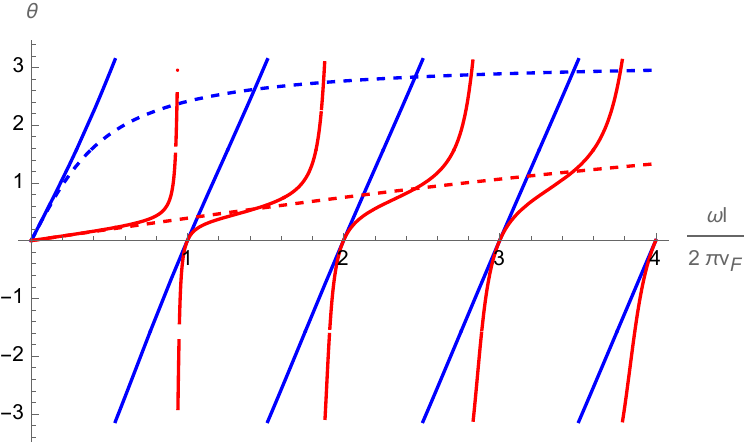}\\
    \includegraphics[width=75mm]{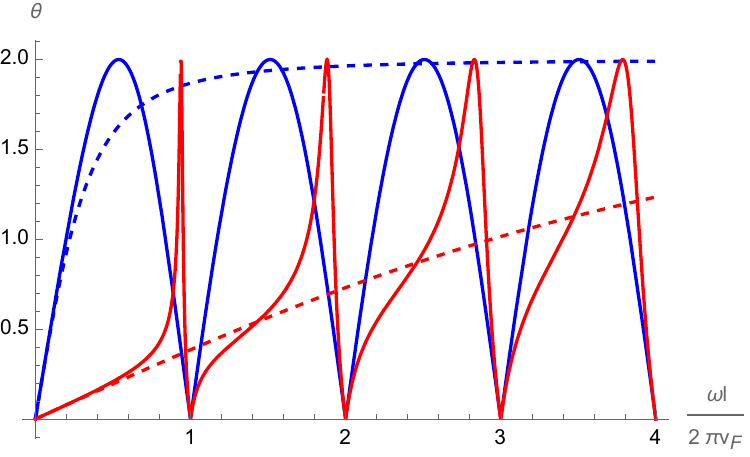}
    \caption{\label{fig/top-gate-impedance} Top panel: Plot of the phase
	$\text{Arg}\left[t(\omega)\right]$ as a function of $\omega l/2\pi
	v_F$ in the voltage locked 
	($\alpha=1/5$, blue full line) and Coulomb blocked 
	($\alpha=15$, red full line) regimes. In each case,
	the angle
	$\text{Arg}\left[1-R_KY(\omega)\right]$ for the corresponding $R_qC_\mu$ circuit is plotted
    in dashed lines with the same
    colors. Bottom panel: plots of $|1-t(\omega)|$ for the same
    examples as in the top panel (coloring identical to the top panel).}
\end{figure}

Figure~\ref{fig/top-gate-impedance} shows the behavior of
$\text{Arg}(t(\omega))$ as well as of $|R_KY(\omega)|=|1-t(\omega)|$ 
as a function of $X=\omega l/v_F$. The former describes how the phase of
the transmission varies with $\omega$ whereas the latter describes the
filtering of the applied voltage $V_g(\omega)$ by the top gate. As can
be seen on the bottom panel of Fig.~\ref{fig/top-gate-impedance}, the
filtering exhibit a texture even at low $\alpha$. However, in this case,
it is very similar to the filtering that would arise from applying 
$V_g(t)$ to the electrons in the $|x|\leq l/2$ region as in 
paragraph~\ref{appendix/top-gate-EMP-scattering/single-particle} whereas
at large $\alpha$, this behavior is only recovered for
$\omega \gtrsim 1/R_KC_\mu\gg v_F/l$ whereas for $\omega 
R_KC_\mu\lesssim 1$, filtering selects
frequencies close to $\omega\simeq 2\pi nv_F/l$ with $n$ being a
positive integer. 

In general, 
the gate
voltage $V_g(t)$ is filtered 
into an effective time dependent voltage $U_{\text{eff}}(t)$ via
\begin{equation}
	\label{eq/capacitive-gate/filtering}
	U_{\text{eff}}(\omega)= \frac{V_g(\omega)}{1+\alpha\,f\left(\omega
		l/v_F\right)}
\end{equation}
The other effect of the top gate is to induce a non linear EMP transmission
phase $\theta(\omega)$ which is responsible for electronic decoherence
beneath the top gate \cite{Cabart-2018-1}. On the other hand, the
single particle model discussed in
paragraph~\ref{appendix/top-gate-EMP-scattering/single-particle} does
not lead to electronic decoherence since, in this case, 
Eq.~\eqref{eq/classical-drive/current-input-output}
corresponds to a dispersionless $\theta(\omega)=\omega l/v_F$.

\section{Two counter propagating edge channels}
\label{appendix/two-channels-EMP-scattering}

In this Appendix, we consider the EMP scattering matrix for a model of
radiation coupler involving two counter-propagating integer Quantum Hall edge
channels, capacitively coupled over a region of length $l$. The
geometric capacitance of this length $l$ capacitor will be denoted by
$C_g$. We assume that these two length $l$ conductors are in total
mutual influence, an hypothesis ensuring maximal coupling between
them.

This modeling has been used in Ref. \cite{Delgard-2021-1} for
a rectangular quantum Hall bar of length $l$ at filling fraction
$\nu=1$. The EMP scattering matrix   can thus be directly
extracted from Ref. \cite{Delgard-2021-2} and we will discuss its
various limiting regimes.

\subsection{The EMP scattering matrix}

Denoting by $b$ the EMP modes of the MZI edge channels and $a$ the EMP
modes of the radiation edge channel, the EMP scattering matrix 
$S(\omega)$ is given by the result of Sec. \MakeUppercase{\romannumeral 3} of 
Ref. \cite{Delgard-2021-2} specialized for $\nu=1$:
\begin{subequations}
    \label{eq/scattering/totally-screened}
    \begin{align}
    \label{eq/scattering/totally-screened/off-diagonal}
    S_{ba}(\omega) &=S_{ab}(\omega)=
    \frac{-\mi Xf(X)}{2+\alpha f(X)}\\
    \label{eq/scattering/totally-screened/diagonal}
    S_{aa}(\omega)&=S_{bb}(\omega)=1-S_{ba}(\omega)\,.
    \end{align}
\end{subequations}
in which $X=\omega l/v_F$ and
$f(X)=(\me^{\mi X}-1)/\mi X$.

The dimensionless coupling 
$\alpha=C_q/C_g$ 
is the inverse ratio of the
geometric capacitance $C_g$ to the quantum capacitance $C_q=e^2l/hv_F$
of a single edge channel of length $l$. It reflects the strength of
Coulomb interactions effects: up to a
numerical factor, it is the ratio
of the electrostatic energy $e^2/2C_g$ for a single electron charge in
a capacitance $C_g$ to the kinetic energy scale
$\hbar v_F/l$ for a single electron added in a length $l$ closed edge
channel. 

The regime of low $\alpha$ ($C_g\gg C_q$) is the regime where
the voltage drop at the capacitor
formed by the two length $l$ facing counter-propagating regions of length
$l$ can be neglected: both channels see the same potential. 
On the other hand, $\alpha\gg 1$ is the regime of
strong coupling in which Coulomb energy is so large that charging these
length $l$ regions is almost impossible. This is the Coulomb blocked
regime. 

We shall now discuss the limiting
forms of the EMP scattering region in these two regimes.
A key point in interpreting the results is the relation between the EMP
scattering matrix and finite frequency impedances. Under the hypothesis
of total screening, there are no current leaks to any external grounded
conductor. Considering the $b_{\text{in}}$ and
$b_{\text{out}}$ as the incoming and outgoing modes for a $B$ lead and
the $a_{\text{in}}$ and $a_{\text{out}}$ modes as the corresponding
modes for the $A$ leads, the finite frequency admittance of this
conductor is 
\begin{equation}
	Y(\omega)=\frac{e^2}{h}\,S_{ba}(\omega)
\end{equation}
At low enough frequency, this dipole can be
viewed as an $RC$ circuit with finite frequency admittance
\begin{equation}
	Y(\omega)=-\mi C_\mu\omega+RC_\mu^2\omega+\mathcal{O}(\omega^3)
\end{equation}
which leads to 
\begin{equation}
	C_\mu=\frac{C_q}{2+\alpha}=\frac{C_gC_q/2}{C_q/2+C_g}
	\quad\&\ R=R_K\,.
\end{equation}
We recover the expression of the quantum capacitance as the series
addition of two single channel quantum capacitances (one for the $a$
modes and the other for the $b$ modes) with the geometric capacitance
$C_g$. The total resistance $R_K$ is the series addition of the two contact
resistances $R_K/2$ for the two folded edge channels.

The total screening hypothesis 
leads to $S_{bb}(\omega)=1-S_{ba}(\omega)$. On the other hand, energy
conservation leads to $|S_{bb}(\omega)|^2+|S_{ab}(\omega)|^2=1$.
Parametrizing 
\begin{subequations}
	\begin{align}
		S_{bb}(\omega)&=\frac{1}{2}\left(1+\xi(\omega)\right)\\
		S_{ba}(\omega)&=\frac{1}{2}\left(1-\xi(\omega)\right)
	\end{align}
\end{subequations}
thus leads to $|\xi(\omega)|=1$. Consequently, the parametric plot of
$S_{ba}(\omega)$ in the complex plane is a circle of radius $1/2$
centered on the point $z=1/2$. Note that this is also the case for an RC circuit
with resistance $R_K$ and capacitance $C_\mu$. 
In the end, the EMP scattering matrix can be fully characterized by the
description of the angle $\vartheta(\omega)=\text{Arg}(\xi(\omega))$.

\begin{figure}
	\includegraphics[width=75mm]{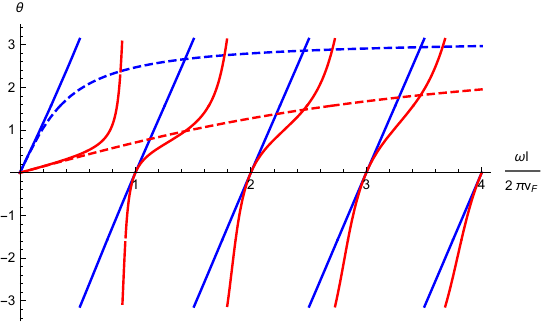}\\
	\includegraphics[width=75mm]{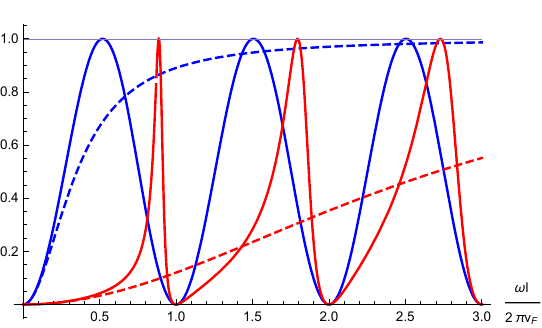}
	\caption{\label{fig/phase-plot} Top panel: Plot of the angle
	$\vartheta(\omega)$ as a function of $\omega l/2\pi v_F$ for two
	different values of the coupling: $\alpha=1/5$ (blue full line) and
	$\alpha=15$ (red full line). For each of these
	values of $\alpha$, the angle for the $R_KC_\mu$ circuit is plotted
	in dashed lines with the same
	colors. Bottom panel: plots of $|S_{ba}(\omega)|^2$ for the same
	examples as in the top panel (coloring identical to the top panel).}
\end{figure}

Fig. \ref{fig/phase-plot} displays the phase $\vartheta(\omega)$ as a
function of $\omega l/v_F$ in units of $2\pi$ for both the voltage
locked and Coulomb blocked regimes
together with the graphs for the $RC$ circuit reproducing
the low frequency behavior of $Y(\omega)$.
The probability $|S_{ab}(\omega)|^2$ for an EMP to be
transmitted from the radiation channel into the MZI upper arm is
plotted in the lower panel. 

\subsection{The voltage locked regime}

The $\alpha\rightarrow 0$ limit of the EMP scattering matrix
given by Eq. \eqref{eq/scattering/totally-screened} is 
\begin{subequations}
    \label{eq/scattering/totally-screened/weak-coupling}
    \begin{align}
    \label{eq/scattering/totally-screened/weak-coupling/off-diagonal}
		S_{ba}(\omega) &=S_{ab}(\omega)=\frac{1}{2}\left(1-\me^{\mi
		X}\right)\\
    \label{eq/scattering/totally-screened/weak-coupling/diagonal}
		S_{aa}(\omega)&=S_{bb}(\omega)=\frac{1}{2}\left(1+\me^{\mi
		X}\right)\,.
    \end{align}
\end{subequations}
whith therefore implies that $\vartheta(\omega)=\omega l/v_F$. This is
consistent with the full blue line behavior displayed on the upper-panel
of Fig.
\ref{fig/phase-plot} ($\alpha=1/5$).

In this regime, the electrochemical capacitance is dominated by the
quantum capacitance's contribution: $C_\mu\simeq C_q/2$ which means that
the corresponding $RC$ time scale is $R_KC_\mu\simeq l/2v_F$. The fact
that even at vanishing $\alpha$, $S_{ba}(\omega)$ is non-zero comes from
the fact that this ``weak coupling limit'' is not a limit without
interactions between the two counter-propagating edge channels: 
it only means that the voltage drop at the capacitance
vanishes and that both edge channels see the same voltage. In this
regime, the transmission probability exhibits sinusoidal oscillations as
a function of $\omega l/v_F$ just like a micro-wave directional coupler
or two strongly coupled copropagating edge channels with local interactions would do
(see Ref.~\cite{Degio-2010-4}). 

Avoiding this physically counter-intuitive behavior in the $\alpha\ll 1$
regime requires considering a model without the total screening
hypothesis. However, note that such a model would ensure a less
efficient coupling between the external quantum radiation and the
electrons propagating within the MZI.

\subsection{The Coulomb blocked regime}

Increasing $\alpha$ introduces non-linearities in the phase
$\vartheta(\omega)$ as shown on the upper panel of Fig.~\ref{fig/phase-plot}. 
The electrochemical capacitance increases since it is not dominated by
the geometric capacitance $C_\mu\sim C_g$ ($C_g\ll C_q$). 
The $R_KC_\mu\sim R_KC_g$ time scale is then much shorter than the free
electron time of flight $R_KC_q=l/v_F$.
As a result $1/R_KC_\mu$ increases thus
explaining the flatter low frequency behavior of $\vartheta(\omega)$
in terms of $\omega l/v_F$ see on the upper panel of
Fig.~\ref{fig/phase-plot}. The lower panel of this figure
shows a strong distortion of the oscillations in the transmission
probability $|S_{ab}(\omega)|^2$ for $\omega R_KC_\mu\lesssim 1$. It
still reaches unity for $\omega l/v_F\equiv 0\pmod{2\pi}$ but with
narrow resonances for frequencies $\omega/2\pi$ close to integer multiples of
$v_F/l$. These resonances become broader with increasing order and  
the typical oscillations of a micro-wave
directional coupler are recovered for
for $\omega R_KC_\mu\gtrsim 1$.

Consequently, at low energy $\omega \lesssim 2\pi l/v_F$, 
there is almost no EMP scattering amplitude in the very large
coupling limit and the EMP transmission amplitude corresponds to a
vanishing time of flight across the length $l$ interaction region. This
is the result of the zero charge constraint at infinite coupling:
every charge density disturbance has to go out of the interaction region
immediately and leads to no charge density change within the interaction
region (hence the vanishing inter-channel EMP scattering amplitude).

At higher $\omega$, the modulus of
decoherence amplitude goes down at each resonance
which becomes broader at higher $\omega$. This comes from the frequency
selectivity of the radiation coupler for
a finite but large
$\alpha$: it can transmit
photons only at the resonances shown on the lower panel of 
Fig.~\ref{fig/phase-plot}. This is an interesting feature for building
a frequency selective detector, the narrowest resonance being the first
one for $\omega/2\pi=l/v_F$. 

When $\omega R_KC_\mu \gtrsim
1$, the directional coupler behavior expected in the voltage
locked regime of the previous Section is recovered. This is the regime
where EMP modes have enough energy to overcome the Coulomb energy
$e^2/2C_g$.


\section{Derivation of the radar equation}
\label{appendix/radar}

In this Appendix, the single radar equation for the outgoing
average electrical current is derived. 
We will first derive it within the framework of time dependent single particle
scattering theory (linear electron quantum optics) using a technique
which can then be adapted to 
the presence of Coulomb interaction effects which
belong to the non-linear regime of electron quantum optics.

\subsection{Time-dependent single particle scattering approach}
\label{appendix/radar/single-particle}

We consider a MZI as depicted on Fig. \ref{fig/simple-radar}-(a) and discussed
in Sec. \ref{sec/radar-physics/simple-model}. Let us recall that the single electron
scattering matrices of the two QPC are assumed to be energy independent
and given by:
\begin{equation}
	\label{eq/1e-scattering/1}
S_\alpha = 
\begin{pmatrix}
\sqrt{T_\alpha} & \mi \sqrt{R_\alpha} \\
\mi \sqrt{R_\alpha} & \sqrt{T_\alpha}
\end{pmatrix}
\end{equation}
in which $T_\alpha$ and $R_\alpha$ respectively denote the transmission
and reflection probabilities at QPC $\alpha=A$ or $B$
($T_\alpha+R_\alpha=1$). 

The main idea is to express the outgoing electron field in branch $1$ 
in terms of incoming fields by back-tracing it from the output to the
input of
the MZI interferometer. Assuming free propagation long the branch $2$,
with time of flight $\tau_2$ leads to
\begin{subequations}
	 \label{eq/1e-scattering/2}
\begin{align}
\psi_{1\mathrm{out}}(t) &=
\sqrt{T_B}\,\psi_{1,B_-}(t)+\mi\sqrt{R_B}\,\psi_{2,B_-}(t)\\
&= \mi\sqrt{R_B}\,\me^{-i\phi_{\mathrm{AB}}/2}\left(
	\sqrt{T_A}\,\psi_{2\mathrm{in}}(t-\tau_2)\right. \nonumber \\ 
&+\mi\left.
\sqrt{R_A}\,\psi_{1\mathrm{in}}(t-\tau_2)\right)\\
&+\sqrt{T_B}\,\me^{\mi\phi_{\mathrm{AB}}/2}\psi_{1,B_-}(t)
	\label{eq:outgoing:1B}
\end{align}
\end{subequations}
in which $\psi_{\alpha,B_-}(t)$ denotes the fermionic field right before
the QPC $B$ and $\psi_{\alpha_{\mathrm{in}}}(t)$ denotes the incoming
fields right before QPC $A$. Equation \eqref{eq:outgoing:1B} involves
the outgoing fermionic field from the radiation coupler. In
the case where propagation within this region can be described by a time dependent linear
scattering, we can relate it linearly to the incoming field 
$\psi_{1,A_+}$ by
\begin{equation}
	\label{eq/1e-scattering/3}
	\psi_{1,B_-}(t)=\int_{\mathbb{R}}R(t,t')\psi_{1,A_+}(t')\,\md t'
\end{equation}
which assumes that no electron can be injected from any other channel
than the branch $1$ of the MZI\footnote{We will discuss other cases later
but for the average current, any other contribution would be
irrelevant.}. 
This enables us to write down the fully general expression for the
outgoing electrical current
$i_{1\mathrm{out}}(t)=-ev_F:(\psi_{1\mathrm{out}}^\dagger\psi_{1\mathrm{out}}):(t)$
in terms of the incoming electronic fields.
We obtain the outgoing current operator $i_{1\mathrm{out}}(t)$ as
\begin{equation}
	i_{1\mathrm{out}}(t) = \widehat{I}_0(t)-e\left(
	\me^{\mi\varphi_{\mathrm{AB}}}\widehat{I}_+(t)
	+\me^{-\mi\varphi_{\mathrm{AB}}}\widehat{I}_-(t)\right)
\end{equation}
in which, at the operator level
\begin{widetext}
\begin{equation}
	\label{eq/1e-scattering/5}
	\widehat{I}_+(t) = -\mi
	\sqrt{T_BR_B}\int_{\mathbb{R}}R(t,t')\left[\sqrt{T_A}\psi_{2_{\text{in}}}^{\dagger}
	-\mi\sqrt{R_A}\psi_{1_{\text{in}}}\right](t-\tau_2)
	\left[\sqrt{T_A}\psi_{1_{\text{in}}}+
	\mi\sqrt{R_A}\psi_{2_{\text{in}}}\right](t')\,\md t'\,.
\end{equation}
\end{widetext}
When computing the average current, only terms that contain the same
numbers of $\psi_{\alpha_{\text{in}}}^\dagger$ and
$\psi_{\alpha_{\text{in}}}$ are retained since
the MZI is fed by two independent electron sources. Consequently
the AB-flux dependent part of the average current is 
$\langle I_+(t)\rangle=-e\sqrt{R_AT_AR_BT_B}\,X_+(t)$ where
\begin{equation}
	\label{eq/1e-scattering/6}
	X_+(t)=\int_{\mathbb{R}}
	R(t,t')\left(\mathcal{G}^{(e)}_{1\text{in}}-
	\mathcal{G}^{(e)}_{2\text{in}}\right)(t'|t-\tau_2)\,
	\md t'\,.
\end{equation}
This is the time domain electron radar equation given by Eq.
\eqref{eq/radar-equation/time-domain}.
Note that, within the single particle scattering formalism, this equation
is valid for any electron source, not necessarily emitting a single
electron excitation.

\subsection{The interacting case}
\label{appendix/radar/non-linear}

\subsubsection{The EMP scattering approach}
\label{appendix:radar:non-linear:EMP-scattering}

Let us now consider the case where Coulomb interactions
cannot be neglected within the radiation coupler.
We shall model it using the EMP scattering formalism discussed in 
Sec.~\ref{sec/radar-equation/radiation coupler}. 

Note that the starting point of Eq.
\eqref{eq/1e-scattering/2} is still valid. But, contrary to the previous paragraph, 
the main challenge is now to backtrack
the fermionic field along the branch $1$ of the MZI. 
Equivalently, we have to express $\psi_{1,B_-}(t)$
in terms of the incoming fields $\psi_{\alpha_{\mathrm{in}}}$ for
$\alpha=1$, $2$. In order to do so, we consider the fermionic field
$\psi_{1,B_-}(t)$, expressed it in terms of the outgoing
EMP modes and use the EMP scattering
matrix to express it in terms of the incoming modes into the radiation coupler. 

To simplify the
notation, the EMP modes along the branch $1$ of the MZI will be denoted
by $b_{\alpha}(\omega)$ with $\alpha=\text{in}$ or $\alpha=\text{out}$
depending whether they are incoming (position $A_+$) or outgoing
(position $B_-$). In  the same way, the electromagnetic modes within the
radiation channel are denoted by
$a_{\alpha}(\omega)$. The scattering matrix describing the coupling
between the edge channel and the electromagnetic modes is 
\begin{equation}
	S(\omega)=\begin{pmatrix}
		S_{bb}(\omega) & S_{ba}(\omega) \\
		S_{ab}(\omega) & S_{aa}(\omega)
	\end{pmatrix}
\end{equation}
so that
\begin{equation}
	\begin{pmatrix}
	b_{\text{out}}(\omega)\\
		a_{\text{out}}(\omega)
	\end{pmatrix}
	=S(\omega)\,
	\begin{pmatrix}
    b_{\text{in}}(\omega)\\
        a_{\text{in}}(\omega)
    \end{pmatrix}
\end{equation}
Using the bosonization formula for the fermionic field, (see Eq. 
\eqref{eq/bosonization/fermion/2}) and back-propagating the bosonic mode operators across the 
radiation coupler, 
$\psi_{1\text{out}}(t)$ can be expressed as
\begin{equation}
	\psi_{1,B_-}(t)=\me^{\mi\Theta}\psi_{1,A_+}(t)\,D_{b_1}[(S_{bb}^*-1)\Lambda(t)]
	\otimes D_a[S_{ba}^*\Lambda(t)]
\end{equation}
where the phase $\Theta$ is independent from $t$. Therefore, we need 
to compute the correlator $\mathcal{G}^{(e)}_{\rho_i,
B_-}(1,t|2,t)$ which is
equal to
\begin{equation}
	\label{eq:big-correlator}
	\langle
	\psi^\dagger_{2,A_+}(t-\tau_2)\psi_{1,A_+}(t)\,D_{b_1}[(S_{bb}^*-1)\Lambda(t)]
	\,D_a[S_{ba}^*\Lambda(t)]\rangle
\end{equation}
in which the correlator is taken over the incoming many-body state
$\rho_i=\rho_S\otimes\rho_{\text{em}}$ which is the tensor product of the many body electronic state
$\rho_S$ injected by the
source by the incoming radiation
state $\rho_{\text{em}}$ for the $a(\omega)$ modes.

\subsubsection{The Franck-Condon factor}

Let us now discuss how the correlator \eqref{eq:big-correlator} can be
evaluated. First of all, the part that depends on the incident radiation
state $\rho_{\text{em}}$ can be singled out thanks to the identity:
\begin{align}
\langle D_a[S_{ba}^*\Lambda(t)]\rangle_{\rho_{\text{em}}}&=
\langle D_a[S_{ba}^*\Lambda(t)]\rangle_{\ket{0}}\nonumber \\ 
&\times 
\langle :D_a[S_{ba}^*\Lambda(t)]:\rangle_{\rho_{\text{em}}}
\end{align}
in which we have introduced the bosonic normal ordering $:\cdots :$.
All the dependence in the incident radiation state $\rho_{\text{em}}$
is thus contained in the average value of the normal ordered
displacement operator $:D_a[S_{ba}^*\Lambda(t)]:$ for the $a(\omega)$
modes. The average value $\langle
D_a[S_{ba}^*\Lambda(t)]\rangle_{\ket{0}}$ is taken over the vacuum state
for the $a(\omega)$ modes which means that it can be reabsorbed into the
correlator given by Eq. \eqref{eq:big-correlator} except that this time
the quantum average is taken over the state $\rho_{S,0}=\rho_S\otimes
\ket{0}\bra{0}$. This can be summarized by
\begin{subequations}
\label{eq/non-linear/factorization/simple}
	\begin{align}
\mathcal{G}^{(e)}_{\rho_i,
		B_-}(1,t|2,t)&=\mathcal{G}^{(e)}_{\rho_{S,0},
B_-}(1,t|2,t)\\ 
		& \times \langle
		:D_a[S_{ba}^*\Lambda(t)]:\rangle_{\rho_{\text{em}}}
		\end{align}
\end{subequations}
since $\rho_{S,0}=\rho_S\otimes \ket{0}\bra{0}$ corresponds to a
situation
where no incident radiation is sent onto the MZI. Therefore, the correlator 
$\mathcal{G}^{(e)}_{\rho_{S,0},
B_-}(1,t|2,t)$ is exactly the one appearing when computing the average
current flowing out of the MZI in the absence of electromagnetic
radiation sent onto it via the $a(\omega)$ modes. This problem
corresponds to the problem of electronic decoherence within the MZI.

In the end, the effect of the radiation injected into the radiation
coupler is
described by the factor
\begin{equation}
	\mathcal{F}_{\rho_{\text{em}}}(t)=
	\langle :D_a[S_{ba}^*\Lambda(t)]:\rangle_{\rho_{\text{em}}}\,
\end{equation}
which is the exact analogous of the Franck-Condon factor that appears in
the spectroscopy of complex molecules \cite{Condon-1926-1}. 

\subsubsection{Electronic propagation contribution}

The electronic coherence $\mathcal{G}^{(e)}_{\rho_{S,0},
B_-}(1,t|2,t)$ is more difficult to evaluate because it corresponds to
the outgoing single electron coherence after propagation across the MZI
in the presence of Coulomb interactions within the branch $1$ of the
MZI. It turns out that simple and physically transparent
expressions can be found in the case where $S$ is an ideal single
electron source. In this case, the state $S$ is of the form
\begin{equation}
	\rho_S=\psi_{1\text{in}}^\dagger[\varphi_e]\ket{F}
	\bra{F}\psi_{1\text{in}}[\varphi_e]
\end{equation}
where $\ket{F}$ corresponds to the Fermi sea with chemical potential $\mu=0$ in the two
incoming electronic channels of the MZI and
$\psi^\dagger_{1\text{in}}[\varphi_e]$ creates a single electron
excitation with with wave function $\varphi_e$ injected into the MZI
(position $A_-$). 

Expressing $\psi_{1\text{in}}^\dagger[\varphi_e]$ 
in terms of $\psi_{1,A_+}^\dagger[\varphi_e]$ and
$\psi_{2,A_+}^\dagger[\varphi_e]$ enables us to show that 
\begin{widetext}
\begin{subequations}
\begin{align}
\mathcal{G}^{(e)}_{\rho_{S,0},
B_-}(1,t|2,t)&= -\mi\sqrt{R_AT_A}\times
\label{eq/MZI-coherence/no-radiation}
\langle 
F,0_a|\psi_{2,A_+}[\varphi_e]\psi^\dagger_{2,A_+}(t-\tau_2)
	\psi_{1,B_-}(t)\psi_{1,A_+}^\dagger[\varphi_e]|F,0_a\rangle\\
&=
-\mi\sqrt{R_AT_A}\,\times 
\langle
F_2|\psi_{2,A_+}[\varphi_e]\psi^\dagger_{2,A_+}(t-\tau_2)|F_2\rangle\times
\langle F_1,0_a|\psi_{1,B_-}(t)\psi_{1,A_+}^\dagger[\varphi_e]|F_1,0_a\rangle
\end{align}
\end{subequations}
\end{widetext}
in which $\ket{F,0_a}$ denotes the tensor product of the Fermi sea in both
branches of the MZI and the ground state for the environmental
modes. Since $\varphi_e$ is an electronic excitation above the Fermi
level, the contribution associated with propagation along the branch $2$
of the MZI can be readily evaluated:
\begin{equation}
\langle \psi_{2,A_+}[\varphi_e]\psi_{2,A_+}(t-\tau_2)\rangle_{\ket{F_2}}
=\varphi_e(t-\tau_2)^*\,.
\end{equation}
We are thus left with evaluating:
\begin{align}
\langle
\psi_{1,B_-}(t)&\psi_{1,A_+}^\dagger[\varphi_e]\rangle_{\ket{F_1,0_a}}
= \nonumber\\
	\int_{\mathbb{R}} &v_F\varphi_e(t')\langle
\psi_{1,B_-}(t)\psi_{1,A_+}(t')\rangle_{\ket{F_1,0_a}}\,\md t'\,.
\end{align}
The time domain amplitude 
\begin{equation}
	\label{eq/Z-amplitude/time-domain/definition}
	\mathcal{Z}_{1}(\tau)=
	v_F\langle
\psi_{1,B_-}(\tau)\psi^\dagger_{1,A_+}(0)\rangle_{\ket{F_1,0_a}}
\end{equation}
is the elastic single electron scattering amplitude across the
branch 1 of the MZI. It is related to the elastic scattering amplitude
$\widetilde{\mathcal{Z}}_{1}(\omega)$ computed in 
Refs.~\cite{Degio-2009-1,Ferraro-2014-2} by a Fourier transform:
\begin{equation}
	\mathcal{Z}_1(\tau)=\int_0^{+\infty}\widetilde{\mathcal{Z}}_{1}(\omega)\,
	\me^{-\mi\omega\tau}
\frac{\md\omega}{2\pi}\,.
\end{equation}
Evaluating 
$\widetilde{\mathcal{Z}}_{1}(\omega)$ can be done using the expressions
given in Ref.~\cite{Ferraro-2014-1,Cabart-2018-1} which we recall here for completeness:
\begin{equation}
	\widetilde{\mathcal{Z}}_{1}(\omega)=1+\int_0^{\omega}B(\omega')\,\md\omega'
\end{equation}
in which $B(\omega)$ is the solution of the integral equation
\begin{equation}
\omega
B(\omega)=S_{bb}(\omega)-1+\int_0^{\omega}B(\omega')(S_{bb}(\omega')-1)\md\omega'\,
\end{equation}
with initial condition $B(0^+)=(\md S_{bb}/\md\omega)(\omega=0^+)$. 
Finally, we obtain:
\begin{equation}
	\langle
	\psi_{1,B_+}(t)\psi_{1,A_+}^\dagger[\varphi_e]\rangle_{\ket{F_1,0_a}}=
\int_{\mathbb{R}}\varphi_e(t')\mathcal{Z}_1(t-t')\,\md t'\,.
\end{equation}
The correlator $\langle
\psi_{2,A_+}[\varphi_e]\psi_{2,B_-}^\dagger(t-\tau_2)\rangle_{\ket{F_2}}$
could also
rewritten in a similar way
\begin{equation}
\langle 
\psi_{2,A_+}[\varphi_e]\psi_{2,B_-}^\dagger(t)\rangle_{\ket{F_2}}
=\int_{\mathbb{R}}\varphi_e(t')^*\mathcal{Z}_2(t-t')^*\,\md t'
\end{equation}
using the elastic scattering amplitude
$\mathcal{Z}_2(\tau)=\delta(\tau-\tau_2)$ corresponding to ballistic
propagation during time of flight $\tau_2$. 

Note that this form of the correlator $\langle
\psi_{2,A_+}[\varphi_e]\psi_{2,B_-}(t)\rangle$ remains valid in the
presence of electronic decoherence along branch $2$ of the MZI provided it
is not caused by direct or indirect interactions (bath mediated) between
the two branches of the MZI. This means that there must be no crosstalk
between the two branches: each of these branch do interact with their
own environment which are prepared in their ground states. In this case,
we should use the elastic scattering amplitude
$\widetilde{\mathcal{Z}}_{2}(\omega)$
for $\omega>0$ to define $\mathcal{Z}_2(\tau)$. The elastic scattering
amplitude $\widetilde{\mathcal{Z}}_{2}(\omega)$ 
can be computed in terms of the finite frequency
admittance of the branch $2$.

\subsubsection{General result}

Let us finally collect the general result for the inter-branch coherence
$\mathcal{G}_{\rho_{S,0},B_-}(1,t|2,t)$ right before the second QPC in the
general situation where each of the branches involves Coulomb
interactions, and possibly a coupling to its own radiation channel fed by
the vacuum state. In the absence of external radiation, the expression 
\begin{widetext}
\begin{equation}
	v_F\mathcal{G}^{(e)}_{\rho_{S,0},B_-}(1,t|2,t)=\int_{\mathbb{R}^2}
v_F\varphi_e(t-\tau_1)\,\varphi_e(t-\tau_2)^*\,\mathcal{Z}_1(\tau_1)\,\mathcal{Z}_2(\tau_2)\,
\md\tau_1\,\md\tau_2\,
\end{equation}
\end{widetext}
has a physically transparent interpretation: each wave packet gets
propagated according to the elastic scattering amplitude
$\mathcal{Z}_\alpha(\tau)$ along the corresponding branch.

In the presence of incoming radiation arriving on each 
radiation coupler of each branch $\alpha$ in independent respective quantum stated
described by the density operators
$\rho_{\text{em},\alpha}$, Eq.~\eqref{eq/non-linear/factorization/simple}
generalizes to
\begin{subequations}
\label{eq/non-linear/factorization/comparator}
\begin{align}
	\mathcal{G}^{(e)}_{\rho_i,B_-}(1,t|2,t)&=\mathcal{G}^{(e)}_{\rho_{S,0},B_-}(1,t|2,t)\times
\nonumber \\
&\prod_{\alpha=1,2}\Big\langle
:D_{a_\alpha}\left[\left(S^{(\alpha)}_{ba}\right)^*\Lambda(t)\right]:\Big\rangle_{\rho_{\text{em},\alpha}}
\end{align}
\end{subequations}
In the end, the interference signal $X_+(t)$ is given by
\begin{align}
	X_+(t)&=v_F\int_{\mathbb{R}^2} 
	\varphi_e(t-\tau_1)\varphi_e(t-\tau_2)^*\times \nonumber \\
	&R_{\text{eff},1}(t,t-\tau_1)
	\,R_{\text{eff},2}(t,t-\tau_2)\,
\md \tau_1,\md \tau_2
\end{align}
in which the effective single particle scattering amplitudes are given
by the product of the elastic scattering amplitude by the Franck-Condon
factor:
\begin{equation}
	R_{\text{eff},\alpha}(t,t')=\mathcal{Z}_\alpha(t-t')\,\mathcal{F}_{\rho_{\text{em},\alpha}}(t)
\end{equation}
thus leading to the single electron radar equation
\eqref{eq/radar-equation/time-domain} which reduces to Eq.
\eqref{eq/radar-equation/time-domain/2} in the case of ballistic
propagation in time $\tau_2$ along branch $2$ and no coupling to
external radiation on this branch.

Last but not least, in the presence of Coulomb interactions, the radar
equation is valid only for an incoming single electron excitation
injected into the electronic MZI! 


\section{Connection to full counting statistics}
\label{appendix/FCS}

\begin{figure}
\centering
\includegraphics[height=5cm]{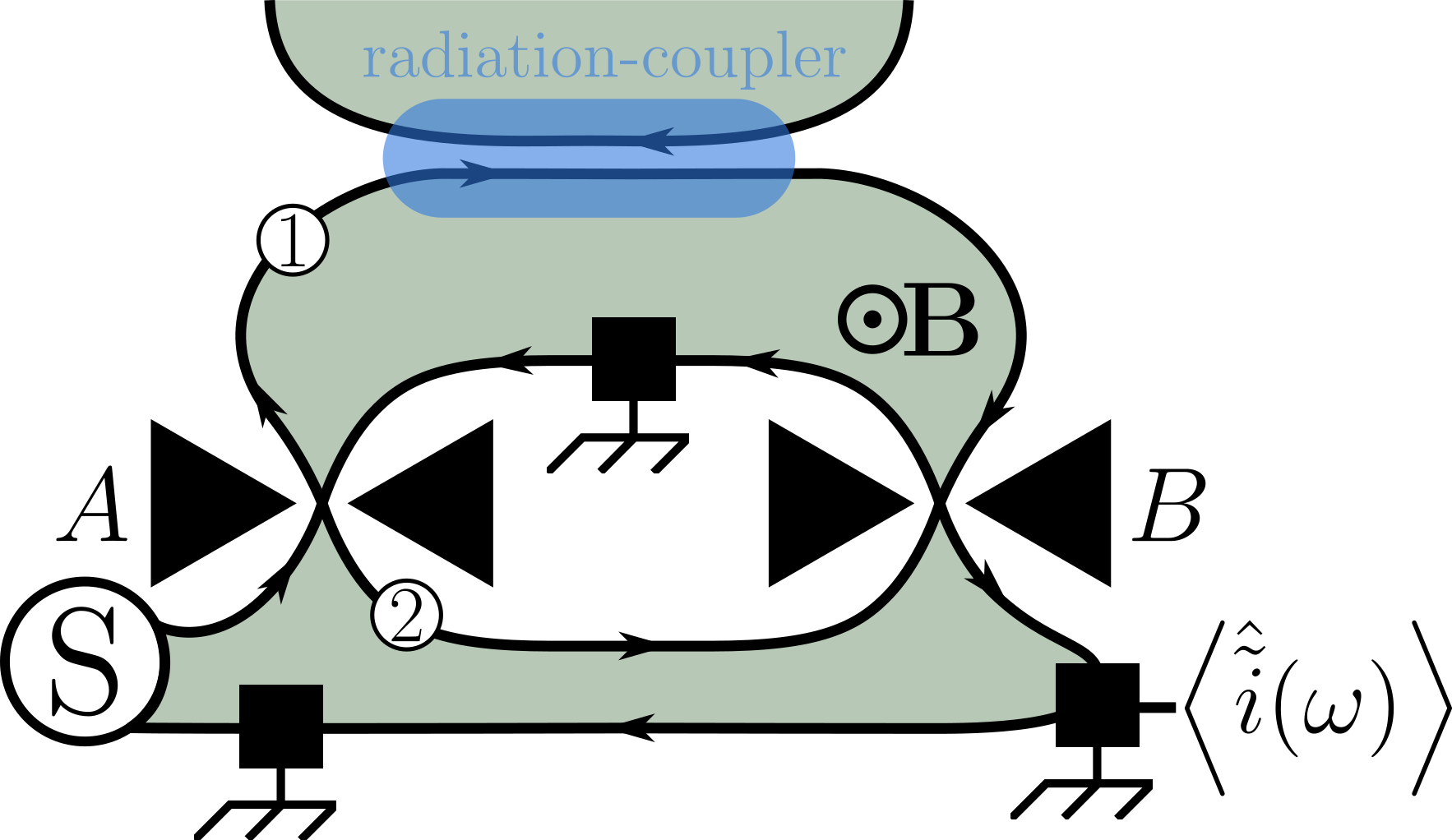}
	\caption{\label{fig/iqhe-radar} (Colors online) Sample design of 
the electron radar using chiral edge channels in
	the integer quantum Hall regime (2DEG in green). 
The blue region
where two counter-propagating quantum Hall edge
channels are facing each other at short ($\lesssim
\SI{100}{\nano\meter}$) distance forms a radiation coupler between
the channel $1$ of the MZI
and the radiation channel in which the radiation to be analyzed is
sent. The quantity of interest is the average outgoing current from
the MZI and more precisely the first harmonic in the Aharonov-Bohm
phase.}
\end{figure}

In the context of experiments performed on AlGaAs/AsGa systems in the
quantum Hall regime, the radiation channel may also be a chiral quantum
Hall edge channel, as depicted on Fig. \ref{fig/iqhe-radar}-(a). In
this case, 
$\mathcal{F}_{\rho_{\text{em}}}(t)$ can also be
interpreted in terms of a form of full counting statistics for 
electronic transport in the radiation channel 
as noticed in Ref.~\cite{Dasenbrook-2016-1}. Since this work relied
on a specific interaction model as well as on single particle
scattering, we present here a full many body derivation independent on
the details of the Coulomb interaction model.

When the radiation channel is a single integer quantum Hall edge
channel,
the incoming electrical current arriving into the interaction
region via this edge channel is, for $\omega>0$:
\begin{equation}
    i_{\text{in}}(\omega)=-e\sqrt{\omega}\,a_{\text{in}}(\omega)
\end{equation}
Introducing
\begin{equation}
	\label{eq/FCS/Gamma}
	\Gamma_{ba}(\tau)=
	\int_0^{+\infty}\frac{S_{ba}(\omega)}{-\mi\omega}\,\me^{-i\omega\tau}\frac{\md\omega}{2\pi}
	+
    \mathrm{c.c.}\,
\end{equation}
the normalized filtered charge operator
\begin{equation}
    N(t)=\int_{\mathbb{R}}\Gamma_{ba}(t-\tau)\,\frac{i_{\text{in}}(\tau)}{-e}\,\md\tau
\end{equation}
associated with the windowing function $\tau\mapsto \Gamma_{ba}(t-\tau)$
can be used to provide a compact expression for
$\mathcal{F}_{\rho_{\text{em}}}(t)$:
\begin{equation}
    \label{eq/radar/non-linear-1e/FCS-expression}
    \mathcal{F}_{\rho}(t)=\Big\langle :\,\me^{2\pi
	\mi N(t)}\,:\Big\rangle_{\rho_{\text{em}}}\,.
\end{equation}
The $2\pi
N(t)$ operator represents the quantum phase kicks associated with the
incoming electrical current felt by a localized electron propagating
within the MZI. Note that the exponential of this quantum phase kick
operator is normal ordered with respect to the bosonic modes
$a_{\text{in}}(\omega)$. One can then connect this ordering
to the time ordering in order to establish the precise connection to
time ordered electrical current correlators.

Finally, let us comment on the interpretation of the windowing function
$\Gamma_{ba}(t-\tau)$.
Since $\Gamma_{ba}(\omega)=\mi S_{ba}(\omega)/\omega$ for $\omega>0$,
the relation between the EMP scattering matrix and finite frequency
admittances \cite{Safi-1999-1,Degio-2010-1,Bocquillon-2013-2,Cabart-2018-1} leads to
\begin{equation}
	\label{eq/radar/FCS/Gamma-definition}
    \Gamma_{ba}(\omega)=\frac{S_{ba}(\omega)}{-\mi
    \omega}=\frac{R_K}{-\mi \omega}\,\frac{\partial
    I_b^{(\text{out})}(\omega)}{\partial V_a^{(\text{in})}(\omega)}
\end{equation}
in which $I_b^{(\text{out}}(\omega)$ denotes the average total current going
out from the interaction region within the MZI interferometer -- the
propagation channel for the $b$ EMP modes -- and
$V_a^{(\text{in})}(\omega)$ denotes a classical voltage drive applied to the
incoming channel of the radiation channel in which the EMP $a$ modes
propagate. This equation thus connects
$\Gamma_{ba}(\omega)$ to the finite frequency admittance of the
electrical dipole associated with the interaction region. Since the
coupling is purely capacitive, one expects this finite frequency
response to be capacitive, that is of the form
\begin{equation}
    \frac{\partial
	I_b^{(\text{out})}(\omega)}{\partial V_a^{(\text{in})}(\omega)}=-\mi
    \omega\, C_{ba}^{(\text{eff})}(\omega)
\end{equation}
in which the effective frequency dependence capacitance
$C_{ba}^{(\text{eff})}(\omega)$ goes to a non zero capacitance in the
low frequency limit. Consequently, $\Gamma_{ba}(\tau)$ appears as the
effective response rate associated with this capacitance
\begin{equation}
	\label{eq/radar/FCS/Gamma-time}
    \Gamma_{ba}(\tau)=\int_0^{+\infty}\,R_KC_{ba}^{(\text{eff})}(\omega)\,
    \me^{-\mi\omega\tau}\frac{\md\omega}{2\pi}+\text{c.c.}
\end{equation}


\section{The radar equation in the frequency domain}
\label{appendix/frequency-domain}

Let us introduce
\begin{equation}
    \widetilde{R}(\omega_+,\omega_-)=\int_{\mathbb{R}^2}
    R(t,t')\,\me^{\mi(\omega_+t-\omega_-t')}\md t\,\md t'
\end{equation}
which is directly proportional to the amplitude for scattering an
incoming electron at energy $\hbar\omega_-$ to the energy
$\hbar\omega_+$. Denoting by $\widetilde{X}_+(\omega)$ the Fourier transform of the
signal $X_+(t)$ at frequency $\omega/2\pi$,
Eq.~\eqref{eq/radar-equation/time-domain} then takes the form
\begin{align}
\label{eq/radar-equation/frequency-domain}  
    \widetilde{X}_+(\omega)=\int_{\mathbb{R}^2}
    &v_F\widetilde{\Delta\mathcal{G}}^{(e)}_S(\omega_-,\omega_+-\omega)\nonumber \\
    &\me^{\mi(\omega-\omega_+)\tau_2}\widetilde{R}(\omega_+,\omega_-)
    \,\frac{\md\omega_+\,\md\omega_-}{(2\pi)^2}
\end{align}
in which 
\begin{equation}
   \label{eq/coherence/frequency-domain}
\Delta\widetilde{\mathcal{G}}^{(e)}_\rho(\omega_+|\omega_-)=
   \int_{\mathbb{R}^2}
   \Delta\mathcal{G}^{(e)}_\rho(t|t')\,\me^{\mi(\omega_+t-\omega_-t')}
   \md t\,\md t'
\end{equation}
denotes the excess
single electron coherence in the frequency domain. 
Eq.~\eqref{eq/radar-equation/frequency-domain} determines the
interference contribution to the finite frequency average electrical current
in the $1_{\text{out}}$ branch of the MZI.
    
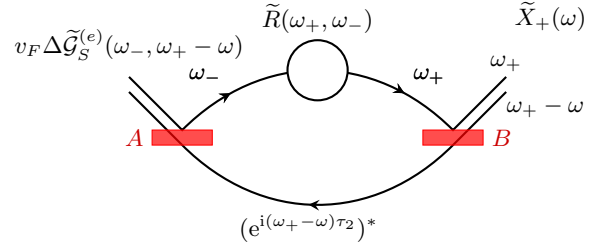
\begin{figure}
    \centering
    \begin{tikzpicture}[beamsplit/.style={red,fill=red,fill opacity=0.7},
		ray/.style={thick},
		markat/.style={
		decoration={
			markings,
			mark={at position #1}},
		postaction={decorate}}]
	\def\bsthick{0.1}
	\def\bswidth{0.4}
	\def\bssplit{1.8}

	\def\raylength{1}
	\pgfmathsetmacro{\ticklength}{0.2*\bswidth}

	\node[,above=+6pt,font=\small] at ($ (-\bssplit, 0) +
	(90+45:\raylength)$) 
		{$v_F \Delta \widetilde{\mathcal{G}}^{(e)}_{S}(\omega_-, \omega_+-\omega)$};
	\node[right,font=\small] at ($ (\bssplit, 0) + (45:\raylength) + (0, 0.9)$)
		{$\widetilde{X}_+(\omega)$};
	\node[above=-1pt,font=\small] at ($ (\bssplit, \bsthick) + (45:\raylength)$)
		{$\omega_+$};
	\node[below right=-1pt and -3pt,font=\small] at ($ (\bssplit, -\bsthick) + (45:\raylength)$)
		{$\omega_+ - \omega$};

	\draw[ray,
		markat={0.3 with {\arrow{stealth}\node[above left] {$\omega_-$};}},
		markat={0.7 with {\arrow{stealth}\node[above right] {$\omega_+$};}}
		]
		(-\bssplit, \bsthick) +(90+45:\raylength) --
		(-\bssplit, \bsthick) .. controls +(45:1.5) and +(90+45:1.5) ..
		(\bssplit, \bsthick) node[midway,fill=white,circle,draw,minimum height=0.8cm] (R) {}
		-- +(45:\raylength)
		;
	\node[above=-1pt] at (R.north) {$\widetilde{R}(\omega_+,\omega_-)$};

	\draw[ray,xscale=-1,
		markat={0.52 with {\arrow{stealth}\node[below] {$(\mathrm{e}^{\mathrm{i} (\omega_+ - \omega) \tau_2})^*$};}}]
		(-\bssplit, -\bsthick) +(+90+45:\raylength) --
		(-\bssplit, -\bsthick) .. controls +(-45:1.5) and +(-90-45:1.5) ..
		(\bssplit, -\bsthick) -- +(45:\raylength);

	\draw[beamsplit]
		(-\bswidth-\bssplit,-\bsthick) rectangle (+\bswidth-\bssplit,+\bsthick); 
	\node[left, color=red!80!black] at (-\bswidth-\bssplit, 0) {$A$};
	\draw[beamsplit]
		(-\bswidth+\bssplit,-\bsthick) rectangle (+\bswidth+\bssplit,+\bsthick); 
	\node[right, color=red!80!black] at (+\bswidth+\bssplit, 0) {$B$};
\end{tikzpicture}
    \caption{\label{fig/electron-radar/interpretation/frequency}
    Physical interpretation of the linear radar equation \eqref{eq/radar-equation/frequency-domain} in
    the frequency domain showing the contribution of incoming single
    electron coherence in the frequency domain to
    $\widetilde{X}_+(\omega)$ for $\omega>0$. Definition of the two paths is the same as in Fig.
\ref{fig/electron-radar/interpretation/time}.}
\end{figure}

Its physical interpretation 
is quite clear: 
for $\omega>0$, the average finite frequency current
$\langle i_{1_{\text{out}}}(\omega)\rangle$ probes the outgoing electronic coherence
beween $\omega_+$ and $\omega_+-\omega$ for all $\omega_+$.
The latter is propagating along
the branch 2 of the MZI with time of flight $\tau_2$, hence picking a phase
$\me^{\mi (\omega_+-\omega)\tau_2}$. Since it
appears in the complex conjugated amplitude, this leads to
the phase factor $\me^{\mi(\omega-\omega_+)\tau_2}$
in Eq. \eqref{eq/radar-equation/frequency-domain}. On branch 1 of the
MZI, the electron enters the radiation coupler with
energy $\hbar\omega_-$ and exits with energy $\hbar\omega_+$ with an
amplitude $\widetilde{R}(\omega_+,\omega_-)$. Therefore, the
contribution of the outgoing coherence between $\omega_+$ and $\omega_+-\omega$
comes from an incoming coherence between $\omega_-$ and
$\omega_+-\omega$ emitted by the source $S$ as summarized on
Fig.~\ref{fig/electron-radar/interpretation/frequency}.

\section{Limiting regimes of the single radar equation}
\label{appendix/limiting-regimes}

We now discuss the limiting regimes of a time or of a frequency
resolved single electron excitation. The resulting limiting form of the
electron radar equation will be relevant whenever the time
(resp. frequency) extension of the probe is much smaller than the
typical time (resp. frequency) length scales of the effective scattering matrix.
As expected, time resolved wave packets are well suited to explore the time
dependence of the scattering whereas energy resolved excitations gives
access to the scattering amplitude in the frequency domain.

\subsection{Energy resolved probes}

We consider an resolved excitation with Gaussian lineshape,
centered at energy $\hbar\omega_e$ and with linewidth $\gamma_e\gg
|\omega_e|$. Its
wave function in the energy domain, defined by
Eq.~\eqref{eq:normalizations:frequency-definition} is:
\begin{equation}
   \widetilde{\varphi}_e(\omega)=\mathcal{N}
	\Theta(\text{sign}(\omega_e)\omega)\,
	\me^{-(\omega-\omega_e)^2/2\gamma_e^2}
\end{equation}
where the normalization condition \eqref{eq:normalizations:frequency}
gives\footnote{The truncation for positive or negative energies can be neglected
provided $\gamma_e\ll |\omega_e|$.}
$\mathcal{N}^2\gamma_e/v_F\sqrt{4\pi}\simeq 1$. For $\omega_e>0$ we are
dealing with an electronic excitation, whereas for $\omega_e<0$ we are
dealing with a hole excitation.
In the limit where $\gamma_e\ll |\omega_e|$ is much smaller than the
scales of variation of $\widetilde{R}(\omega_+,\omega_-)$,
the electron radar signal in the frequency domain given by
Eq. \eqref{eq/radar-equation/frequency-domain} then gives
\begin{equation}
	\label{eq/limiting-regimes/X/frequency-resolved-in-appendix}
   \widetilde{X}_{+}(\omega;\tau_2)\simeq \frac{\gamma_e}{\sqrt{\pi}}
   \me^{-\mi \omega_e \tau_2} \widetilde{R}(\omega +
   \omega_e,\omega_e)\,.
\end{equation}
Whenever $\omega_e$ and $\omega+\omega_e$ have the same sign, we are
accessing the electron to electron or hole to hole effective scattering whereas
whenever $\omega_e$ and $\omega+\omega_e$ do not have the same sign, we
are accessing electron to hole or hole to electron effective scattering.

\subsection{Time resolved probes}

We consider a normalized Gaussian single electron wave packet:
\begin{equation}
\label{eq/probe/gaussian/time}
\varphi_\text{tr}(t)=\frac{1}{\sqrt{v_F \sigma_e\sqrt{\pi}}}
	\me^{-(t-t_e)^2/2\tau^2}
\end{equation}
with $t_e$ the emission time and $\tau_e$ the duration of the
single electron wave packet. 
Using Eq. \eqref{eq/radar-equation/time-domain} to compute the average
time dependent current and brutally taking the limit $\sigma_e\rightarrow 0$ leads
to, at first order in $\sigma_e$:
\begin{equation}
   X_{+}(t)\simeq \tau_e R(t_e+\tau_2,t_e)\,\delta(t-\tau_2-t_e)
\end{equation}
The condition $t=t_e+\tau_2$ comes from ballistic propagation along the
reference arm of the MZI and, $X_+(t)$ is then proportional to the
amplitude for the electron to enter branch $1$ at $t_e$ and exit it at
time $t$. 

However, this naive computation does not take into account the
constraint that the incoming single electron excitation must be
restricted to positive energies due to the presence of the Fermi sea
which is not the case for the wave packet considered in
Eq.~\eqref{eq/probe/gaussian/time}.
Introducing a finite energy shift via the factor
$\me^{-\mi\omega_e(t-t_e)}$ in front of the r.h.s of
Eq.~\eqref{eq/probe/gaussian/time} with $\omega_e\gg 2\pi v_F/\tau_e$
leads to a wave packet with
negligible weight at negative energies. But this introduces a rapidly
oscillating term $\me^{-\mi\omega_e(t-t'-\tau_2)}$ in front of 
$R(t,t')=\mathcal{Z}_1(t-t')\mathcal{F}_{\rho_{\text{em}}}(t)$ 
in the radar equation \eqref{eq/radar-equation/time-domain}.
This means that we are probing electronic decoherence at energies close to
$\hbar\omega_e$.
But, as shown in Ref.~\cite{Ferraro-2014-1}, electronic decoherence 
is expected to be worse at high energies and may indeed kill the
interference signal.

In order to mitigate this problem, the proper approach consists of
introducing electronic wave packets that are close to the Fermi surface
and well localized in space such as the Leviton. This is
discussed in Appendix
\ref{appendix/levitons}.


\section{Fringe contrast for Levitons}
\label{appendix/levitons}

Let us consider the case of a Leviton wave-packet of width $\tau_e$. Its
wave function is Lorentzian in the time domain and exponential in the
frequency domain \cite{Grenier-2013-1}. We are
interested
in the interference contribution to the outgoing electrical current for
a Leviton of duration $\tau_e$ injected at time $t_e$. 

Because of the simple expression for the Leviton wave packet in the frequency
domain, the electron radar equation in the frequency domain
(see Eq.~\eqref{eq/radar-equation/frequency-domain}) 
gives us a convenient form for $X_+^{(\text{dc})}$ suitable
for numerical evaluations:
\begin{equation}
	\label{eq/Leviton/X+dc}
    X_+^{(\text{dc})}=\int_{\mathbb{R}}
    \widetilde{\mathcal{F}}_{\rho_{\text{em}}}(\Omega)\,
	\me^{-\mi\Omega(t_e+\tau_2)}\,f_{\tau_e,\tau_2}(\Omega)
    \frac{\md\Omega}{2\pi}
\end{equation}
in which the filter
\begin{equation}
	\label{eq/Leviton/filter}
    f_{\tau_e,\tau_2}(\Omega)=4\pi \tau_e\int_{|\Omega|/2}^{+\infty}
    \widetilde{\mathcal{Z}}_1\left(\omega-\frac{\Omega}{2}\right)\,
    \me^{-2\omega\tau_e}\me^{-\mi(\omega-\frac{\Omega}{2})\tau_2}
    \frac{\md\omega}{2\pi}
\end{equation}
contains all the effects of electronic decoherence along branch $1$ and ballistic
propagation along branch $2$ of the interferometer. 

Because the elastic
scattering amplitude $\widetilde{Z}_1(\omega)$ tends to decrease in modulus with
increasing $\omega$, $f_{\tau_e,\tau_2}(\Omega)$ is expected to
be a low pass filter. The limit $\tau_e\rightarrow 0^+$ may even lead to
vanishing signal since, in the case of the two counter-propagating edge
channel model considered in Appendix~\ref{appendix/two-channels-EMP-scattering}, 
a very short Leviton may experience
fractionalization as it flies across the radiation coupler. For short
wave packets, this would
kill the electronic interference signal.
In the absence of external radiation, since
$\mathcal{F}_{\ket{0}}(t)=1$, we obtain
\begin{equation}
	\label{eq/Leviton/base-signal}
\left[X_+^{(\text{dc})}\right]_{\ket{0}}
=4\pi\tau_e\int_0^{+\infty}\widetilde{\mathcal{Z}}_1(\omega)\,\me^{-2\omega\tau_e}\me^{-\mi\omega\tau_2}
\frac{\md\omega}{2\pi}\,.
\end{equation}
This quantity, which does not anymore depend on $t_e$ but depends on
$\tau_2$ represents the base interference contribution for the dc
outgoing current from the MZI. 

\section{Squeezing detection}
\label{appendix/squeezing-detection}

This appendix presents the computations of the Franck-Condon factor
in the case of a time periodic noise, with specialization to narrow band
squeezed noise. We first briefly recall the structure of a single mode
squeezed vacuum for pedagogy and then proceed to the discussion of
time-periodic noise. In Sec.~\ref{appendix/squeezing/narrowband}, we
show that such a noise can exhibit sub-vacuum fluctuations coming from
two-mode squeezing in a narrow band around a given frequency $\omega_0/2\pi$.
The harmonic structure of the resulting Franck-Condon factor is then
discussed in Sec.~\ref{appendix/squeezing/harmonics}.

\subsection{Squeezed vacuum in a single mode}
\label{appendix/squeezing-detection/squeezed-vacuum}

In this Appendix, we recall the basics needed to describe the squeezed
vacuum for a single mode\footnote{Note that
$\ket{\text{Sq}_0}=\ket{0}$.}:
\begin{equation}
    \label{eq/squeezing/1}
    \ket{\text{Sq}_z}=\me^{z(a^\dagger)^2-z^*a^2}\ket{0}
\end{equation}
where $a$ and $a^\dagger$ are the creation and destruction operators for
a single mode and $z\in\mathbb{C}$. The squeezing operator $\mathbb{S}_z$
equal to $\me^{z(a^\dagger)^2-z^*a^2}$, performs a Bogoliubov
transformation on the original mode operators:
\begin{equation}
        \label{eq/squeezing/2}
    \mathbb{S}_z^\dagger
    a\,\mathbb{S}_z=\cosh(2|z|)\,a+\me^{\mi\varphi}\sinh(2|z|)\,a^\dagger\,.
\end{equation}
This enables us to compute the expectation value of any products of $a$
and $a^\dagger$ in the state $\ket{\text{Sq}_z}$ as the expectation
value of the same expression in terms of the Bogoliubov transformed
operators 
\begin{subequations}
        \label{eq/squeezing/3}
    \begin{align}
        a_z &= \cosh(2|z|)\,a+\me^{\mi\varphi}\sinh(2|z|)\,a^\dagger\\
        a^\dagger_z &=\cosh(2|z|)\,a^\dagger+\me^{\mi\varphi}\sinh(2|z|)\,a
    \end{align}
\end{subequations}
where $\varphi=\mathrm{Arg}(z)$.
More precisely
\begin{equation}
        \label{eq/squeezing/4}
    \langle \mathcal{O}[a,a^\dagger]\rangle_{\ket{\text{Sq}_z}}
    =\langle \mathcal{O}[a_z,a^\dagger_z]\rangle_{\ket{0}}
\end{equation}
Applying this relation to the quadrature
\begin{equation}
        \label{eq/squeezing/5}
    X_\vartheta=\frac{1}{\sqrt{2}}\left(\me^{\mi\vartheta}a+\me^{-\mi\vartheta}a^\dagger\right)
\end{equation}
leads to $\langle X_\vartheta\rangle_{\ket{\text{Sq}_z}}=0$ and
\begin{subequations}
        \label{eq/squeezing/6}
    \begin{align}
        \langle (\Delta X_\vartheta)^2\rangle_{\ket{\text{Sq}_z}}
        &= \frac{1}{2}+\sinh^2(2|z|)\label{eq/squeezing/6b}\\
        &+\cosh(2|z|)\sinh(2|z|)\,\cos(2\vartheta+\varphi)
    \end{align}
\end{subequations}
which shows that the fluctuations are anisotropic in the Fresnel plane
and
that $\text{Arg}(z)$ determines the principal
axes of the ellipsoid of fluctuations. Its extrema are:
\begin{subequations}
        \label{eq/squeezing/7}
\begin{align}
    \max_{\vartheta}\langle (\Delta X_\vartheta)^2\rangle_{\ket{\text{Sq}_z}} &=
    \frac{1}{2}\,\me^{4|z|}\\
    \min_{\vartheta}\langle (\Delta X_\vartheta)^2\rangle_{\ket{\text{Sq}_z}} &=
    \frac{1}{2}\,\me^{-4|z|}
\end{align}
\end{subequations}
The vacuum fluctuations being given by $1/2$, the state
$\ket{\text{Sq}_z}$ appears as squeezed with sub-vacuum fluctuations
compressed at most by a factor $\me^{-4|z|}<1$ as soon as $|z|\neq 0$.
Note that with $\vartheta=\omega_0t$, we recover the $2\omega_0$
oscillations of the
fluctuations of harmonic mode of energy $\hbar\omega_0$.

Since the isotropic part of the fluctuations corresponds to
$\frac{1}{2}+\langle a^\dagger a\rangle_{\ket{\text{Sq}_z}}$,
Eq.~\eqref{eq/squeezing/6b} determines the average photon number in the squeezed
vacuum $\ket{\text{Sq}_z}$:
\begin{equation}
        \label{eq/squeezing/8}
    \langle a^\dagger a\rangle_{\ket{\text{Sq}_z}}=\sinh^2(2|z|)
\end{equation}
Thinking of this harmonic mode as an optical mode, the above discussion
shows that increasing the squeezing parameter $z$ increases the
average number of photons. A squeezed vacuum is thus on average more
noisy than the true vacuum even if increasing $|z|$
decreases its minimal fluctuations.

\subsection{Narrow band time periodic noise}
\label{appendix/squeezing/narrowband}

Let us consider a time periodic current noise with period
$\pi/\omega_0$ which, as discussed in the previous paragraph,
is natural for squeezing around a frequency
$\omega_0/2\pi$.
This is more natural in the spirit of mesoscopic physics where
the radiation source may be a periodically driven conductor at
a given
frequency which would, by time translation invariance,
generate a radiation whose correlators have the same time periodicity.

We then consider the simplest correlation structure for the creation and
destruction operators
$a(\omega)$ and $a^\dagger(\omega)$ that satisfies both the 
requirement of $\pi/\omega_0$
periodicity as well as the requirement that its time-averaged power
spectrum is concentrated in a narrow band around $\omega_0$. This leads
to \cite{Rebora-2023-1} ($\omega_\pm>0$):
\begin{subequations}
	\label{eq/squeezing/10}
	\begin{align}
		\langle a^\dagger(\omega_-)\,a(\omega_+)\rangle &=
		\delta(\omega_+-\omega_-)\,\overline{n}(\omega) \\
		\langle a(\omega_-)\,a(\omega_+)\rangle &= \delta
		(\omega_++\omega_--2\omega_0)\,\xi\left(\frac{\omega_+-\omega_-}{2}\right)
	\end{align}
\end{subequations}
where $\overline{n}(\omega)\geq 0$ denotes the average photon number in
the mode at $\omega$ and $\xi(\Omega)$ denotes the average photon
pair correlation between the modes at frequencies $\omega_0+
\Omega$ and $\omega_0-\Omega$. 
In the narrow band situation where the
emitted power density $\hbar\omega\,\overline{n}(\omega)$ is
concentrated near $\omega_0$, $\overline{n}(\omega)\neq 0$ only for
$|\omega-\omega_0|\lesssim \gamma_0/2$ and $|\xi(\Omega)|\neq 0$
only for $|\Omega|\lesssim \gamma_0/2$. 
Note that the $\delta$-function
constraints in these equations arises from time periodicity as well as
from Cauchy Schwartz inequalities (see
Ref.~\cite{Roussel-2020-1} for analogous considerations in the context
of electron quantum optics). 

Substituting these correlators in
Eq.~\eqref{eq/targets/squeezing/Gaussian-result} leads to:
\begin{equation}
	\label{eq/squeezing/11}
	\left|\mathcal{F}_{\rho_{\text{em}}}(t)\right|
		= \me^{
		\Re\left[\xi_{\text{eff}}\,\me^{-2\mi\omega_0t}\right]-N_{\text{eff}}}
\end{equation}
in which
\begin{subequations}
	\label{eq/squeezing/12}
	\begin{align}
		N_{\text{eff}} &=\int_0^{+\infty}\frac{|S_{ba}(\omega)|^2}{\omega}
		\,\overline{n}(\omega)\,\md\omega\\
		\xi_{\text{eff}} &=\int_{-\omega_0}^{+\omega_0}
		\frac{S_{ba}(\omega_0+\Omega)\,S_{ba}(\omega_0-\Omega)}{\sqrt{\omega_0^2-\Omega^2}}
		\,\xi(2\Omega)\,\md\Omega\,.
	\end{align}
\end{subequations}
In the narrow band approximation in which 
$\overline{n}(\omega)=\bar{n}$ for $|\omega-\omega_0|\leq \gamma_0/2$,
$\xi(\Omega)=\xi$ for $|\Omega|\leq \gamma_0/2$ and where the frequency
dependence of $S_{ba}(\omega)$ around $\omega_0$ is neglected, we obtain
\begin{subequations}
	\label{eq/squeezing/13}
	\begin{align}
	N_{\text{eff}} &= |S_{ba}(\omega_0)|^2 \frac{\gamma_0}{\omega_0}\,\overline{n}\\
		\xi_{\text{eff}} &= |S_{ba}(\omega_0)|^2
		\frac{\gamma_0}{\omega_0}\,
	\me^{2\mi\text{Arg}(S_{ba}(\omega_0))} 
		\xi
	\end{align}
\end{subequations}
Two mode squeezing between $\omega_0+\Omega/2$ and $\omega_0-\Omega/2$
for all $|\Omega|\leq \gamma_0$ leads to
\begin{subequations}
	\label{eq/squeezing/14}
	\begin{align}
		\overline{n} &= \sinh^2(2|z|)\\
		|\xi| &= \sinh(4|z|)/2
	\end{align}
\end{subequations}
where $z$ is the squeezing parameter, assumed to be the same for all
these pairs of modes. Substituting this into
Eqs.~\eqref{eq/squeezing/13} and the result into \eqref{eq/squeezing/11} 
then leads to Eq.~\eqref{eq/examples/squeezing/Franck-Condon/narrowband}.

\subsection{Harmonic decomposition}
\label{appendix/squeezing/harmonics}

We now decompose the time periodic Franck-Condon factor given by
Eq.~\eqref{eq/examples/squeezing/Franck-Condon/narrowband} in Fourier series:
\begin{align}
	\mathcal{F}_{\text{Sq}_z}(t)\simeq\, & 
    \me^{-\Lambda\sinh^2(2|z|)}\times \nonumber\\
	&\sum_{n\in\mathbb{Z}}I_{|n|}\left(
    \Lambda\cosh(2|z|)\sinh(2|z|)\right)\,\me^{-2\mi n\omega_0(t-\tilde{t}_e)}
    \label{eq/examples/squeezing/FC/Fourier}
\end{align}
where $\Lambda=|S_{ba}(\omega_0)|^2/Q_0$
and $I_n$ denotes the modified Bessel function of order $n$.
The time $\tilde{t}_e$ arises from the phase of $S_{ba}(\omega_0)$ as
well as of the squeezing parameter. We will forget it in the following
since it can be absorbed in a redefinition of the Leviton's injection
times.

Because $|S_{ba}(\omega_0)|^2\leq 1$ and $Q_0$ is significantly larger
than $1$, $\Lambda\ll 1$. Remembering also that
experimentally reachable squeezing factors are not very high ($|z|\simeq
0.1755$ for $\SI{3}{\decibel}$ squeezing) and
therefore, $\Lambda\cosh(2|z|)\sinh(2|z|)\lesssim 1$.
Moreover, $I_n(x)\simeq (x/2)^n/n!$ for $n\geq 1$. Consequently, 
the most important contribution comes from
the first harmonics: $n=0$ and $n=\pm 1$ in
Eq.~\eqref{eq/examples/squeezing/FC/Fourier}. Retaining only these
leads to:
\begin{equation}
    \label{eq/examples/squeezing/first-harmonics/result}
    \mathcal{F}_{\text{Sq}_z}(t)\simeq \mathcal{F}_0(z,\Lambda)+
    \me^{-2i\omega_0 t}\mathcal{F}_1(z,\Lambda)+
    \me^{2i\omega_0 t}\mathcal{F}_{-1}(z,\Lambda)
\end{equation}
where, at the lowest non-trivial order in $\Lambda$:
\begin{subequations}
    \label{eq/examples/squeezing/first-harmonics/coefficients}
    \begin{align}
        \mathcal{F}_0(z,\Lambda)&=\me^{-\Lambda\sinh^2(2|z|)}
        I_0\left(\Lambda\cosh(2|z|)\sinh(2|z|)\right)\nonumber \\
        &\simeq
        1-\Lambda\sinh^2(2|z|)+\mathcal{O}(\Lambda^2)\\
        \mathcal{F}_{\pm 1}(z,\Lambda)&=\me^{-\Lambda\sinh^2(2|z|)}
        \,I_1\left(\Lambda\cosh(2|z|)\sinh(2|z|)\right)
        \nonumber\\
        &\simeq
        \frac{\Lambda}{2}\,\cosh(2|z|)\sinh(2|z|)+\mathcal{O}(\Lambda^2)\,.
    \end{align}
\end{subequations}
The $n=0$ harmonic contains the contrast obtained when averaging over
the emission time $t_e$:
\begin{equation}
    \label{eq/examples/squeezing/time-averaged-contrast}
    \overline{\left[X_+^{(\text{dc})}\right]}^{t_e}_{\mathrm{Sq}(z)}=
    \left[X_+^{(\text{dc})}\right]_{\ket{0}}\,\mathcal{F}_0(z,\Lambda)\,.
\end{equation}
Since $|\mathcal{F}_0(z,\Lambda)|<1$, this $t_e$-independent measurement is lower than the
one when only vacuum injected in the radiation channel. It also
represents the average contrast for Levitons of width $\tau_e$ injected
at random emission and it is lower than the vacuum baseline. This is the effect of
the average number of photons in a squeezed vacuum given by
Eq.~\eqref{eq/squeezing/8}.

\section{Single EMP detection}
\label{appendix/single-EMP}

We now discuss the evaluation of the parameter 
$x(t)=2\pi |\braket{\chi|S_{ba}^*\Lambda_t}|^2$
that appears in the evaluation of the Franck-Condon factor
for Fock states (see Eq.~\eqref{eq/targets/Fock/result}). 
The qualitative behavior of this quantity can be understood by using a
time/frequency representation of the current noise. The case of a
narrow band EMP then leads to analytical approximated expressions for $x(t)$.

\subsection{Time frequency analysis of current noise}
\label{appendix/single-EMP/Wigner}

Using $\widehat{\imath}(\omega)
=-e\sqrt{\omega}\,b(\omega)$ ($\omega>0$) for a single chiral integer
quantum Hall edge channel, 
$x(t)=2\pi |\braket{\chi|S_{ba}^*\Lambda_t}|^2$ can be expressed
in terms of the excess current noise $\Delta
S^{(i)}_{\ket{1;\chi}}(t,t')$ of a single quantum EMP in the mode
$\chi$:
\begin{equation}
   \label{eq/targets/Fock/parameter-result}
x(t)=\frac{2\pi^2}{e^2}\int_{\mathbb{R}^2}
   \Gamma_{ba}(t-t_+)\,\Gamma_{ba}(t-t_-)^*\Delta
S^{(i)}_{\ket{1;\chi}}(t_+,t_-)
\md t_+\md t_-\,.
\end{equation}
in which $\Gamma_{ba}$ denotes the function defined by
Eq.~\eqref{eq/FCS/Gamma}.
The physical meaning of this expression is better understood in terms of
the Wigner function $\Delta W_\rho^{(i)}(t,\omega)$ of the excess
current noise
\begin{equation}
   \label{eq/targets/Fock/W_i}
   \Delta W^{(i)}_\rho(t,\omega)=\int_{\mathbb{R}}
   \Big\langle :\widehat{\imath}\left(t-\frac{\tau}{2}\right)
   \widehat{\imath}\left(t+\frac{\tau}{2}\right):
   \Big\rangle_{\rho,c}
   \me^{\mi\omega\tau}
   \md\tau
\end{equation}
in which $\langle A\,B \rangle_c$ denotes the connected correlator
$\langle A\,B\rangle - \langle
A\rangle\langle B\rangle$. Introducing
the Ville transform associated with $\Gamma_{ba}(t)$:
\begin{equation}
   \label{eq/targets/Fock/W_Gamma}
   W_{\Gamma_{ba}}(t,\omega)=\int_{\mathbb{R}}
   \Gamma_{ba}\left(t+\frac{\tau}{2}\right)\,
   \Gamma_{ba}\left(t-\frac{\tau}{2}\right)
   \,\me^{\mi\omega\tau}\md\tau\,.
\end{equation}
leads to
\begin{equation}
   \label{eq/targets/Fock/parameter-Wigner-filtering}
   x(t)=\frac{2\pi^2}{e^2}\int_{\mathbb{R}^2}
   W_{\Gamma_{ba}}(t-\tau,\omega)\,\Delta
   W^{(i)}_{\ket{1;\chi}}(\tau,\omega)\,
   \frac{\md\omega\md \tau}{2\pi}\,.
\end{equation}
This shows that the radiation coupler's response function
$\Gamma_{ba}(\tau)$ leads to time-frequency filtering of
of the excess quantum current noise of the single EMP state
$\ket{1,\chi}$. The time resolution for single EMP detection is thus
limited by the duration of the excess current noise associated with the
single EMP as well as by the response time of the radiation coupler, typically the
$RC$-time scale appearing in $\Gamma_{ba}$ (see Appendix
\ref{appendix/FCS}).

\subsection{Current noise of a single EMP}
\label{appendix/single-EMP/power-spectrum}
   
For the single plasmon with wave-function $\chi$, the average excess
current $\langle i(t)\rangle_{\ket{1;\chi}}$ is zero and
the Wigner function of the excess current noise
defined in Eq.~\eqref{eq/targets/Fock/W_i}
is given by
\begin{subequations}
    \label{eq/1-plasmon/Wigner-noise}
\begin{align} 
\Delta
W^{(i)}_{\ket{1;\chi}}(t,&\omega)=\frac{e^2}{2\pi}\int_{|\Omega|\leq
2|\omega|}
\me^{-\mi
\Omega t}\sqrt{\omega^2-\frac{\Omega^2}{4}}\nonumber\\
&\times
\chi\left(|\omega|+\frac{\Omega}{2}\right)\,\chi\left(|\omega|-\frac{\Omega}{2}\right)^*
\frac{\md\Omega}{2\pi}\,.
\end{align}
\end{subequations}
In the case of a narrowband
single plasmon centered at the energy $\hbar\omega_0$ with Lorentzian linewidth
$\gamma_0$ 
\begin{equation}
    \label{eq/1-plasmon/lorentzian}
   \chi(\omega)=\frac{\sqrt{\gamma_0}\,\Theta(\omega)}{\omega-\omega_0+\frac{\mi\gamma_0}{2}}\,.
\end{equation}
such that $\gamma_0\ll \omega_0$, the Wigner function of the excess
current
noise 
can be approximated by the usual expression for such energy 
resolved excitations:
\begin{equation}
    \label{eq/1-plasmon/lorentzian/Wigner}
\Delta
    W^{(i)}_{\ket{1;\chi}}(t,\omega)\simeq\Theta(t)\,\frac{e^2\omega_0}{2\pi}\,
    4\gamma_0t\,\sinc\left(2(|\omega|-\omega_0)t\right)\,\me^{-\gamma_0 t}\,.
\end{equation}
The general expression of the heat current
$\mathcal{J}_Q(t)$ in terms of the current $i(t)$ leads to
\begin{equation}
    \label{eq/current-Wigner/instantaneous-Joule}
    \langle \mathcal{J}_{Q}(t)\rangle_\rho=\frac{R_K}{2}\left(\langle i(t)\rangle_\rho^2+
    \int_{\mathbb{R}} \Delta
    W^{(i)}_\rho(t,\omega)\,\frac{\md\omega}{2\pi}\right)\,.
\end{equation}
For a 
single energy resolved EMP with energy $\hbar\omega_0$ and
Lorentzian lineshape of width $\gamma_0$, this enables us to compute the
instantaneous heat current just from the excess noise since $\langle
i(t)\rangle_{\ket{1;\chi}}=0$. Using
Eq.~\eqref{eq/1-plasmon/lorentzian/Wigner}, we find:
\begin{equation}
    \label{eq/1-plasmon/lorentzian/heat-current}
    \langle
    \mathcal{J}_Q(t)\rangle_{\ket{1;\chi}}=\hbar\omega_0\Theta(t)
    \,\gamma_0\,\me^{-\gamma_0t}\,.
\end{equation}
This heat current carries an average energy $\int_{\mathbb{R}}\langle
\mathcal{J}_Q(t)\rangle_{\ket{1;\chi}}\md t=\hbar\omega_0$ as
expected.

\subsection{Filtering of current noise}
\label{appendix/single-EMP/iltering}

If the EMP has a narrow band compared to the typical scale of variation
of $\Gamma_{ba}(\omega)$, the variation of $S_{ba}(\omega)$ around
$\omega_0$ can be neglected in Eq.~\eqref{eq/targets/Fock/result/1}. 
One then recognizes in $x(t)$ the modulus square
\begin{equation}
	\frac{e^2}{2\pi}
	\left|
    \int_0^{+\infty}\sqrt{\omega} \chi(\omega)\,\me^{-\mi\omega 
    t}\frac{\md\omega}{\sqrt{2\pi}}\right|^2
	=\int_{\mathbb{R}}\Delta W^{(i)}_{\ket{1;\chi}}(t,\omega)\,\frac{\md\omega}{2\pi}
\end{equation}
where the r.h.s. directly follows from
Eq.~\eqref{eq/1-plasmon/Wigner-noise}.
Eq.~\eqref{eq/current-Wigner/instantaneous-Joule} then shows that this 
is related to the average instantaneous heat current
carried by the single EMP in state $\chi$. This finally leads to
Eq.~\eqref{eq/target/Fock/analytical-expression/b}.



%

\end{document}